\documentclass[12pt]{article}

\usepackage[T2A]{fontenc}
\usepackage[utf8]{inputenc}
\usepackage[english]{babel}
\usepackage{cmap}
\usepackage{indentfirst}

\usepackage{amsmath, amsfonts, amssymb, amsthm, mathtools}
\usepackage{icomma}
\usepackage{bm}

\usepackage{graphicx}
\usepackage{wrapfig}
\usepackage{multirow}
\usepackage{caption}
\usepackage{subcaption}
\graphicspath{{noiseimages/}}

\usepackage{array, tabularx, tabulary, booktabs} 
\usepackage{longtable} 
\usepackage{multirow} 

\usepackage{extsizes}
\usepackage{float} 
\usepackage{geometry}
\usepackage{hyperref}
\usepackage[capitalize,nameinlink]{cleveref}

\crefname{equation}{Eq.}{Eqs.} 
\Crefname{equation}{Equation}{Equations}
\hypersetup{
    colorlinks=true,
    citecolor=black,
    filecolor=black,
    linkcolor=black,
    urlcolor=black
}

\usepackage{xcolor}
\usepackage{titlesec}
\usepackage{authblk}
\usepackage{cite}
\date{}

\renewcommand{\epsilon}{\varepsilon}
\renewcommand{\phi}{\varphi}
\renewcommand{\kappa}{\varkappa}

\newcommand{\<}{\langle}
\renewcommand{\>}{\rangle}

\usepackage{color}
\definecolor{ascol}{rgb}{0.7,0, 0}
\definecolor{tmcol}{rgb}{0, 0.7, 0}

\usepackage{tikz}
\usetikzlibrary{arrows.meta, positioning, fit, calc}

\title{Automated Prediction of Thermodynamic Properties\\
via Bayesian Free-Energy Reconstruction from Molecular Dynamics}

\author[1,2,3]{Ekaterina Spirande\footnote{spirande.ek@phystech.edu}}
\author[3]{Timofei Miryashkin}
\author[1,2]{Andrei Kolmakov}
\author[3,2]{Alexander Shapeev\footnote{a.shapeev@skoltech.ru}}

\affil[1]{Moscow Institute of Physics and Technology, Dolgoprudny, Russia}
\affil[2]{Digital Materials LLC, Odintsovo, Russia}
\affil[3]{Skolkovo Institute of Science and Technology, Moscow, Russia}

\begin{document}

\maketitle

\begin{abstract}
Accurate free-energy calculations are essential for predicting thermodynamic properties and phase stability, but existing methods are limited: phonon-based approaches neglect anharmonicity and liquids, while molecular dynamics (MD) is computationally demanding, neglects low-temperature quantum effects, and often requires manual planning and post-processing of simulations. We present a unified workflow that reconstructs the Helmholtz free-energy surface from MD data using Gaussian Process Regression (GPR), augmented with zero-point energy corrections from harmonic/quasi-harmonic theory. The framework propagates statistical uncertainties, mitigates finite-size effects, and employs active learning to optimize sampling in the volume-temperature space. It applies seamlessly to both crystalline and liquid phases. We demonstrate the methodology by computing heat capacities, thermal expansion, isothermal and adiabatic bulk moduli, and melting properties for nine elemental FCC and BCC metals using 20 classical and machine-learned interatomic potentials, with all predictions accompanied by quantified confidence intervals. Automated, general, and uncertainty-aware, the workflow advances high-throughput thermodynamics and provides a systematic benchmark for interatomic potentials.

\end{abstract}

\section{Introduction}

Recent advances in high-throughput simulation, multi-scale modeling, and artificial intelligence are fundamentally accelerating materials design by enabling rapid prediction, efficient screening, and integrative experiment-theory workflows, driving a paradigm shift toward data-driven materials innovation \cite{intro_1,intro_2,intro_3,intro_4,intro_5}.
Notably, machine-learning interatomic potentials (MLIPs) now make it possible to perform long, finite-temperature molecular dynamics simulations of large systems with \textit{ab initio}-level accuracy, providing the necessary foundation for practical free-energy calculations
~\cite{behler2007generalized, bartok2010gaussian, zhang2018deep, batatia2022mace, yang2024mattersim}. 
These advances have enabled quantitative insights into diffusion \cite{qi2024-diffusion}, phase transformations \cite{Michalchuk2024}, transport \cite{rybin2024thermophysical}, and mechanical properties \cite{mortazavi2021-mech}, bridging the gap between electronic-structure theory and experimentally relevant length and time scales.

Within this broader context, the computation of thermodynamic free energies has emerged as a particularly critical application. Free energy $F(V,T)$ encodes the equilibrium properties of a material, and free energy derivatives yield key quantities such as heat capacity, thermal expansion, and bulk moduli. Accurate free-energy evaluation is essential for predicting phase stability, finite-temperature phase diagrams, and materials performance under operating conditions.

Free energy can be reconstructed from phonon calculations in the harmonic (HA) or quasi-harmonic approximation (QHA)\cite{born1996dynamical,dove1993introduction,wallace1972thermodynamics,baroni2001phonons,parlinski1997first}. These approaches are computationally efficient, provide excellent performance at low temperatures, and are implemented in a wide range of software packages~\cite{otero2011gibbs2a,otero2011gibbs2b,qin2019qha,carrier2007first,erba2015differently,nath2016high}. However, HA and QHA neglect anharmonic contributions~\cite{allen2015anharmonic,allen2020theory,erba2015differently}, which limits their accuracy at elevated temperatures and in systems with soft or unstable modes. Quasiparticle extensions of phonon theory~\cite{allen2015anharmonic,allen2020theory,blancas2024} provide a more accurate description by systematically incorporating higher-order anharmonic effects, and have proven powerful in many cases. Nonetheless, they remain subject to important limitations: (i) they do not capture free-energy contributions from configurational disorder or low-frequency vibrational modes such as hindered rotations, (ii) their reliance on static equilibrium geometries can lead to deviations at high temperature, and (iii) anharmonic effects in zero-point energies cannot be directly included. In addition, phonon-based methods in general are restricted to crystalline solids and cannot be applied to liquids or amorphous phases.

In contrast to phonon-based approaches, classical molecular dynamics (MD) naturally captures anharmonic effects and can be applied to any phase. In MD-based workflows, the free energy $F(V,T)$ is typically obtained by sampling at multiple temperatures and volumes, followed by thermodynamic integration along a chosen path~\cite{kirkwood1935statistical,frenkel2023understanding,tuckerman2023statistical,chipot2007free,torrie1977nonphysical,kumar1992weighted,kastner2011umbrella,kone2005selection,abrams2013enhanced}. A central limitation of classical MD is its treatment of nuclei as classical particles, neglecting quantum effects that play a key role in low-temperature thermodynamics. In addition, the accuracy of thermodynamic integration depends sensitively on the placement and density of sampling points $(V,T)$, which often necessitates manual convergence testing and parameter tuning.

In this work, we address the limitations of both phonon- and MD-based approaches by introducing a workflow that captures anharmonic effects while retaining low-temperature quantum accuracy. Specifically, we reconstruct the free-energy surface $F(V,T)$ from irregularly sampled MD trajectories using Gaussian Process Regression (GPR)~\cite{miryashkin2023bayesian}, and incorporate quantum effects through a zero-point energy (ZPE) correction derived from HA/QHA. The GPR framework propagates statistical uncertainties from MD sampling into predicted thermodynamic properties, thereby reducing systematic errors and mitigating finite-size effects. Because the model operates on irregularly spaced samples, MD simulations can be performed at arbitrary state points rather than on a fixed $(V,T)$ grid. This flexibility also enables direct treatment of liquids and other disordered phases, which are inaccessible to phonon-based methods. Finally, we implement an active learning strategy that adaptively selects new $(V,T)$ points, rendering the workflow fully automated and highly efficient.

We demonstrate the workflow by computing thermodynamic properties of nine elemental FCC and BCC metals across the full temperature range, employing a total of 20 interatomic potentials (EAM, MEAM, and MTP). For each system, we evaluate heat capacity, linear thermal expansion, isothermal and adiabatic bulk moduli, as well as melting properties, including the enthalpy of fusion and the volume change upon melting, all with quantified uncertainties.

Because our approach is fully automated and transferable across different systems, it aligns naturally with broader efforts toward automation in materials design. Platforms such as AiiDA~\cite{huber2020aiida,muy2023aiida}, atomate/atomate2~\cite{mathew2017atomate,ganose2025atomate2}, FireWorks~\cite{jain2015fireworks}, and pyiron~\cite{pyiron-paper} provide general workflow automation, while initiatives such as the JARVIS-Leaderboard and Matbench Discovery enable large-scale benchmarking of models and methods across diverse properties~\cite{choudhary2024jarvis,riebesell2023matbench,riebesell2025framework}. In parallel, the OpenKIM project offers an open repository of interatomic potentials together with a framework for standardized testing and evaluation~\cite{tadmor2011potential,karls2020openkim}. Within this landscape, our workflow plays a complementary role: it delivers thermodynamic properties automatically, with quantified uncertainties, and can serve as a validation tool by systematically comparing predictions from interatomic potentials against reference calculations or experimental data.

The remainder of this article is organized as follows. Section~\ref{sec:methodology} introduces the methodology for reconstructing the free-energy surface and extracting thermodynamic properties. Section~\ref{sec:results_aluminum} illustrates the workflow using aluminum as a representative case, while Section~\ref{sec:results_all} extends the analysis to a broader set of cubic metals with different interatomic potentials. Section~\ref{sec:limitations} discusses the limitations of the employed interatomic models, and Section~\ref{sec:conclusion} concludes with a summary and outlook.

\section{Methodology}
\label{sec:methodology}

Here we present the developed methodology for reconstructing the Helmholtz free-energy surface and obtaining thermodynamic properties such as heat capacity, thermal expansion, and bulk moduli from the results of molecular dynamics simulations.
Our approach follows the general strategy previously applied for phase diagram reconstruction~\cite{Ladygin,miryashkin2023bayesian}, and therefore we provide only a concise description here, while detailed derivations of the relevant expressions are given in the Supplementary Information (Section S3).

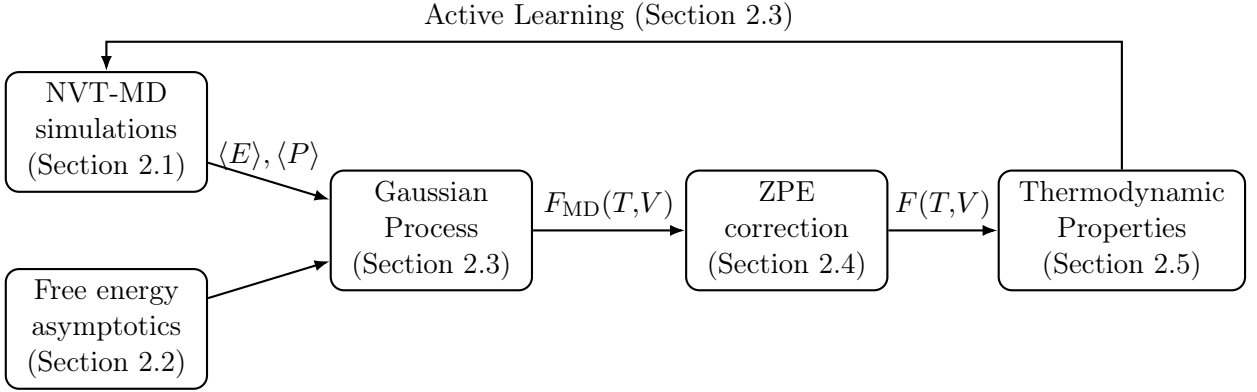
\begin{figure}[H]
\centering
\begin{tikzpicture}[
  block/.style={
    rectangle, draw=black, thick,
    minimum height=1.2cm,
    minimum width=2.4cm,
    align=center,
    font=\small,
    text width=2.4cm,
    rounded corners=5pt
  },
  blockwide/.style={
    block,
    minimum width=3.0cm,  
    text width=3.0cm      
  },
  arrow/.style={-{Latex[length=2mm]}, thick}
]

\node[block] (md) {NVT-MD\\ simulations\\ (Section \ref{sec:free_energy})};
\node[block, below=of md] (asympt) {Free energy\\ asymptotics\\ (Section \ref{sec:f_asymptotics})};

\coordinate (midL) at ($(md)!0.5!(asympt)$);

\def\xsep{4.5cm}

\node[block]     (gpr) at ($(midL)+(0.96*\xsep,0)$) {Gaussian\\ Process\\ (Section \ref{subsec:gpr})};
\node[block]     (zpe) at ($(midL)+(2.01*\xsep,0)$) {ZPE \\correction\\ (Section \ref{subsec:zpe})};
\node[blockwide] (td)  at ($(midL)+(3*\xsep,0)$) {Thermodynamic\\ Properties\\ (Section \ref{subsec:td_properties})};

\node[above=2.5cm of gpr, align=center, font=\small] (al) at ($(gpr)!0.5!(zpe)$) {Active Learning (Section \ref{subsec:gpr})};

\draw[arrow] (md) -- node[above, font=\small]{$\langle E \rangle, \langle P \rangle$} (gpr);
\draw[arrow] (asympt) -- node[above, font=\small]{} (gpr);
\draw[arrow] (gpr) -- node[above, font=\small]{$F_{\mathrm{MD}}(T,V)$} (zpe);
\draw[arrow] (zpe) -- node[above, font=\small]{$F(T,V)$} (td);

\draw[arrow] (td.north) |- ++(0,1.7) -| (md.north);

\end{tikzpicture}
\caption{
Workflow for computing thermodynamic properties. MD simulations at each $(T, V)$ point provide mean energy $\langle E \rangle$ and pressure $\langle P \rangle$, which are fitted using Gaussian Process Regression (accounting for free energy asymptotics). The resulting $F_{\text{MD}}(T, V)$ is corrected for zero-point energy, and thermodynamic properties follow from its derivatives. Our Bayesian approach also allows us to optimally select the next simulation point via active learning.
}
\label{fig:workflow}
\end{figure}

The central idea is that from NVT-MD simulations (one or several trajectories at given $(V,T)$ points), we extract ensemble-averaged potential energies and pressures. These averages provide direct access to derivatives of the Helmholtz free energy $F(V,T)$, which is reconstructed using Gaussian Process Regression. Once $F(V,T)$ is available, all target thermodynamic properties are obtained from its derivatives, with uncertainties propagated consistently through the GPR framework. The details of our workflow are provided below, and an illustrative summary is shown in Fig.~\ref{fig:workflow}.

In addition, we develop a simplified NPT-based workflow in which the system is simulated at zero external pressure while varying only the temperature. This computationally efficient protocol yields a subset of properties, including the constant-pressure heat capacity and melting-point properties.

\subsection{Free Energy}
\label{sec:free_energy}

We define the Helmholtz free energy as 
\begin{equation}
\label{eq:free_energy_def}
    \hat{F} = -T \log \int_{\hat{V}^N} \exp\left(-\frac{\hat{E}(x)}{T}\right) dx,
\end{equation}
where $\hat{E}(x)$ is the total potential energy of a configuration $x$ consisting of $N$ atoms within volume $\hat{V}$. 
We set the Boltzmann constant $k_{\mathrm B}=1$, so the temperature $T$ is expressed in electron-volts (eV).
Here and in what follows by $\hat{\vphantom{t}\bullet}$ we denote extensive quantities; thus, for example, per-atom free energy is $F = \hat{F} / N $.

The average of any quantity $\hat{f}$ over the NVT ensemble is given by:
\begin{equation*}
\label{eq:average_nvt}
\langle \hat{f} \rangle = \frac{\int_{\hat{V}^N} \hat{f} \exp{(-\hat{E}(x)/T) \,{\rm d}x}}{{\int_{\hat{V}^N} \exp{(-\hat{E}(x)/T) \,{\rm d}x}}}.
\end{equation*}

Differentiating the free energy in \eqref{eq:free_energy_def} with respect to volume and temperature, and using the definition of the ensemble average, we obtain the following expressions for the derivatives:

\begin{equation*}
    \frac{\partial F}{\partial V} = \< P \>,
    \qquad
    \frac{\partial (F/T)}{\partial T} = -\frac{\< E\>}{T^2},
\end{equation*}
where $P$ is the virial pressure. The averages over $E$ and $P$ are estimated by sampling the NVT ensemble. We reconstruct the free energy by integrating it from derivatives via Gaussian Process Regression algorithm. In the subsequent sections, we first describe how the physical knowledge is incorporated into the fitting and then describe the GPR in detail.  

\subsection{Free Energy Asymptotics}
\label{sec:f_asymptotics}

We include physics in our model by utilizing the reference free energy $F_\text{ref}(T, V)$ which expresses the asymptotic limit for the free energy as $T \rightarrow 0$. The reference free energy improves the fit, as subtracting it from $F(T, V, N)$ removes the divergent terms, for example, $T\log(N V)$ as $N \rightarrow \infty$. We rewrite the equation \eqref{eq:free_energy_def}:
\begin{equation}
\label{eq:free_energ_ref}
    F(T, V, N) = F_{\text{ref}}(T, V, N) - T \hat{S}(T, V, N)\,
\end{equation}
where the newly introduced quantity $S$ is called the \emph{entropy}.
There exist different ``entropies'' (conventional entropy, excess entropy) depending on the choice of $F_{\text{ref}}(T, V, N)$.
In this work we choose $F_{\text{ref}}(T, V, N)$ for the convenience of using GPR, and still call $S$ the entropy.

We adopt the reference free energy from Ladygin~\textit{et al.}~\cite{Ladygin}; the detailed formulas are given in the Supplementary Information (Section S1).
We thus reduce the problem of reconstructing free energy (with its divergent terms) to reconstructing the entropy from its derivatives over $T$ and $V$.

\subsection{Gaussian Process Regression}
\label{subsec:gpr}

We reconstruct the entropy of each phase using Gaussian Process Regression (GPR) \cite{rasmussen2006-book-gaussian}, which naturally incorporates the uncertainty in the input data. As the simulation data are approximately normally distributed, the Gaussian process provides an appropriate probabilistic model for the entropy.

In GPR, the entropy $ S(T, V, N) $ is modeled as a Gaussian process, where the covariance between two input points $(T_1, V_1, N_1)$ and $(T_2, V_2, N_2)$ is defined via a kernel function:
\[
	\begin{array}{r}
k[(T_1, V_1, N_1), (T_2, V_2, N_2)]
\hspace{12em}\mathstrut
\\
:= {\rm Cov}[S(T_1, V_1, N_1), S(T_2, V_2, N_2)].
\end{array}
\]
Assuming that the entropy is a smooth function of $T$, $V$, and $N^{-1}$, we adopt a squared exponential kernel. For the solid phase, we use:
\begin{align}
\label{eq:cryst_ker}
\begin{split}
    k_{\rm cryst}&:=  \theta_0^2 + \theta_f^2  
    \exp \left( -\frac{(T_1 - T_2)^2}{2 \theta_T^2}\right) 
    \exp \left( -\frac{(V_1 - V_2)^2}{2 \theta_V^2} \right) 
    \\&\phantom{:=\mathstrut} 
    \cdot \exp  \left( -\left(\frac{1}{N_1} - \frac{1}{N_2} \right)^2 \frac{\theta_N^2}{2} \right),
\end{split}
\end{align}
where $\theta_i$ are learnable hyperparameters. Notably, this kernel is constructed to remain well-defined in the limit $N \to \infty$, enabling extrapolation to the thermodynamic limit based on simulations at finite sizes.

For the liquid, it is important to explicitly take into account that
the energy is defined up to an additive constant (an arbitrary energy shift).  Therefore we allow the entropy to have a linear shift, $S \sim \theta_1/T$, which features the last term in our kernel for the liquid phase:
\begin{align*}
\label{eq:liq_ker}
    k_{\rm liq} &:=  \theta_0^2 + \theta_f^2  
    \exp \left( -\frac{(T_1 - T_2)^2}{2 \theta_T^2}\right) 
    \exp \left( -\frac{(V_1 - V_2)^2}{2 \theta_V^2} \right)  
\\ &\mathstrut\phantom{:=\mathstrut}\mathstrut
    \cdot \exp  \left( -\left(\frac{1}{N_1} - \frac{1}{N_2} \right)^2 \frac{\theta_N^2}{2} \right)
	+ \frac{\theta_{1}^2}{T_1 T_2}.
\end{align*}

A key characteristic of Gaussian processes is that any linear functional of a Gaussian process is also normally distributed. This is essential in our case, as the input data include not only entropy values but also their derivatives. For instance, the covariance between the entropy and its temperature derivative is given by:
\[
{\rm Cov}\left[\frac{\partial }{\partial T_1} S(T_1, V_1), S(T_2, V_2)\right] = \frac{\partial }{\partial T_1} k[(T_1, V_1), (T_2, V_2)].
\]
This allows us to use the full power of Gaussian Processes and reconstruct the free energy together with its derivatives (e.g., heat capacity) from ensemble-averaged energy and pressure, and moreover estimate the predictive uncertainly of any quantity.
In the limit of ``big data'', when uncertainties are small, we can linearize nonlinear functionals (like equilibrium volume at a given temperature) and apply the GPR machinery to them as well.

We follow Rasmussen \cite{rasmussen2006-book-gaussian} and make regression predictions as detailed in the Supplementary Information (Section S2). The hyperparameters of the algorithm, such as ${\bm \theta} = (\theta_0, \theta_f, \theta_T, \theta_V, \theta_N)$ in \eqref{eq:cryst_ker}, are systematically optimized by maximizing the marginal likelihood. 

\subsection{Zero Point Energy Correction}
\label{subsec:zpe}

In this work, we go beyond the classical MD in order to account for quantum contributions to nuclear motion, which become significant at low temperatures. To that end, we apply a correction using the quasi-harmonic approximation (QHA). 
This correction is applied only to the solid phase, since the QHA is meaningful only for crystals. For the systems with relatively heavy atoms and high melting points considered in this work, quantum effects in the liquid phase are expected to be negligible. Although the applied zero-point energy correction represents an approximate treatment of nuclear quantum effects, it captures their dominant contribution to the vibrational free energy of solids, at least at low temperatures.

We obtain a corrected free energy $F(V,T)$ by combining the MD-based free energy with QHA results. In essence, we replace the purely classical vibrational contribution with a quantum vibrational contribution. Formally, we write the corrected free energy as
\begin{equation}
\label{eq:fe_correction}
    F = F_{\text{MD}} - F^{\text{QHA}}_{\text{cl}} + F^{\text{QHA}}_{\text{qm}},
\end{equation}
where $F_{\text{MD}}$ is the classical MD free energy (including all anharmonic effects), $F_{\text{cl}}^{\text{QHA}}$ is the classical free energy in the quasi-harmonic approximation, and $F_{\text{qm}}^{\text{QHA}}$ is the vibrational free energy within QHA (including zero-point energy). 

The MD term $F_{\text{MD}}(V,T)$ is reconstructed via GPR from MD simulations as described in Section \ref{subsec:gpr}.
The classical QHA term is given by
$$
F^{\text{QHA}}_{\text{cl}}(V,T) = E_0(V) - {\textstyle\frac{3}{2}} T \log(2\pi T) + \frac{T}{2N} \log \det H(V),
$$
where $E_0(V)$ is the potential energy at zero temperature for the given volume, and $H(V)$ is the Hessian matrix of the potential energy (evaluated at the equilibrium configuration for each volume). The determinant of the Hessian can be expressed through the vibrational frequencies as
${\textstyle\frac{1}{2}}\log \det H(V) = \sum_i \log(\sqrt{m_i},\omega_i(V))$. 
Here, we once again utilize the GPR machinery to fit the volume dependence of $E_0(V)$ and $\log \det H(V)$. Once fitted, $F^{\text{QHA}}_{\text{cl}}$ is represented by a Gaussian process and its uncertainty is further propagated to the free energy $F$.

Finally, the quantum QHA term is
$$
F^{\text{QHA}}_{\text{qm}}(V,T) = E_0(V) + \sum_{\mathbf{k}, p} \left( \tfrac{1}{2}\hbar \omega_{\mathbf{k}p}(V) + T \ln \left( 1 - e^{-\frac{\hbar \omega_{\mathbf{k} p}(V)}{T}} \right) \right),
$$
where $\omega_{kp}(V)$ are the phonon frequencies for the mode index $p$ in $k$-space. In the $T\to0$ limit, $F_{\text{vib}}$ approaches the total zero-point energy of the crystal, and at higher $T$ it incorporates the quantum statistics of phonons. By subtracting $F_{\text{cl}}^{\text{QHA}}$ and adding $F_{\text{ZPE}}^{\text{QHA}}$, we correct the MD free energy to include quantum zero-point contributions while retaining the classical anharmonic effects. All further analysis of the solid phase uses the ZPE-corrected free energy.

The motivation behind \eqref{eq:fe_correction} is that $F_{\text{MD}}$ is accurate at large temperatures, while at small temperatures it is close to $F^{\text{QHA}}_{\text{cl}}$ and hence \eqref{eq:fe_correction} restores\footnote{This idea was communicated to the last author by J\"org Neugebauer} the correct asympotics at $T \to 0$.

\subsection{Thermodynamic Properties}
\label{subsec:td_properties}

After reconstructing the Helmholtz free-energy surface $F(V,T)$ via GPR (and applying the QHA correction for solids), we can analytically compute all the desired thermodynamic properties from the derivatives of $F(V, T)$. 
In this work, we focus on properties at zero external pressure (the $P = 0$ case), meaning that at each temperature $T$ the system volume is taken to minimize $F(V,T)$ (defining the equilibrium volume $V_{\mathrm{eq}}(T)$). 

Below, we summarize how each property is obtained from the free energy derivatives. 
The detailed derivations of these relations are provided in the Supplementary Information (Section S3), and the corresponding expressions for uncertainty propagation are given in Section S4.

\begin{itemize}
    \item \textit{Equilibrium volume} at a given temperature, $V_{\text{eq}}(T)$, is obtained by minimizing the free energy with respect to $V$: 
    \begin{equation}
    \label{eq:eqv_condition}
    \left.\frac{\partial F}{\partial V}\right|_{V = V_{\text{eq}}(T)} = 0;
    \end{equation}
    
    \item Melting properties are determined as the difference of between thermodynamic quantities in solid and liquid phases. The melting point temperature is computed with confidence intervals by the coexistence algorithm introduced by Klimanova~\textit{et al.}~\cite{meltingpoint}.
    
    \begin{itemize}
        \item \textit{Volume change upon melting} is defined as the difference between the equilibrium volumes of the solid and liquid phases:
        \begin{equation}
        \label{eq:delta_v}
        \Delta V = V_{\rm liq}(T_{\rm melt}) - V_{\rm sol}(T_{\rm melt});
        \end{equation}
        \item \textit{Enthalpy of fusion} (melting enthalpy change):
        \begin{equation}
        \label{eq:delta_h}
        \Delta H = H_{\rm liq}(T_{\rm melt}) - H_{\rm sol}(T_{\rm melt});
        \end{equation}
    \end{itemize}
    
    \item \textit{Linear thermal expansion coefficient} for an isotropic material:
    \begin{equation}
    \label{eq:alpha}
    \alpha(T) = \frac{1}{3V_\mathrm{eq}(T)} \left( \frac{\partial V_{\mathrm{eq}}(T)}{\partial T} \right);
    \end{equation}
    
    \item \textit{Constant-volume heat capacity}: 
    \begin{equation}
    \label{eq:cv}
    C_V(T) = -T \left( \frac{\partial^2 F}{\partial T^2} \right)_{V=V_{\mathrm eq}};
    \end{equation}
    
    \item \textit{Constant-pressure heat capacity} at zero pressure:
    \begin{equation}
    \label{eq:cp}
        C_{p}(T) = C_V(T) - T \frac{\partial^{2} F}{\partial T\,\partial V} (T, V_{\text{eq}}(T)) \cdot
        \frac{\partial V_{{\rm eq}}}{\partial T} (T).
    \end{equation}

    \item \textit{Isothermal bulk modulus} at zero pressure:
    \begin{equation}
    \label{eq:kt}
    K_T(T) = V_{{\rm eq}}(T) \left. \frac{\partial^2 F}{\partial V^2} \right|_{V = V{{\rm eq}}(T)};
    \end{equation}

    \item \textit{Adiabatic bulk modulus}:
    \begin{equation}
    \label{eq:ks}
    K_S(T) = \left. K_T(T) \cdot \frac{C_P(T)}{C_V(T)} \right|_{V = V{{\rm eq}}(T)};
    \end{equation}

\end{itemize}

\subsection{Methodology for NPT Ensemble}
\label{sec:npt_methodology}

In addition to the NVT approach, we also introduce an NPT-based framework. In the NPT approach, we do not reconstruct the free-energy surface $F(V,T)$; instead, molecular dynamics is performed directly at zero external pressure (i.e., at the equilibrium volume), and ensemble-averaged quantities $\langle E\rangle$ and $\langle V\rangle$ are collected as functions of temperature. While the NVT approach requires simulations over both volume and temperature to extract all thermodynamic quantities from $F(V,T)$ and its derivatives, the NPT approach varies only the temperature, making it significantly more computationally efficient.

Compared to the NVT case, the GPR model is simpler because its kernel has no dependence on volume. 
We perform MD simulations at \(P = 0\) for various temperatures and system sizes \(N\), collecting the equilibrium volume \(\langle V \rangle\) and energy \(\langle E \rangle\). 
These quantities are fitted with GPR as functions of \((T, N)\), and the thermodynamic limit \(N \rightarrow \infty\) is then taken. 
For the energy, this yields the enthalpy \(H(T) = \langle E \rangle(T, N \rightarrow \infty)\).

Melting-point quantities --- such as the volume change \(\Delta V_{\rm fus}\) and the enthalpy of fusion \(\Delta H_{\rm fus}\) --- are determined from the discontinuities in the equilibrium volume and enthalpy at the melting temperature \(T_{\rm melt}\). 
The constant-pressure heat capacity is computed as the temperature derivative of the enthalpy, \(C_P = dH(T)/dT\).

A zero-point energy (ZPE) correction can still be applied via the harmonic approximation, enabling accurate prediction of the quantum-corrected heat capacity in the solid phase at low temperatures. 
This correction is applied by replacing the classical harmonic contribution $3k_\mathrm{B}$ in the MD-derived $C_P$ with the value obtained from phonon calculations. 
However, in this framework, there is no straightforward way to correct thermal expansion and bulk moduli, as the harmonic approximation yields zero contribution by construction.

Thus, the NPT approach provides a fast and reliable alternative for selected thermodynamic properties --- melting-related quantities and $C_P$ --- particularly when computational resources are limited or when fast screening is required. In contrast, the NVT approach enables full thermodynamic reconstruction, including quantum-corrected thermal expansion and bulk modulus, at the cost of a more expensive MD dataset.

\section{Application to Aluminum}
\label{sec:results_aluminum}

In order to demonstrate how our pipeline works, we first applied the methodology described in Section~\ref{sec:methodology} to face-centered cubic (FCC) aluminum using the semi-empirical embedded atom model (EAM) potential of Zope and Mishin~\cite{zope2003interatomic}. Calculations were performed in two statistical ensembles: 
the canonical NVT ensemble, in which the Helmholtz free energy $F(V,T)$ is reconstructed from MD data using Gaussian Process Regression, 
and the isothermal-isobaric NPT ensemble, in which thermodynamic properties are extracted directly from the temperature dependence of equilibrium observables. We emphasize that the free energy, along with all derived properties, is computed in the infinite-atom limit ($N\to\infty$).

For each ensemble, we considered two versions of the results: 
the purely classical predictions based on the MD-derived free energy or observables; 
and the predictions with a ZPE correction, in which the solid phase is adjusted using the quasi-harmonic or harmonic approximation.
This dual evaluation enables a direct quantification of the impact of the ZPE correction.

In the NVT ensemble, the full set of thermodynamic properties is accessible: melting-point discontinuities in enthalpy $\Delta H_{\mathrm{fus}}$ and volume $\Delta V_{\mathrm{fus}}$, linear thermal expansion coefficient $\alpha(T)$, constant-pressure heat capacity $C_P(T)$, and isothermal and adiabatic bulk moduli $K_T(T)$ and $K_S(T)$.
In the NPT ensemble, only the melting-point properties ($\Delta H_{\mathrm{fus}}$, $\Delta V_{\mathrm{fus}}$) and the heat capacity $C_P(T)$ can be obtained, since the volume derivatives of the free energy required for $\alpha(T)$ and $K_T(T)$ are not directly accessible.

All results are compared against experimental reference data, allowing both validation of the methodology and accuracy assessment of the chosen interatomic potential. The following subsections present the computational workflow and results for the NVT and NPT ensembles, followed by a comparison of the two approaches in terms of accuracy and computational efficiency.

\label{sec:results}
\subsection{Workflow in NVT Ensemble}

In the NVT ensemble, molecular dynamics simulations were performed for both solid and liquid aluminum over a range of volumes and temperatures using LAMMPS~\cite{LAMMPS}.
For each state point $(V,T)$, the average internal energy $\langle E \rangle$ and pressure $\langle P \rangle$ were obtained by averaging over multiple independent trajectories to reduce statistical noise and estimate standard errors.
To assess and remove finite-size effects, simulations were carried out for several system sizes ($N = 128$, 256, 864, 2048 atoms), and the GPR kernel incorporated an explicit $N^{-1}$ dependence \eqref{eq:cryst_ker} to extrapolate to the thermodynamic limit.

The Helmholtz free energy $F(V,T)$ was reconstructed from these data using the GPR procedure described in Section~\ref{subsec:gpr}, yielding a smooth, differentiable surface with uncertainty estimates for all derived quantities.
The quasi-harmonic ZPE correction (Section~\ref{subsec:zpe}) was then applied to the free energy for the solid phase.

From the classical and ZPE-corrected free energies, we computed $\Delta H_{\mathrm{fus}}$, $\Delta V_{\mathrm{fus}}$, $\alpha(T)$, $C_P(T)$, $K_T(T)$, and $K_S(T)$ over the relevant temperature range.
In the plots, results from the classical free energy are shown as \textit{Classical MD} (blue), while those with the solid-phase ZPE correction are labeled \textit{Classical MD + ZPE correction} (purple).

\subsubsection{Equilibrium Volume}

The equilibrium volume $V_{\mathrm{eq}}(T)$ at each temperature was obtained by minimizing the Helmholtz free energy $F(V,T)$ with respect to volume \eqref{eq:eqv_condition}. This procedure was carried out separately for the classical and the ZPE-corrected versions of the free energy.

\begin{figure}[H]
    \centering
    \includegraphics[width=0.75\linewidth]{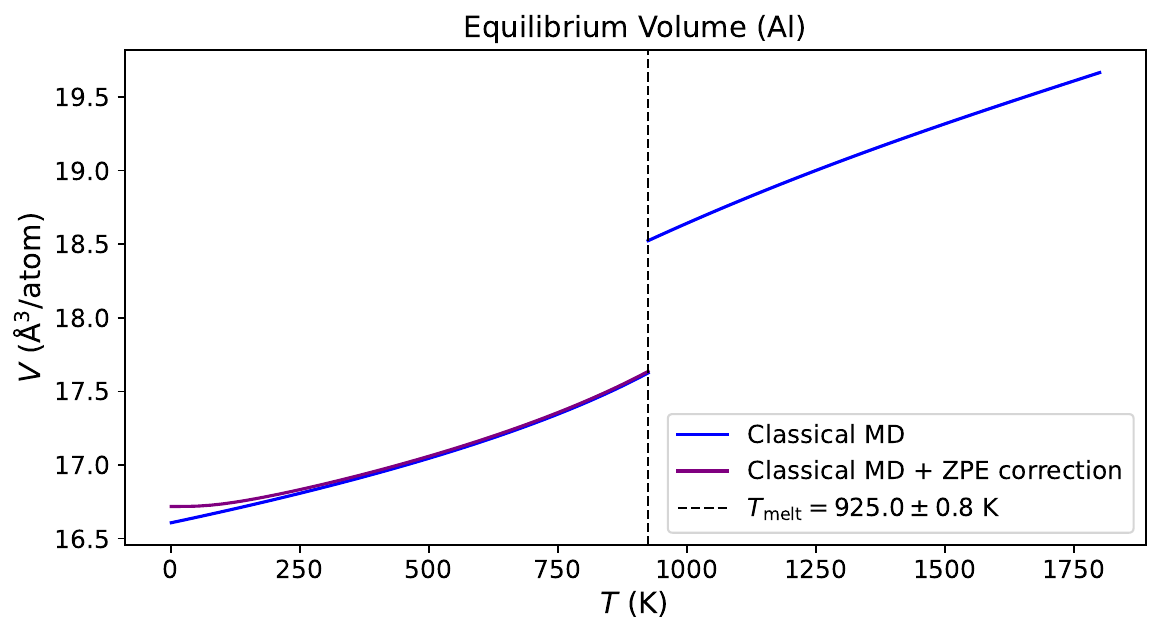}
    \caption{Equilibrium volume for FCC aluminum as a function of temperature, with and without the ZPE correction. Values are shown with $2\sigma$ confidence intervals obtained from the GPR model; for $V_{\mathrm{eq}}(T)$ these intervals are of the order $10^{-3}$ and therefore not visible.}
    \label{fig:vol_zpe}
\end{figure}

Figure~\ref{fig:vol_zpe} shows the temperature dependence of the equilibrium volume. At high temperatures, both classical and corrected results exhibit close agreement, as quantum effects become negligible. At low temperatures ($T < 300$~K), the ZPE-corrected equilibrium volume remains nearly constant, in agreement with the physical expectation of negligible thermal expansion in the quantum regime. In contrast, the classical MD curve shows a decrease in volume as $T \to 0$, which is corrected by the inclusion of quantum effects.

\subsubsection{Melting-Point Properties}
\label{sec:results_melting_nvt}

The melting point of aluminum was determined using the solid-liquid coexistence method~\cite{meltingpoint}, yielding a melting temperature 
$T_{\mathrm{melt}} = 925.0 \pm 0.8$~K.  
At $T_{\mathrm{melt}}$, we evaluated the GPR-derived free energies for the solid and liquid phases to compute the enthalpy of fusion $\Delta H_{\mathrm{fus}}$ and the volume change upon fusion $\Delta V_{\mathrm{fus}}$, propagating uncertainties from the GPR predictions as well as from the melting temperature and the equilibrium volume at that temperature.  
All melting-point properties were evaluated using the uncorrected classical free energies of both phases, as nuclear quantum effects should not be strong at melting temperatures, and their contributions to the free energy, enthalpy, and equilibrium volume may partially cancel out.

The results are as follows:  
\[
\begin{aligned}
\Delta H_{\rm fus} &= 92.4 \pm 0.3\ \text{meV/atom}, \\
\Delta V_{\rm fus} &= 0.888 \pm 0.003\ \text{\AA}^3/\text{atom}.
\end{aligned}
\]

For comparison, the recommended experimental values (at the measured $T_{\mathrm{melt}} = 933.6$~K) are $\Delta H_{\text{fus}} = 109.7 \pm 1.6$~meV/atom from Desai~\cite{desai1987aluminium} and $\Delta V_{\text{fus}} \approx 1.21$~\AA$^3$/atom from Saeger Jr.~\textit{et~al.}~\cite{saeger1932method}.  
The computed $\Delta H_{\mathrm{fus}}$ is lower than the experimental value by $\approx 17.3$~meV/atom, corresponding to about $10\sigma$ given the combined uncertainties. The EAM potential of Zope and Mishin~\cite{zope2003interatomic} thus predicts a melting temperature for aluminum that is lower than experiment, which in turn leads to underestimated enthalpy and volume changes at melting relative to the measured values.

\subsubsection{Thermal Expansion}

We computed the linear thermal expansion coefficient $\alpha(T)$ both in the purely classical case and with the ZPE correction applied to the solid phase.  
The coefficient was obtained directly from the final expression in \eqref{eq:alpha}, with uncertainties estimated from the GPR prediction variance.

\begin{figure}[H]
    \centering
    \includegraphics[width=0.75\linewidth]{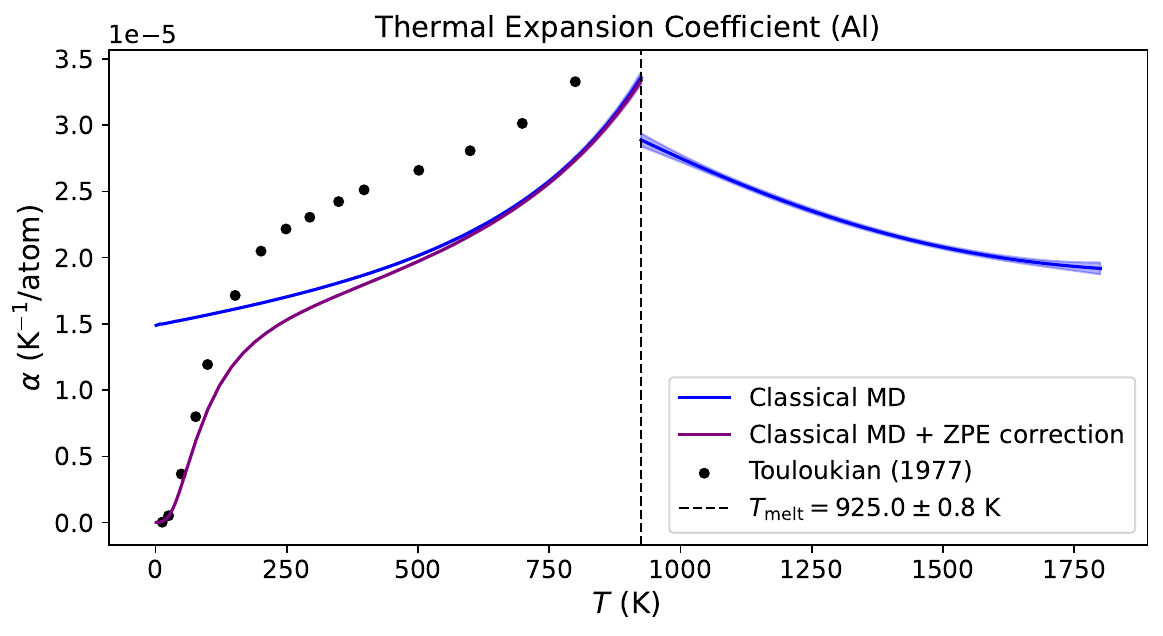}
    \caption{Linear thermal expansion coefficient for FCC aluminum. Classical MD predictions and ZPE-corrected values for the solid phase are compared with experimental data from \cite{touloukian1977_alpha}. The shaded region indicates the $2\sigma$ confidence interval of the predictions.}
    \label{fig:alpha}
\end{figure}

The resulting values are shown in Fig.~\ref{fig:alpha}.  
The ZPE-corrected results are in a qualitative agreement with the experimental data of Touloukian~\textit{et~al.} \cite{touloukian1977_alpha}, reproducing the expected near-zero thermal expansion in the quantum regime, but underestimating the experimental values.

In their original study, Zope and Mishin reported that for this EAM potential, calculations reproduce experimental $\alpha(T)$ at low temperatures, while at high temperatures the values are slightly lower than experiment \cite{zope2003interatomic}.  
Our results, obtained independently using the present methodology, confirm this trend: we also find close agreement with experiment in the low-$T$ region and an underestimation of $\alpha(T)$ in the high-temperature regime.

\subsubsection{Heat Capacity}

\begin{figure}[H]
    \centering
    \includegraphics[width=0.8\linewidth]{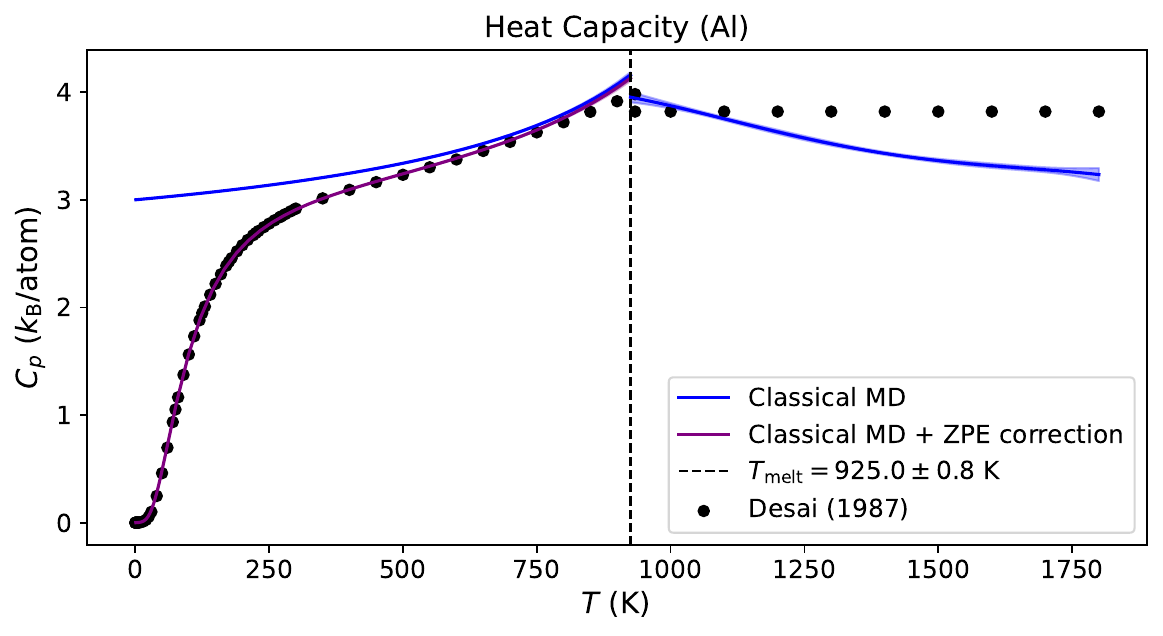}
    \caption{Temperature dependence of the constant-pressure heat capacity for FCC aluminum. Classical MD prediction (blue) and ZPE-corrected values (purple) for the solid phase are compared with experimental data from \cite{desai1987aluminium}. The shaded region indicates the $2\sigma$ confidence interval of the predictions.}
    \label{fig:cp}
\end{figure}

In the NVT ensemble, the constant-pressure heat capacity $C_P(T)$ was obtained from the reconstructed free energy at the equilibrium volume $V_{\mathrm{eq}}(T)$ using \eqref{eq:cp}.  
The results are shown in Fig.~\ref{fig:cp}.  
In the classical case, $C_P$ tends to the value $3k_\mathrm{B}$ per atom as $T \to 0$, as expected for a purely classical model.  
Including the ZPE correction for the solid phase removes this unphysical low-temperature limit and yields predictions in very good agreement with the experimental data of Desai~\cite{desai1987aluminium} across the solid region.  

For the liquid phase, the recommended experimental values of $C_P$ remain approximately constant with temperature, whereas our results show a gradual decrease at higher $T$.  

\subsubsection{Bulk Modulus}

\begin{figure}[H]
\centering
\includegraphics[width=0.8\linewidth]{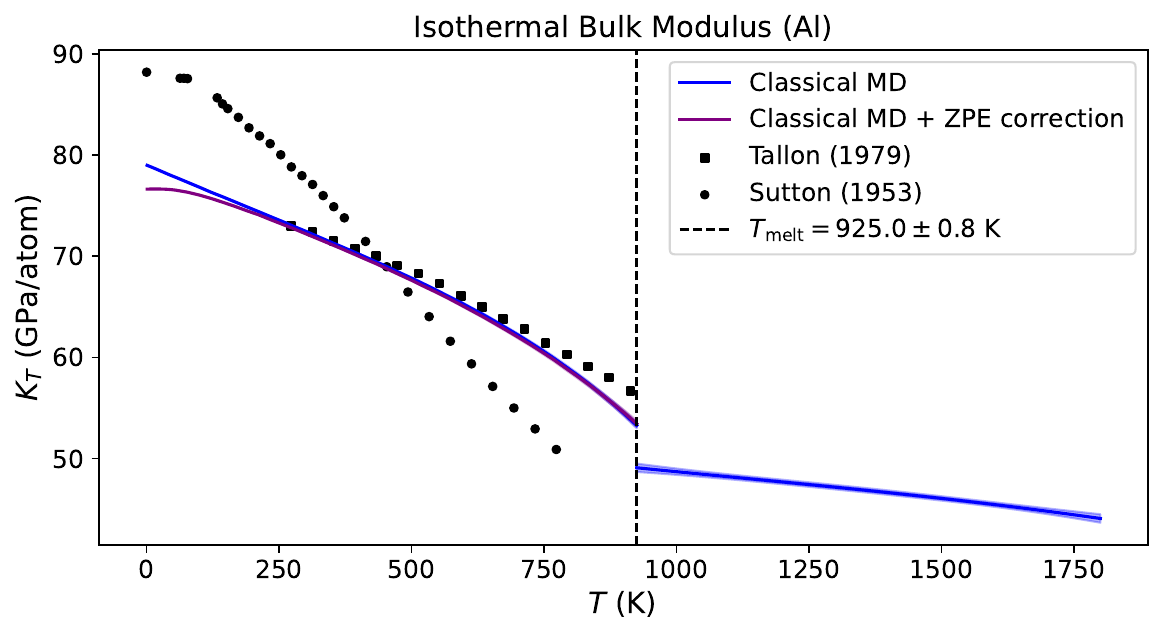}
\caption{Isothermal bulk modulus $K_T(T)$ of FCC aluminum. Classical MD predictions (blue) and ZPE-corrected values (purple) are compared with experimental data from \cite{tallon1979temperature, Sutton}. The shaded region indicates the $2\sigma$ confidence interval.}
\label{fig:bulk_isoth}
\end{figure}

The isothermal bulk modulus $K_T(T)$ was obtained using \eqref{eq:kt} as the second derivative of the Helmholtz free energy with respect to volume, evaluated at the equilibrium volume $V_{\mathrm{eq}}(T)$.
The resulting $K_T(T)$ is shown in Fig.~\ref{fig:bulk_isoth}.  
The ZPE-corrected results reproduce the general experimental trend: the modulus remains nearly constant at low temperatures before decreasing with increasing $T$.  
In the classical case, $K_T$ decreases monotonically from the lowest temperatures.  
At high temperatures, both predictions approach each other.

\begin{figure}[H]
\centering
\includegraphics[width=0.8\linewidth]{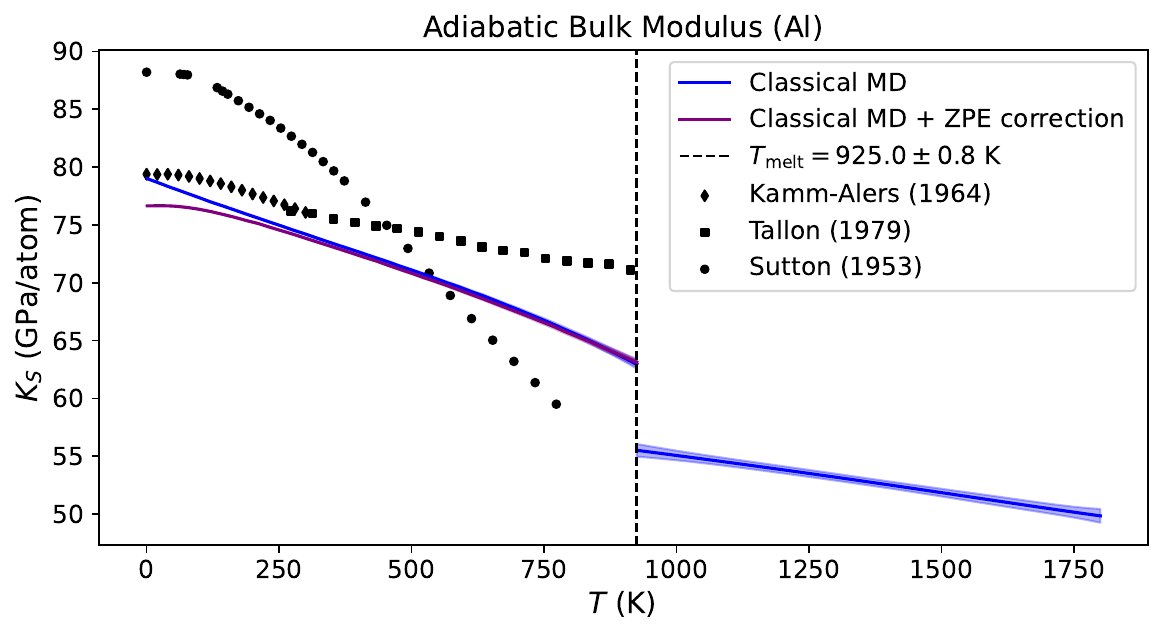}
\caption{Adiabatic bulk modulus $K_S(T)$ of FCC aluminum. Classical MD predictions (blue) and ZPE-corrected values (purple) are compared with experimental data from \cite{tallon1979temperature, Sutton,kamm1964low}. The shaded region indicates the $2\sigma$ confidence interval.}
\label{fig:bulk_adiab}
\end{figure}

The adiabatic bulk modulus $K_S(T)$ was calculated from $K_T(T)$ using \eqref{eq:ks}.  
The $K_S(T)$ values, shown in Fig.~\ref{fig:bulk_adiab}, follow the same temperature dependence as $K_T(T)$ and are consistently higher due to the $C_P/C_V$ factor.  
The corrected results reproduce the expected temperature behavior, while the classical predictions decrease monotonically from the lowest temperatures.

\subsection{Workflow in NPT Ensemble}

In the NPT ensemble, simulations were performed directly at zero pressure ($P=0$), allowing equilibrium observables $\langle H(T) \rangle$ and $\langle V(T) \rangle$ to be obtained as functions of temperature without explicit volume scans.
This enables direct evaluation of $C_P(T)$ from the derivative of $\langle H(T) \rangle$ and determination of melting-point properties $\Delta H_{\mathrm{fus}}$ and $\Delta V_{\mathrm{fus}}$ with significantly lower computational cost compared to the NVT workflow.

For the solid phase, a ZPE correction via the harmonic approximation was applied to account for quantum contributions to $C_P(T)$. However, in the NPT framework the thermal expansion coefficient $\alpha(T)$ cannot be obtained with a ZPE correction, and the isothermal and adiabatic bulk moduli $K_T(T)$ and $K_S(T)$ cannot be computed, as the free energy is not reconstructed as a function of volume.

Statistical averaging, error estimation, and finite-size extrapolation to the thermodynamic limit were performed using the same procedure as in the NVT calculations.

\subsubsection{Melting-Point Properties}

Thermodynamic discontinuities at melting were obtained by fitting the enthalpy and volume as functions of temperature and evaluating their jumps at the melting point $T_\mathrm{melt}$, determined from the solid-liquid coexistence method~\cite{meltingpoint}. 
At the melting temperature $T_\mathrm{melt} = 925.0 \pm 0.8$~K, the computed discontinuities are:
\[
\begin{aligned}
\Delta H_\mathrm{fus} &= 92.9 \pm 0.4\ \text{meV/atom}, \\
\Delta V_\mathrm{fus} &= 0.902 \pm 0.003\ \text{\AA}^3/\text{atom}.
\end{aligned}
\]

For comparison, the NVT results (Section~\ref{sec:results_melting_nvt}) are  
$\Delta H_{\rm fus}^{\rm NVT} = 92.4 \pm 0.3$~meV/atom and  
$\Delta V_{\rm fus}^{\rm NVT} = 0.888 \pm 0.003$~\AA$^3$/atom.  
The NPT and NVT values differ by $1.0\,\sigma$ for $\Delta H_{\rm fus}$ and by $3.3\,\sigma$ for $\Delta V_{\rm fus}$, given the combined uncertainties.
In theory, the NPT and NVT results should agree within statistical uncertainty in the limit $N\to\infty$. In practice, the enthalpy discontinuity agrees within $1.0\,\sigma$, while the volume discontinuity shows a deviation of $3.3\,\sigma$, which is larger than the expected $1\sigma$ and may indicate that our GPR slightly underestimates the true uncertainty.

\subsubsection{Heat Capacity}

The temperature dependence of $C_P$ was obtained by differentiating $\langle H(T) \rangle$ fitted via Gaussian Process Regression. The results are shown in Fig.~\ref{fig:cp}. At high temperatures, the classical MD curve agrees well with experimental data. At low temperatures, the classical model underestimates heat capacity due to missing quantum contributions. After applying the harmonic ZPE correction, the results agree well with the experimental values from Desai \cite{desai1987aluminium} across the entire temperature range.

\begin{figure}[H]
\centering
\includegraphics[width=0.8\linewidth]{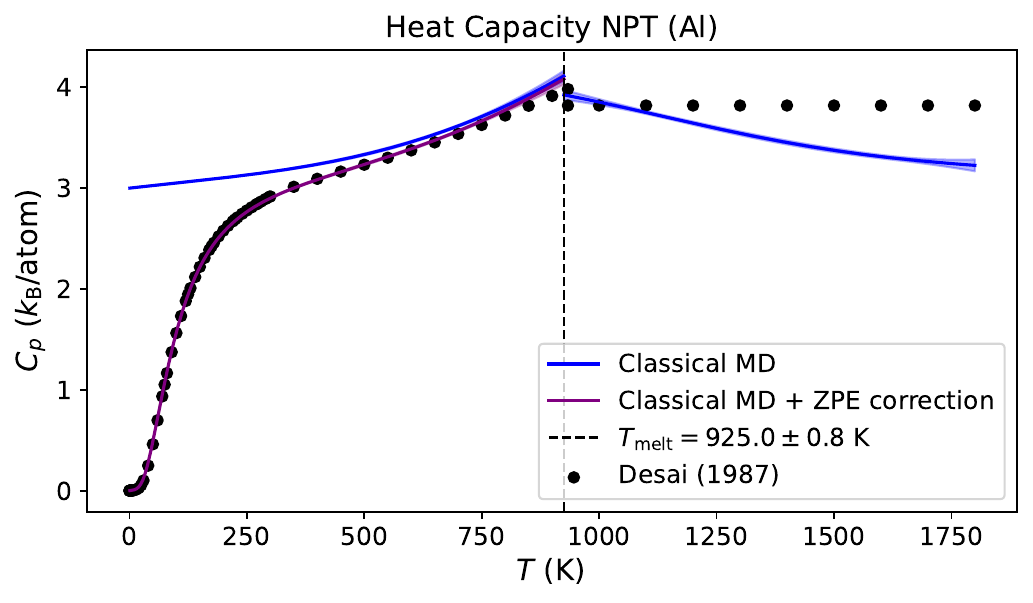}
\caption{Temperature dependence of the constant-pressure heat capacity $C_P$ computed from NPT simulations. The classical MD results (blue) and ZPE-corrected values (purple) are compared with experimental data from Desai~\cite{desai1987aluminium}. The shaded region indicates the $2\sigma$ confidence interval.}
\label{fig:cp_npt}
\end{figure}

\subsection{Comparison of Computational Cost between NVT and NPT Ensembles}

While both ensembles yield consistent predictions for melting-point properties and heat capacity, the NPT ensemble offers a substantial advantage in terms of computational efficiency. This stems from its simpler data requirements: in the NPT approach, simulations are performed only as a function of temperature at zero pressure, whereas the NVT ensemble requires sampling across volumes and temperatures. Table~\ref{tab:npt_nvt_efficiency} summarizes the number of MD data points and the corresponding total CPU time required for simulations in each ensemble.

\begin{table}[H]
\centering
\caption{Summary of data points and total CPU $\cdot$ hours  used in NVT and NPT simulations for aluminum.} 
\label{tab:npt_nvt_efficiency}
\begin{tabular}{llcc}
\toprule
\textbf{Ensemble} & \textbf{Phase} & \textbf{MD points} & \textbf{CPU $\cdot$ hours} \\
\midrule
\multirow{3}{*}{NVT}  
  & Solid  & 177 & 97 \\
  & Liquid & 145 & 83 \\
  & \textbf{Total} & \textbf{322} & \textbf{180} \\
\midrule
\multirow{3}{*}{NPT} 
  & Solid  & 26 & 20 \\
  & Liquid & 23 & 19 \\
  & \textbf{Total} & \textbf{49} & \textbf{39} \\
\bottomrule
\end{tabular}
\end{table}

In total, the full NVT dataset required approximately 180 CPU $\cdot$ hours (per core), while the NPT dataset required only 39 CPU $\cdot$ hours. Despite this significant difference, the NPT approach still enables accurate determination of melting-point properties and $C_P(T)$. However, it does not provide access to quantum-corrected thermal expansion or bulk moduli, which depend on the volume derivatives of the free energy and can only be obtained through the NVT ensemble.

\subsection{Active Learning for Thermodynamic Properties}

The active learning procedure described in Supplementary Information (Section S2) can be applied to any thermodynamic property in either the NVT or NPT ensemble.
Here, we illustrate its application to predicting the constant-pressure heat capacity $C_P(T)$ for the solid phase in the NVT ensemble.

The key idea is to identify the simulation point that would most effectively reduce the uncertainty. This is achieved by introducing the information function $H(T,V)$, which measures how much the predicted uncertainties of thermodynamic properties (such as $C_P(T)$) would decrease if an additional MD simulation were performed at a given $(T, V)$ point. At each step, the point with the highest $H(T,V)$ is selected, a molecular dynamics simulation is then performed, the Gaussian Process (GP) model is retrained, and the target property is updated.

\begin{figure}[H]
\centering
\begin{tikzpicture}[node distance=0.6cm, every node/.style={inner sep=0}, arrow/.style={->, thick}]
    \node (img1) {\includegraphics[width=0.3\textwidth]{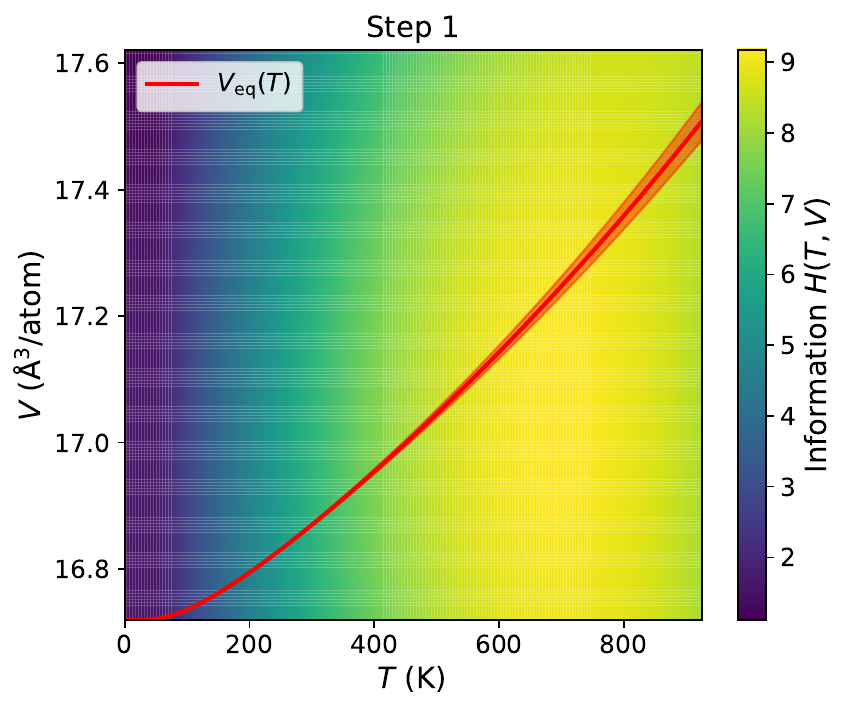}};
    \node[right=of img1] (img2) {\includegraphics[width=0.3\textwidth]{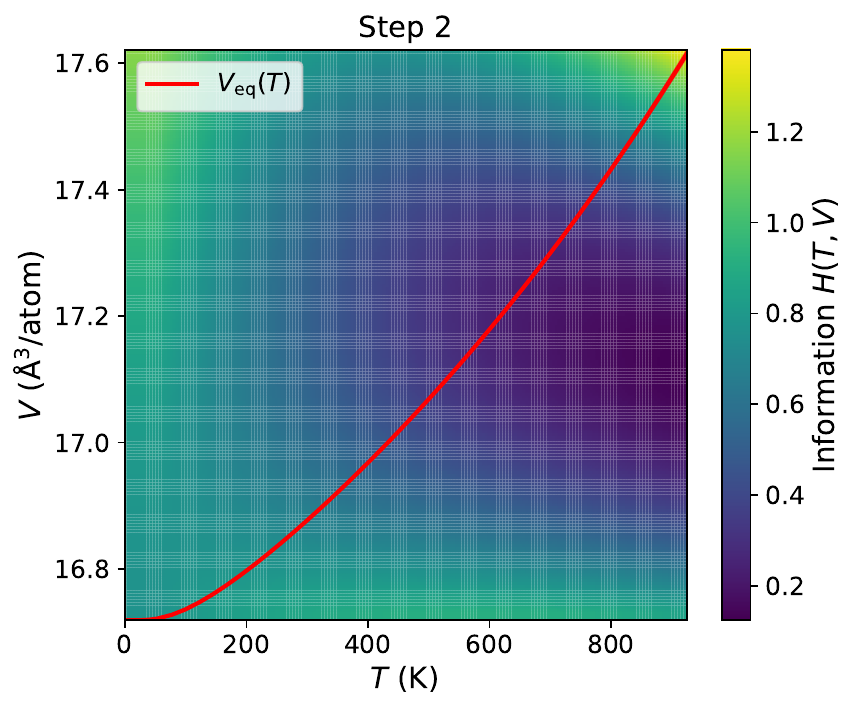}};
    \node[right=of img2] (img3) {\includegraphics[width=0.3\textwidth]{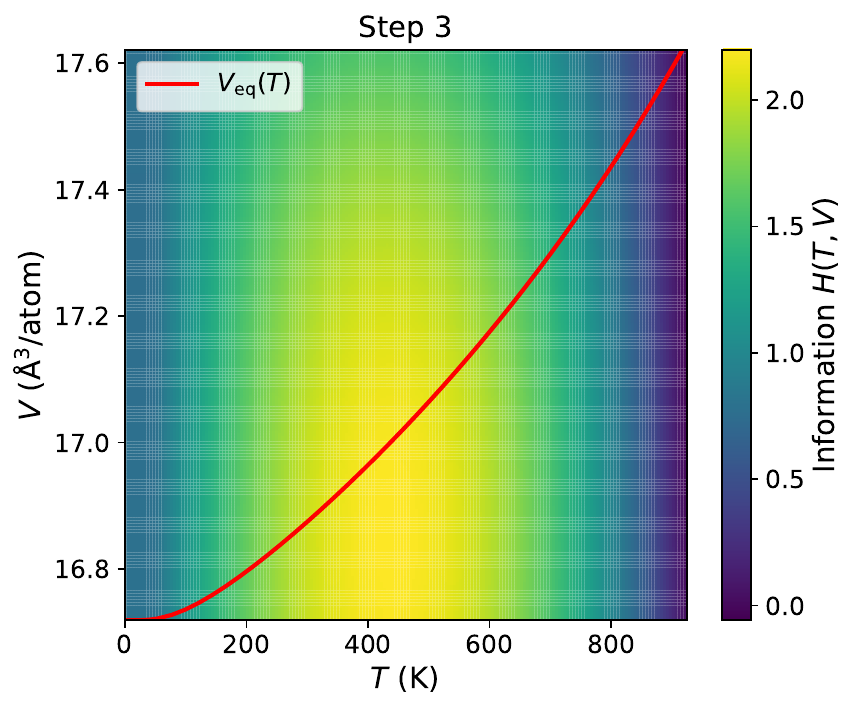}};

    \draw[arrow] (img1.east) -- (img2.west);
    \draw[arrow] (img2.east) -- (img3.west);
\end{tikzpicture}
\caption{    
Heatmaps of the information function $H(T, V)$ at active learning steps 1, 2, and 3. The red line denotes the predicted equilibrium volume $V_{\mathrm{eq}}(T)$.}
\label{fig:heatmaps}
\end{figure}

Figure~\ref{fig:heatmaps} shows how the information landscape evolves during the first three iterations.
Initially, $H(T, V)$ exhibits large values in poorly sampled regions --- most prominently at high temperatures --- indicating high model uncertainty.
As new simulations are added, these peaks flatten, reflecting increased model confidence.
The most informative points tend to lie near the equilibrium volume $V_{\mathrm{eq}}(T)$, especially in regions of sparse data.

\begin{figure}[H]
\centering
\begin{tikzpicture}[
  node distance=0.2cm and 1.2cm,
  box/.style={inner sep=0pt, outer sep=0pt},
  arrow/.style={->, thick}
]

\node[box] (s0) {\includegraphics[width=0.42\linewidth]{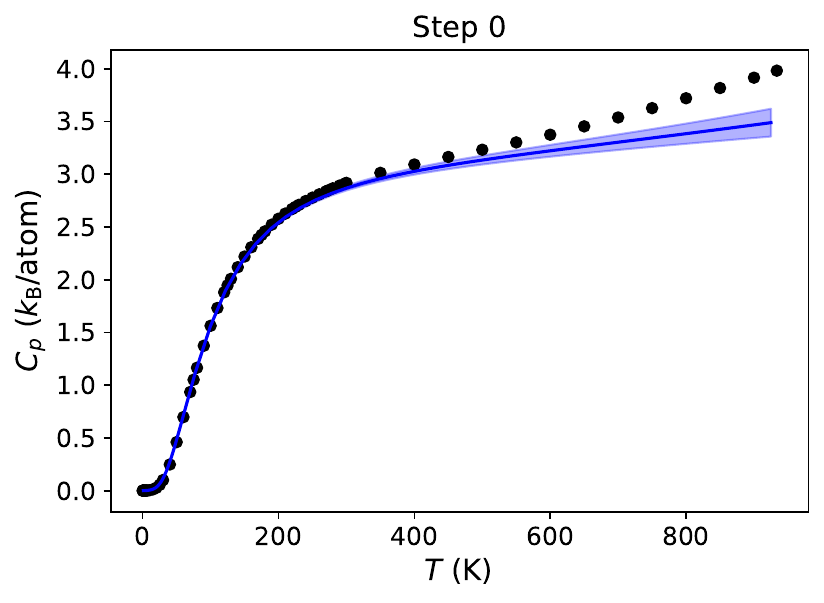}};
\node[box, right=of s0] (s1) {\includegraphics[width=0.42\linewidth]{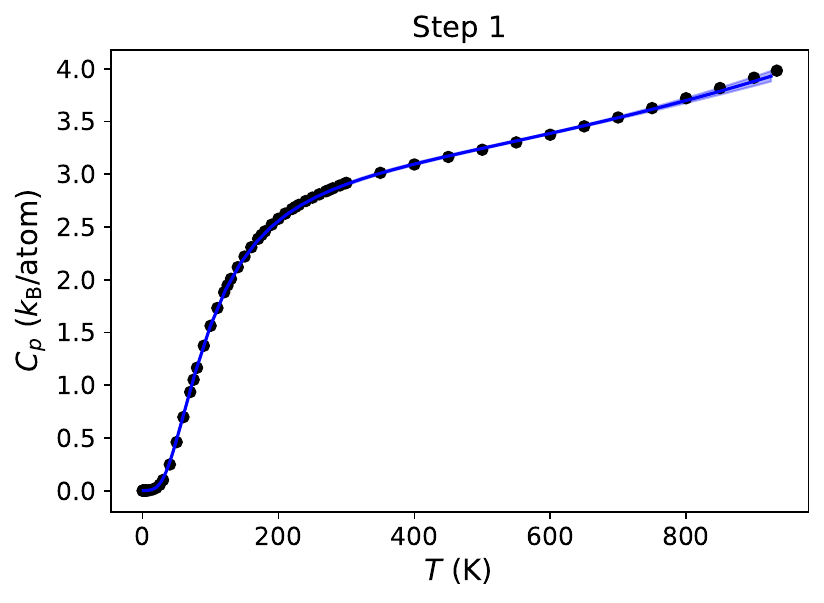}};
\node[box, below=of s0] (s2) {\includegraphics[width=0.42\linewidth]{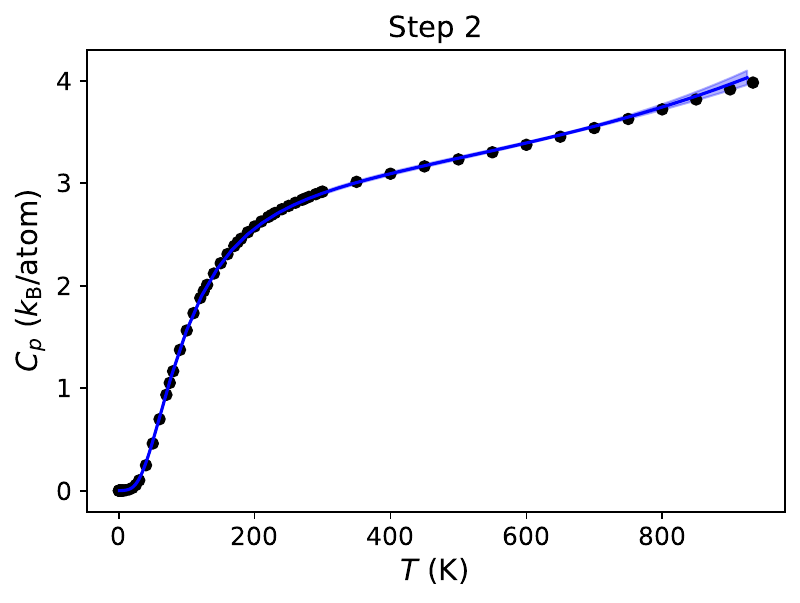}};
\node[box, right=of s2] (s3) {\includegraphics[width=0.42\linewidth]{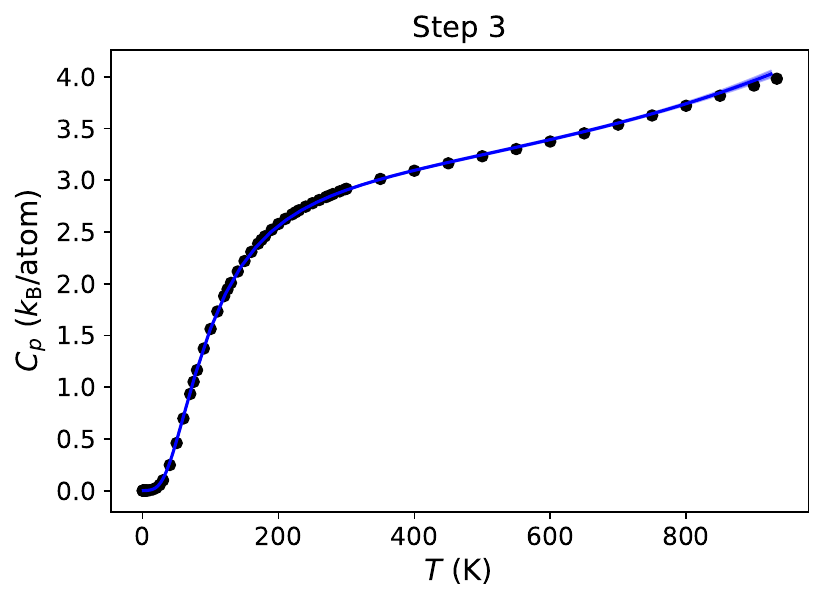}};

\draw[arrow] (s0) -- (s1);
\draw[arrow] (s1) -- (s2);
\draw[arrow] (s2) -- (s3);

\end{tikzpicture}
\caption{Predicted heat capacity $C_P(T)$ at successive steps of active learning. The blue curves show the GPR prediction with $2\sigma$ confidence intervals; black dots represent experimental data from Desai~\cite{desai1987aluminium}. Already after the first step, the prediction improves significantly; further updates yield only minor refinements.}
\label{fig:cp_al_steps}
\end{figure}
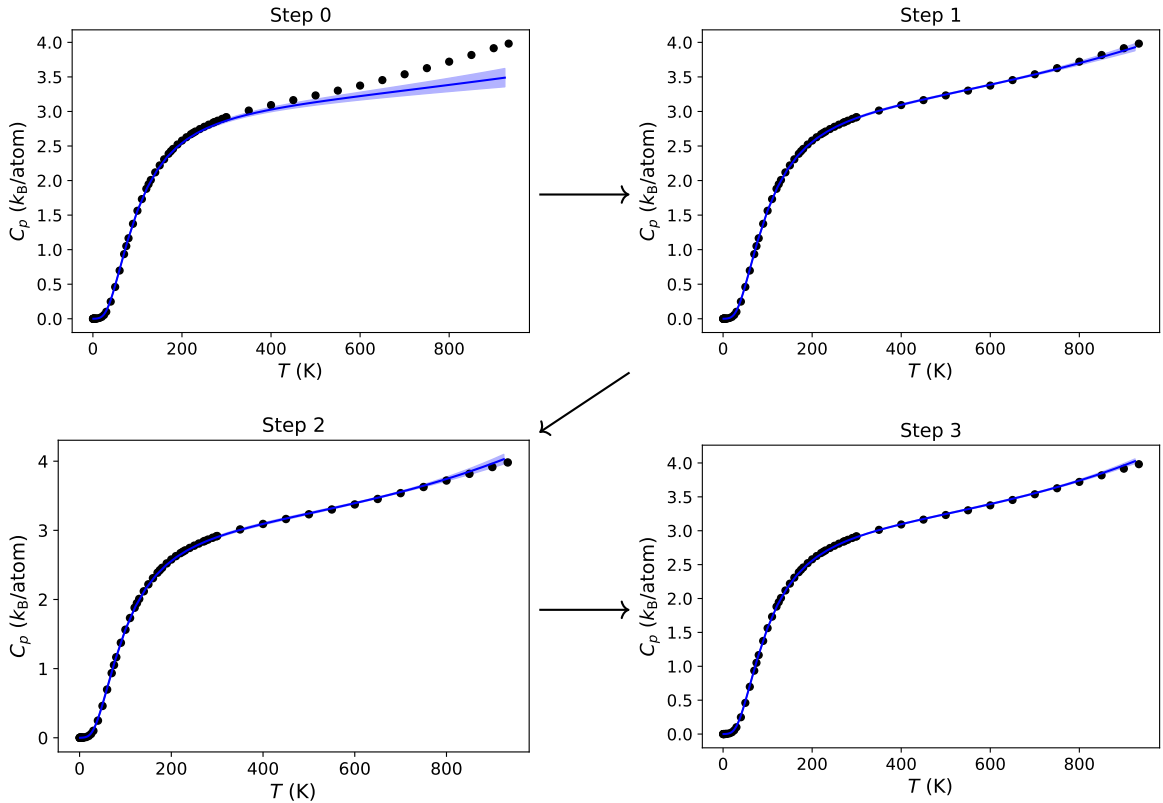

Figure~\ref{fig:cp_al_steps} shows the evolution of $C_P(T)$ after each iteration.
Already after the first update, the prediction aligns closely with the experimental data at high temperatures, where the initial uncertainty was largest.
Subsequent steps yield only minor refinements, indicating that active learning quickly identifies and corrects the most uncertain regions of the model.
It should be noted that at Step 0, when only limited data are available, the GPR's predictive uncertainty is overly optimistic: the true curve lies outside the $2\sigma$ interval. This indicates that predictive uncertainty should be treated with caution when used as a quantitative error estimate.

This example demonstrates that the active learning approach efficiently reduces model uncertainty with a minimal number of additional simulations.
Although shown here for $C_P(T)$ in the NVT ensemble, the same procedure can be directly applied to other quantities --- such as thermal expansion, bulk modulus, or melting-point properties --- in either ensemble.

\section{Application to Other Materials}
\label{sec:results_all}

To assess the generality of our approach, we applied it to a broad set of elemental metals with either face-centered cubic (FCC) or body-centered cubic (BCC) structures. For each element, we selected multiple interatomic potentials from the literature, including the Embedded Atom Model (EAM), the Modified Embedded Atom Model (MEAM), and the Moment Tensor Potential (MTP).

Specifically, we studied: Au, Ag, Pd, and Pt (FCC) using EAM from Zhou~\textit{et al.}~\cite{zhou2004misfit}; Ag and Cu using EAM from Williams~\textit{et al.}~\cite{Williams_2006}; Cu, Ni, and Fe using MEAM from Asadi~\textit{et al.}~\cite{Asadi_meam} and from Etesami~\textit{et al.}~\cite{ETESAMI201861}; Mo using MEAM from Kim~\textit{et al.}~\cite{KIM2017131}; Al, Cu, and Pt using machine-learned MTPs~\cite{mtp}.

For all systems, we performed NVT molecular dynamics simulations, reconstructed the Helmholtz free energy using Gaussian Process Regression in the infinite-atom ($N\to\infty$) limit, and applied zero-point energy corrections for the solid phases. Thermodynamic properties were computed from the corrected free-energy surface, following the same protocol used for aluminum.

We evaluated the melting-point properties (melting temperature $T_\mathrm{m}$, enthalpy of fusion $\Delta H$, volume change $\Delta V$), the constant-pressure heat capacity $C_P(T)$, the linear thermal expansion coefficient $\alpha(T)$, and the isothermal and adiabatic bulk moduli $K_T(T)$ and $K_S(T)$ across a wide temperature range. All calculated results are compared directly to experimental references, allowing us to assess the accuracy of each potential for both low- and high-temperature thermodynamic behavior.

\subsection{Melting-Point Properties}

We determined the melting-point properties for all the studied metals. 
First, we obtained the melting temperature $T_\mathrm{m}$ using the algorithm of Ref.~\cite{meltingpoint}. 
Then, at the corresponding equilibrium volume ($P = 0$), we calculated the discontinuities of atomic volume $\Delta V_\mathrm{fus}$ and enthalpy $\Delta H_\mathrm{fus}$ with their statistical uncertainties. 
Table~\ref{tab:melting_all} summarizes the calculated values together with the available experimental data. 

\begin{table*}[ht]
\centering
\scriptsize
\caption{
Melting properties at $P=0$ from NVT ensemble compared to experimental values from \cite{lide2009crc}.
Listed are the melting temperature $T_{\rm m}$, enthalpy of fusion $\Delta H$, and volume change $\Delta V$. 
}
\label{tab:melting_all}
\begin{tabular}{llccccc}
\toprule
Material & Potential & \multicolumn{2}{c}{$T_{\rm m}$ (K)} & \multicolumn{2}{c}{$\Delta H$ (meV/atom)} & \multicolumn{1}{c}{$\Delta V$ (\AA$^3$/atom)} \\
\cmidrule(lr){3-4}\cmidrule(lr){5-6}
         &           & Calc & Exp & Calc & Exp & Calc \\
\midrule
Au & EAM \cite{zhou2004misfit}      & 1376.5 $\pm$ 1.2 & 1337.3 &  93.9 $\pm$ 0.7 & 130 & 0.628 $\pm$ 0.005 \\
Ag & EAM \cite{zhou2004misfit}      & 1143.5 $\pm$ 3.1 & 1234.9 &  99.2 $\pm$ 1.0 & 117 & 0.802 $\pm$ 0.005 \\
   & EAM \cite{Williams_2006}       & 1259.8 $\pm$ 1.1 &        & 127.5 $\pm$ 0.4 &     & 1.156 $\pm$ 0.003 \\
Pd & EAM \cite{zhou2004misfit}      & 1591.4 $\pm$ 3.3 & 1828.0 & 134.9 $\pm$ 0.9 & 173 & 0.602 $\pm$ 0.002 \\
Cu & EAM \cite{Williams_2006}       & 1328.3 $\pm$ 1.8 & 1357.8 & 130.9 $\pm$ 0.7 & 137 & 0.642 $\pm$ 0.003 \\
   & MEAM \cite{Asadi_meam}         & 1395.6 $\pm$ 0.7 &        & 155.5 $\pm$ 0.4 &     & 0.855 $\pm$ 0.002 \\
   & MEAM \cite{ETESAMI201861}      & 1345.5 $\pm$ 0.7 &        & 116.0 $\pm$ 0.8 &     & 0.589 $\pm$ 0.005 \\
   & MTP  \cite{mtp}                & 1246.6 $\pm$ 3.7 &        & 121.5 $\pm$ 1.2 &     & 0.676 $\pm$ 0.003 \\
Ni & MEAM \cite{Asadi_meam}         & 1756.1 $\pm$ 1.3 & 1728.2 & 228.6 $\pm$ 0.5 & 181 & 0.996 $\pm$ 0.001 \\
   & MEAM \cite{ETESAMI201861}      & 1704.9 $\pm$ 0.9 &        & 190.2 $\pm$ 0.6 &     & 0.801 $\pm$ 0.002 \\
Fe & MEAM \cite{Asadi_meam}         & 1795.5 $\pm$ 1.3 & 1811.2 & 135.6 $\pm$ 0.7 & 143 & 0.433 $\pm$ 0.002 \\
   & MEAM \cite{ETESAMI201861}      & 1809.9 $\pm$ 1.7 &        & 115.8 $\pm$ 0.7 &     & 0.412 $\pm$ 0.001 \\
Mo & MEAM \cite{KIM2017131}         & 2937.6 $\pm$ 2.1 & 2896.2 & 215.4 $\pm$ 0.9 & 388 & 0.583 $\pm$ 0.002 \\
Al & MTP  \cite{mtp}                &  887.0 $\pm$ 0.7 &  933.2 & 104.5 $\pm$ 0.3 & 110 & 1.273 $\pm$ 0.004 \\
Pt & EAM  \cite{zhou2004misfit}     & 1458.5 $\pm$ 1.9 & 2041.2 & 146.3 $\pm$ 0.5 & 230 & 0.386 $\pm$ 0.001 \\
   & MTP  \cite{mtp}                & 1527.4 $\pm$ 2.4 &        & 179.9 $\pm$ 0.8 &     & 1.027 $\pm$ 0.004 \\
\bottomrule
\end{tabular}
\end{table*}

The tested potentials generally agree with experimental melting data.
For most metals, the calculated melting temperatures differ from experiment by less than 10\%, and the enthalpy of fusion by about 10--25\%.
Larger deviations are observed for some potentials, such as the EAM for Pt, which underestimates both $T_\mathrm{m}$ and $\Delta H_\mathrm{fus}$ by about 30--35\%, and the MEAM for Mo, which gives a much lower $\Delta H_\mathrm{fus}$ than experiment.
On the other hand, the three results obtained with the MTPs, trained on DFT data computed using the GGA-PBE functional and PAW pseudopotentials~\cite{kresse1993ab,kresse1996efficiency,kresse1996efficient}, follow a consistent trend of slightly underestimating both the melting point and the enthalpy of fusion, which is generally attributed to the PBE functional underestimating the true binding between atoms \cite{meltingpoint}.

It is interesting to look at the Cu section of the table, which includes three empirical (one EAM \cite{Williams_2006} and two MEAM \cite{Asadi_meam, ETESAMI201861}) and one machine-learning potential (MTP \cite{mtp}). The EAM and MTP were fitted only to DFT energies (without any input from experimental data) and give very similar values for all the quantities, consistent with the underbinding trend mentioned above. In contrast, the two MEAMs were parameterized using the experimental melting point, which they reproduce rather well, but not the enthalpy of fusion, whose values deviate significantly from each other, as well as from the EAM, MTP, and experimental results.

These differences highlight the strong dependence of the results on the underlying interatomic potential --- in particular, on the way in which it was constructed --- while our approach provides a consistent framework for validating how well different interatomic models reproduce thermodynamic properties.

\subsection{Thermal Expansion Coefficient}

\begin{figure}[H]
    \centering
    \begin{subfigure}[t]{0.32\linewidth}
        \includegraphics[width=\linewidth]{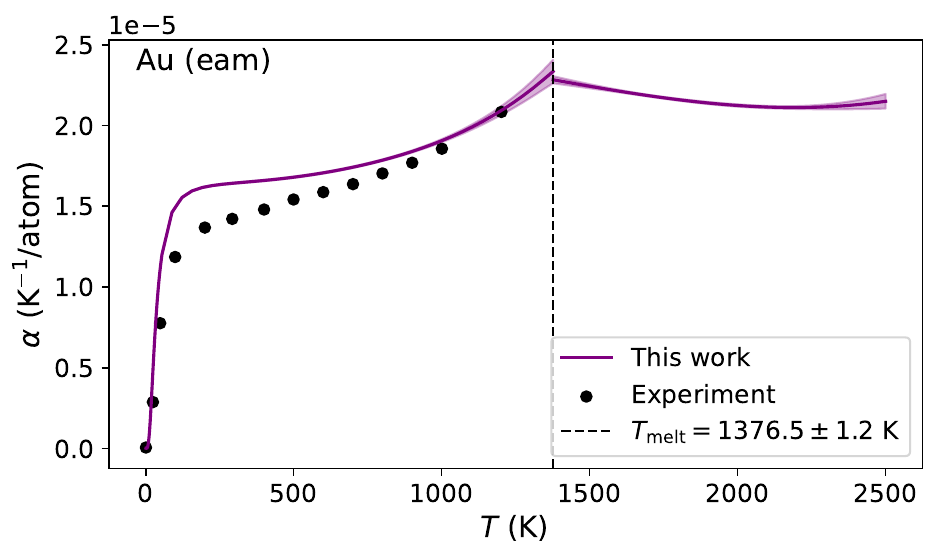}
        \caption{Au (fcc), EAM \cite{zhou2004misfit} vs exp. from\cite{touloukian1977_alpha}}
        \label{fig:alpha-au-zhou}
    \end{subfigure}
    \hfill
    \begin{subfigure}[t]{0.32\linewidth}
        \includegraphics[width=\linewidth]{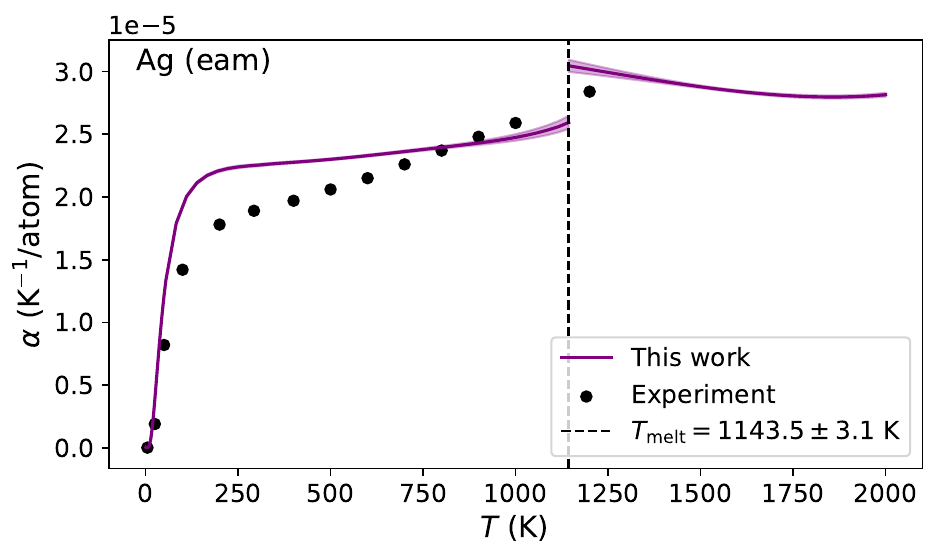}
        \caption{Ag (fcc), EAM \cite{zhou2004misfit} vs exp. from \cite{touloukian1977_alpha}}
        \label{fig:alpha-ag-zhou}
    \end{subfigure}
    \hspace{0.001\linewidth}
    \begin{subfigure}[t]{0.32\linewidth}
        \includegraphics[width=\linewidth]{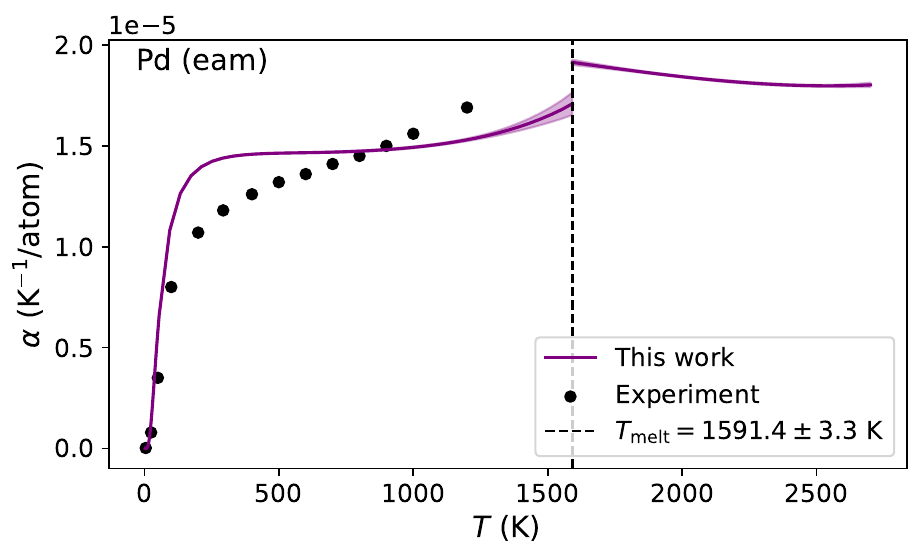}
        \caption{Pd (fcc), EAM \cite{zhou2004misfit} vs exp. from \cite{touloukian1977_alpha}}
        \label{fig:alpha-pd-zhou}
    \end{subfigure}

    \vspace{0.1cm}

    \begin{subfigure}[t]{0.32\linewidth}
        \includegraphics[width=\linewidth]{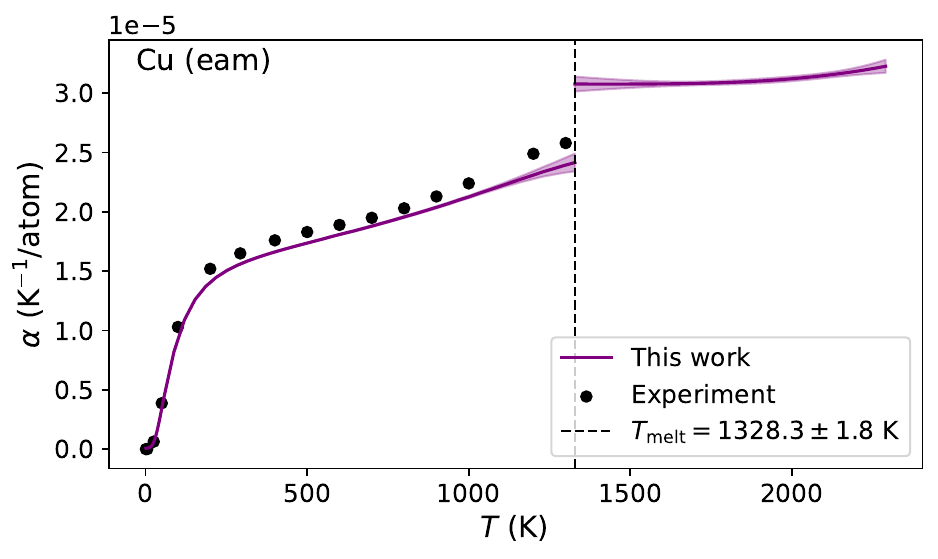}
        \caption{Cu (fcc), EAM \cite{Williams_2006} vs exp. from \cite{touloukian1977_alpha}}
        \label{fig:alpha-cu-williams}
    \end{subfigure}
    \hspace{0.001\linewidth}
    \begin{subfigure}[t]{0.32\linewidth}
        \includegraphics[width=\linewidth]{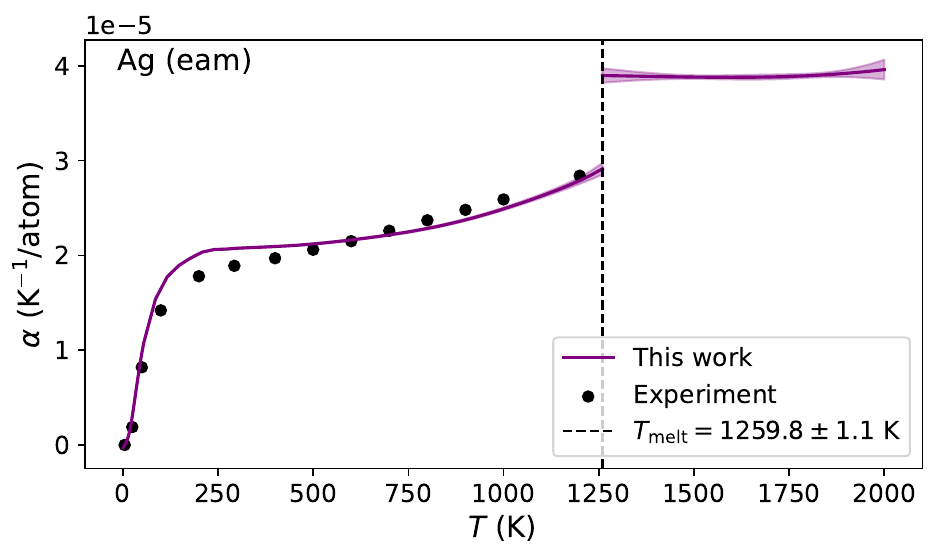}
        \caption{Ag (fcc), EAM \cite{Williams_2006} vs exp. from \cite{touloukian1977_alpha}}
        \label{fig:alpha-ag-williams}
    \end{subfigure}
    \hspace{0.001\linewidth}
    \begin{subfigure}[t]{0.32\linewidth}
        \includegraphics[width=\linewidth]{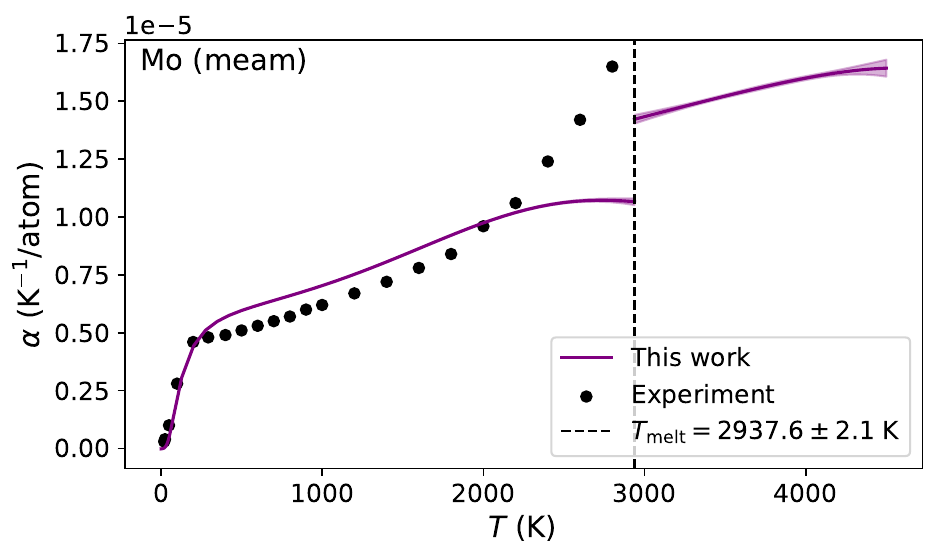}
        \caption{Mo (bcc), MEAM \cite{KIM2017131} vs exp. from \cite{touloukian1977_alpha}}
        \label{fig:alpha-mo}
    \end{subfigure}

    \vspace{0.1cm}

    \begin{subfigure}[t]{0.32\linewidth}
        \includegraphics[width=\linewidth]{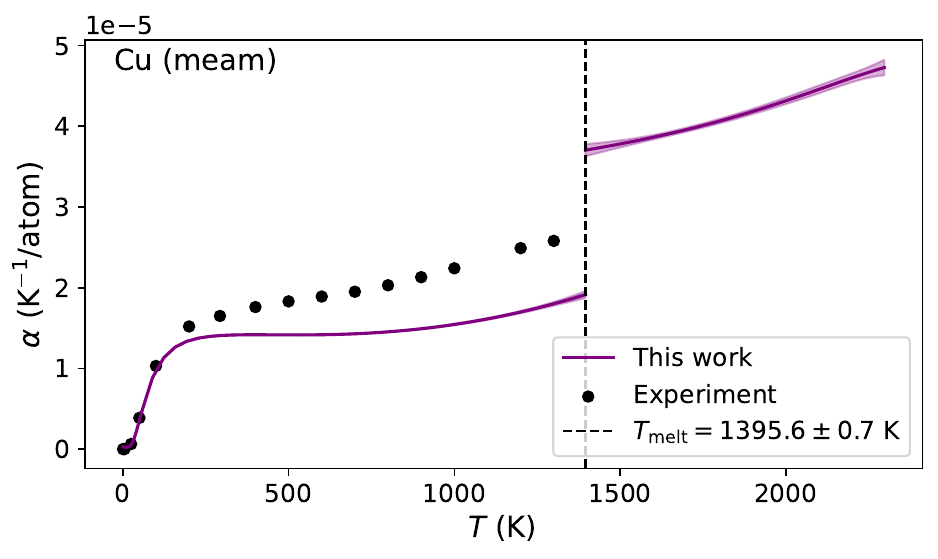}
        \caption{Cu (fcc), MEAM \cite{Asadi_meam} vs exp. from \cite{touloukian1977_alpha}}
        \label{fig:alpha-cu-asadi}
    \end{subfigure}
    \hspace{0.001\linewidth}
    \begin{subfigure}[t]{0.32\linewidth}
        \includegraphics[width=\linewidth]{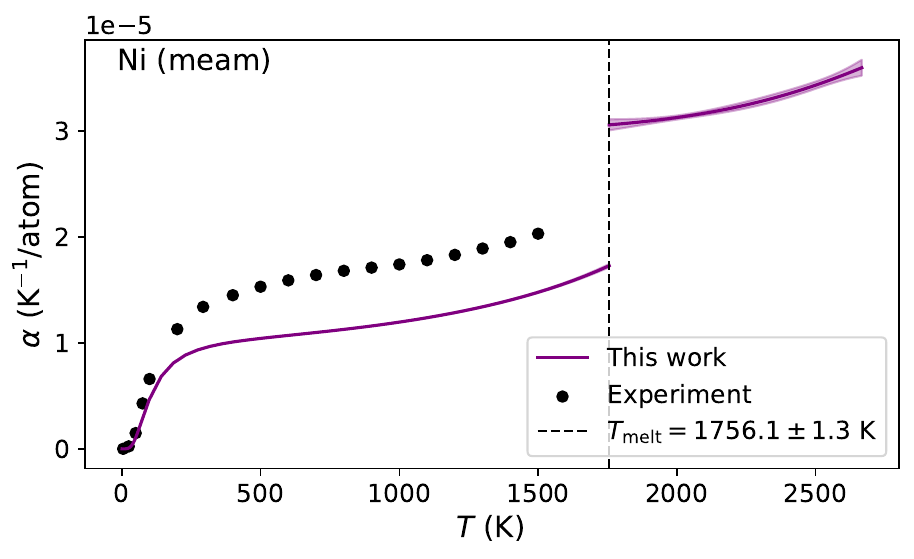}
        \caption{Ni (fcc), MEAM \cite{Asadi_meam} vs exp. from \cite{touloukian1977_alpha}}
        \label{fig:alpha-ni-asadi}
    \end{subfigure}
    \hspace{0.001\linewidth}
    \begin{subfigure}[t]{0.32\linewidth}
        \includegraphics[width=\linewidth]{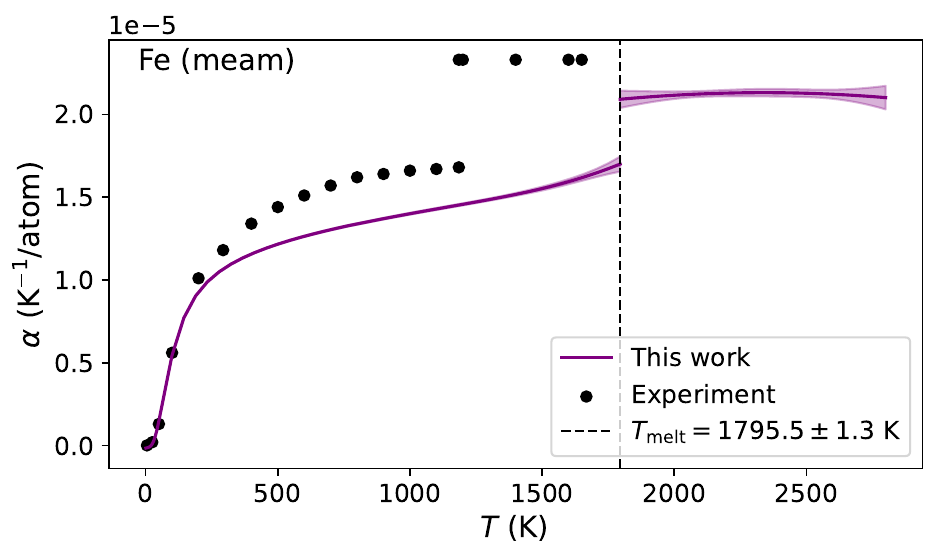}
        \caption{Fe (bcc), MEAM \cite{Asadi_meam} vs exp. from \cite{touloukian1977_alpha}}
        \label{fig:alpha-fe-asadi}
    \end{subfigure}

    \begin{subfigure}[t]{0.32\linewidth}
        \includegraphics[width=\linewidth]{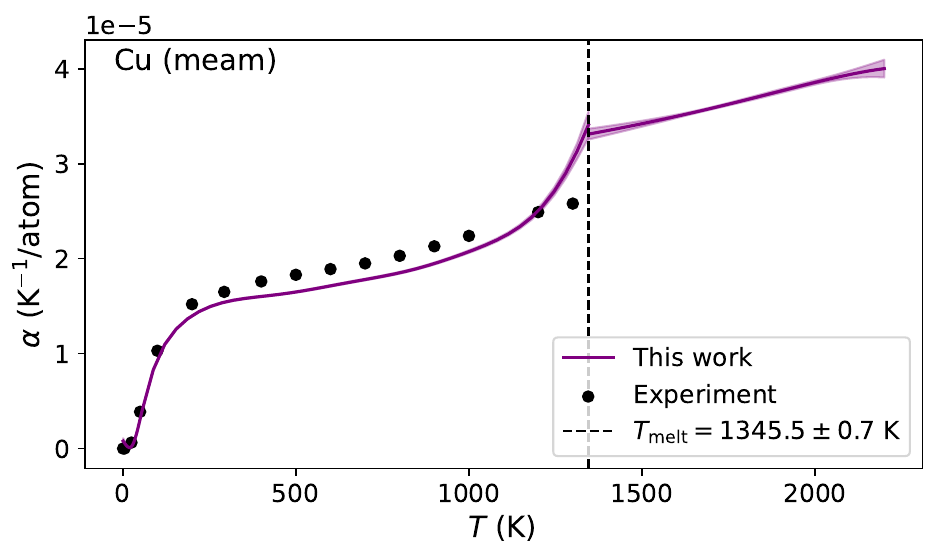}
        \caption{Cu (fcc), MEAM \cite{ETESAMI201861} vs exp. from \cite{touloukian1977_alpha}}
        \label{fig:alpha-cu-etesami}
    \end{subfigure}
    \hspace{0.001\linewidth}
    \begin{subfigure}[t]{0.32\linewidth}
        \includegraphics[width=\linewidth]{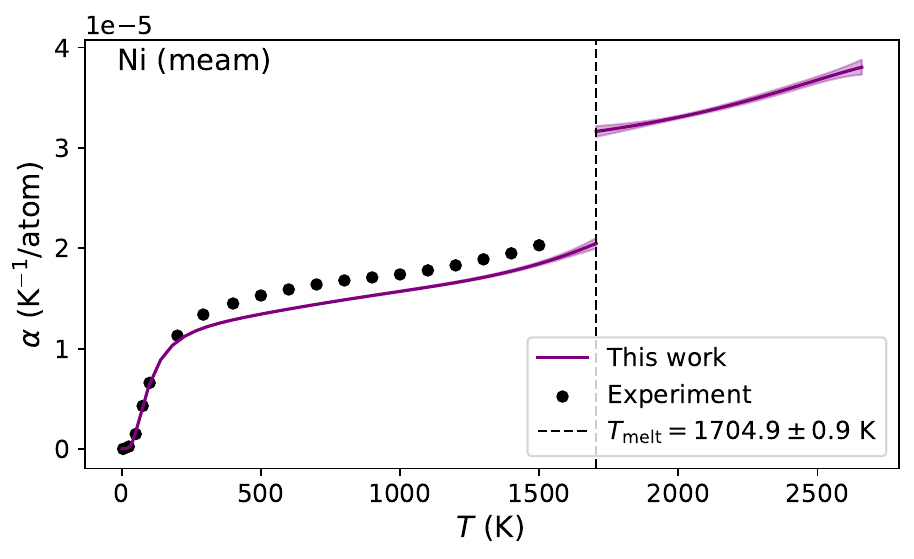}
        \caption{Ni (fcc), MEAM \cite{ETESAMI201861} vs exp. from \cite{touloukian1977_alpha}}
        \label{fig:alpha-ni-etesami}
    \end{subfigure}
    \hspace{0.001\linewidth}
    \begin{subfigure}[t]{0.32\linewidth}
        \includegraphics[width=\linewidth]{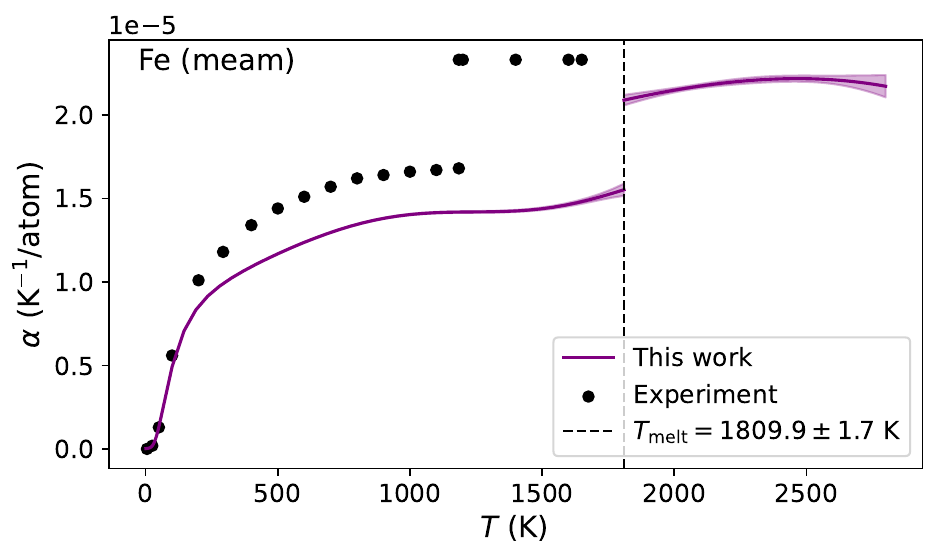}
        \caption{Fe (bcc), MEAM \cite{ETESAMI201861} vs exp. from \cite{touloukian1977_alpha}}
        \label{fig:alpha-fe-etesami}
    \end{subfigure}

    \vspace{0.1cm}

    \vspace{0.1cm}
    \begin{subfigure}[t]{0.32\linewidth}
        \includegraphics[width=\linewidth]{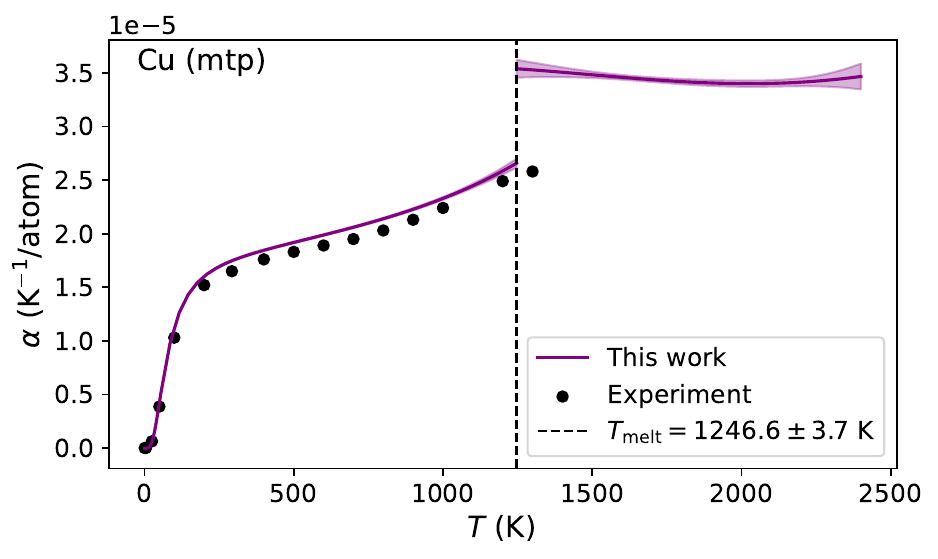}
        \caption{Cu (fcc), MTP vs exp. from \cite{touloukian1977_alpha}}
        \label{fig:alpha-cu-mtp}
    \end{subfigure}
    \hspace{0.001\linewidth}
    \begin{subfigure}[t]{0.32\linewidth}
        \includegraphics[width=\linewidth]{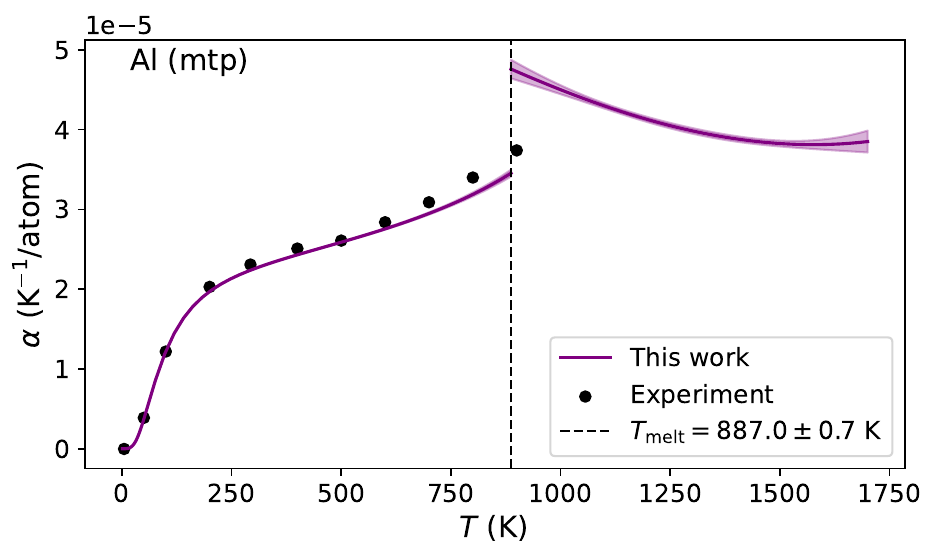}
        \caption{Al (fcc), MTP vs exp. from \cite{touloukian1977_alpha}}
        \label{fig:alpha-al-mtp}
    \end{subfigure}
    \hspace{0.001\linewidth}
    \begin{subfigure}[t]{0.32\linewidth}
        \includegraphics[width=\linewidth]{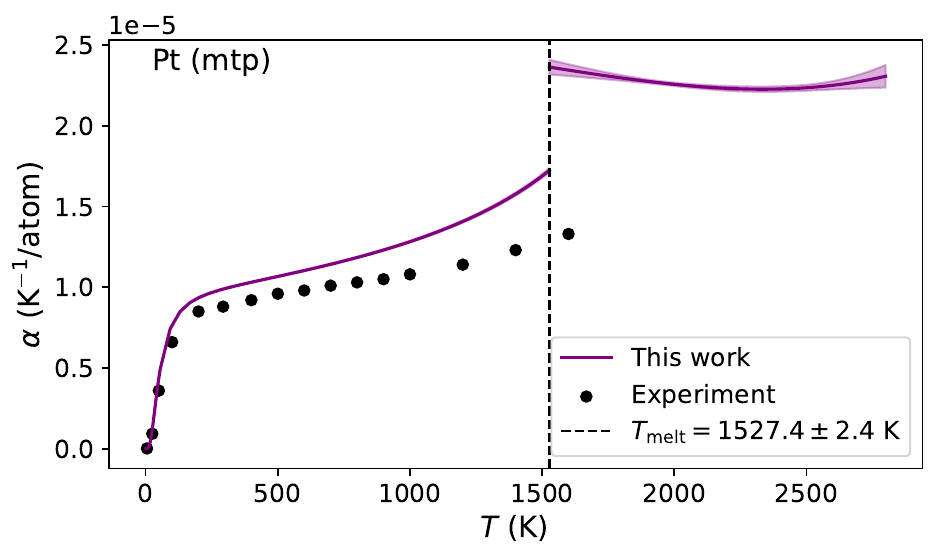}
        \caption{Pt (fcc), MTP vs exp. from \cite{touloukian1977_alpha}}
        \label{fig:alpha-pt-mtp}
    \end{subfigure}

    \caption{Temperature dependence of the linear thermal expansion coefficient $\alpha(T)$ for FCC and BCC metals calculated using EAM, MEAM, and MTP interatomic potentials. Experimental data are shown for comparison where available.}
    \label{fig:thermal-expansion-other}
\end{figure}

The linear thermal expansion coefficient $\alpha(T)$ was obtained from the temperature derivative of the equilibrium volume extracted from NVT simulations, with the solid-phase results corrected for zero-point energy. The computed temperature dependence for all studied metals is shown in Figure~\ref{fig:thermal-expansion-other}, together with experimental data for comparison.

At low temperatures, the ZPE correction ensures that all curves smoothly approach zero as $T \to 0$, reproducing the expected quantum behavior. With our approach, we achieve qualitative agreement with experiment across the full temperature range, while the quantitative accuracy depends on the chosen interatomic potential. This also enables us to assess how well each potential reproduces the thermal expansion behavior. 

Among the considered potentials, only \cite{Williams_2006} and \cite{ETESAMI201861} explicitly included thermal-expansion information during fitting. These two models reproduce the temperature dependence of $\alpha(T)$ more accurately than the others, although the agreement with experiment is still not perfect. Namely, the MEAM potential of Etesami~\cite{ETESAMI201861}, which incorporates additional high-temperature elastic constants, provides an improved description of $\alpha(T)$ compared to the earlier MEAM of Asadi~\cite{Asadi_meam}, while the EAM potential of Williams~\cite{Williams_2006} provides the best overall agreement among the empirical models. The MTP potentials trained on DFT/PBE data, despite the absence of experimental input, reproduce realistic trends as well. 

\subsection{Heat Capacity}  

\begin{figure}[H]
    \centering

    \begin{subfigure}[t]{0.32\linewidth}
        \includegraphics[width=\linewidth]{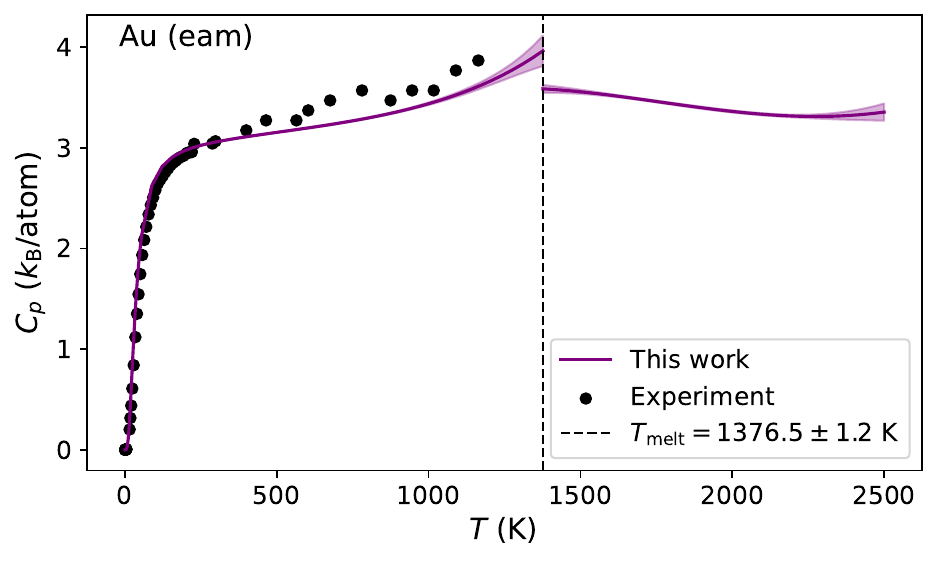}
        \caption{Au (fcc), EAM \cite{zhou2004misfit} vs exp. from \cite{corak1955_au, butler1958_au, geballe1952_au} as compiled in  \cite{touloukian1970_cp}}
        \label{fig:cp-au-zhou}
    \end{subfigure}
    \hspace{0.001\linewidth}
    \begin{subfigure}[t]{0.32\linewidth}
        \includegraphics[width=\linewidth]{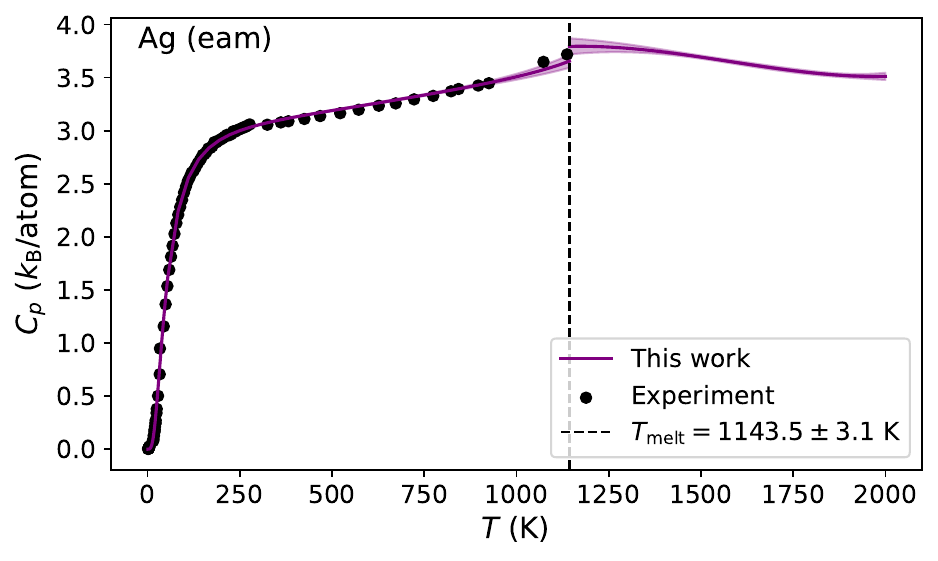}
        \caption{Ag (fcc), EAM \cite{zhou2004misfit} vs exp. from \cite{meads1941_ag, moser1936_ag, keesom1934_ag} as compiled in  \cite{touloukian1970_cp}}
        \label{fig:cp-ag-zhou}
    \end{subfigure}
    \hspace{0.001\linewidth}
    \begin{subfigure}[t]{0.32\linewidth}
        \includegraphics[width=\linewidth]{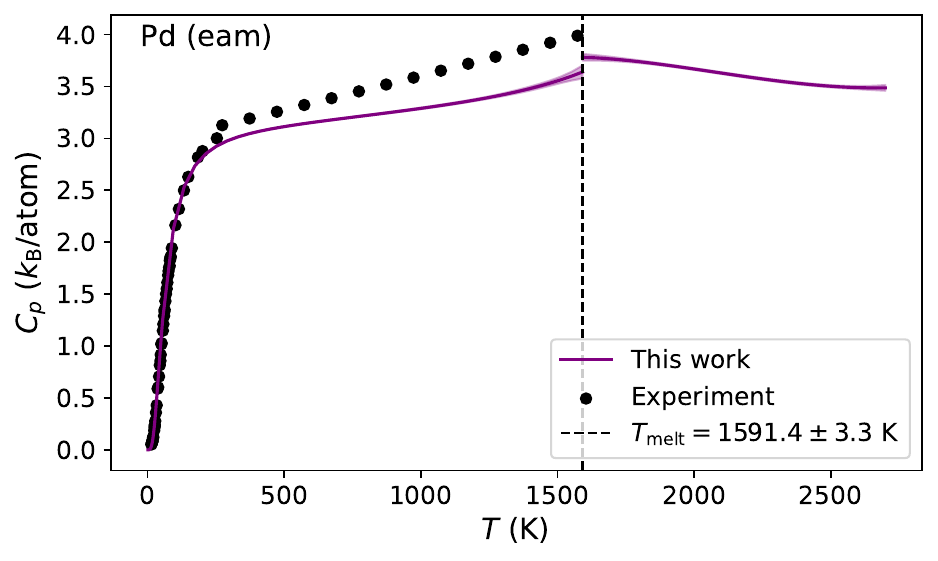}
        \caption{Pd (fcc), EAM \cite{zhou2004misfit} vs. exp. from \cite{pd_clusius1947_cp, pd_jaeger1936_cp, pd_mitacek1963_cp}  as compiled in  \cite{touloukian1970_cp}}
        \label{fig:cp-pd-zhou}
    \end{subfigure}

    \vspace{0.1cm}

    \begin{subfigure}[t]{0.32\linewidth}
        \includegraphics[width=\linewidth]{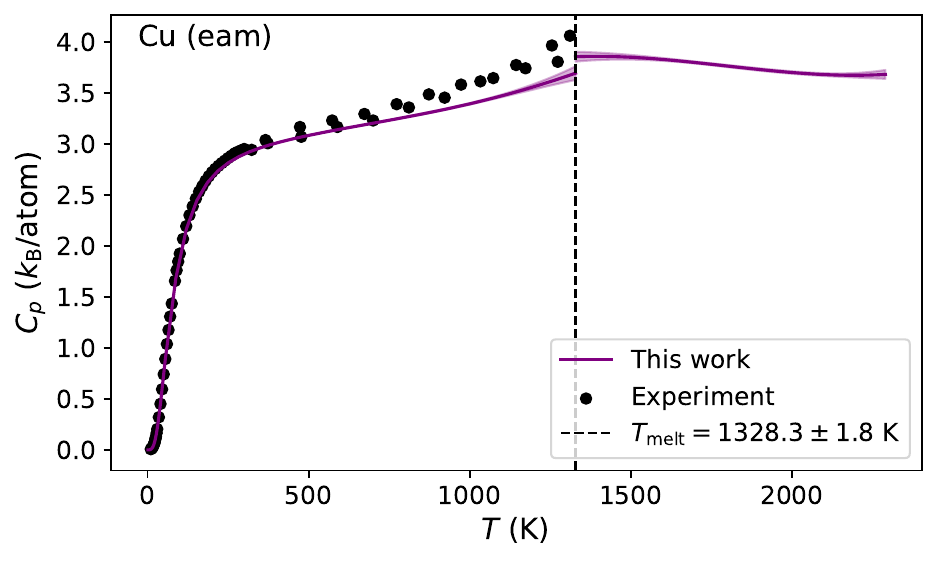}
        \caption{Cu (fcc), EAM \cite{Williams_2006} vs exp. from \cite{cu_marun1960_cu, cu_howse1961_cp, cu_lyustemik1959_cp, cu_franck1961_cp} as compiled in  \cite{touloukian1970_cp}}
        \label{fig:cp-cu-williams}
    \end{subfigure}
    \hspace{0.001\linewidth}
    \begin{subfigure}[t]{0.32\linewidth}
        \includegraphics[width=\linewidth]{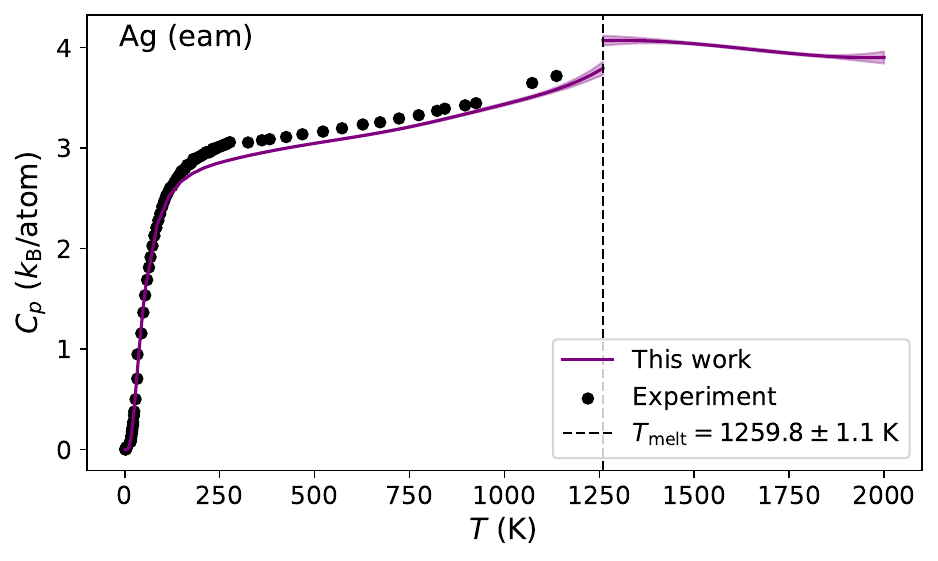}
        \caption{Ag (fcc), EAM \cite{Williams_2006} vs exp. from \cite{meads1941_ag, moser1936_ag, keesom1934_ag} as compiled in  \cite{touloukian1970_cp}}
        \label{fig:cp-ag-williams}
    \end{subfigure}
    \hspace{0.001\linewidth}
    \begin{subfigure}[t]{0.32\linewidth}
        \includegraphics[width=\linewidth]{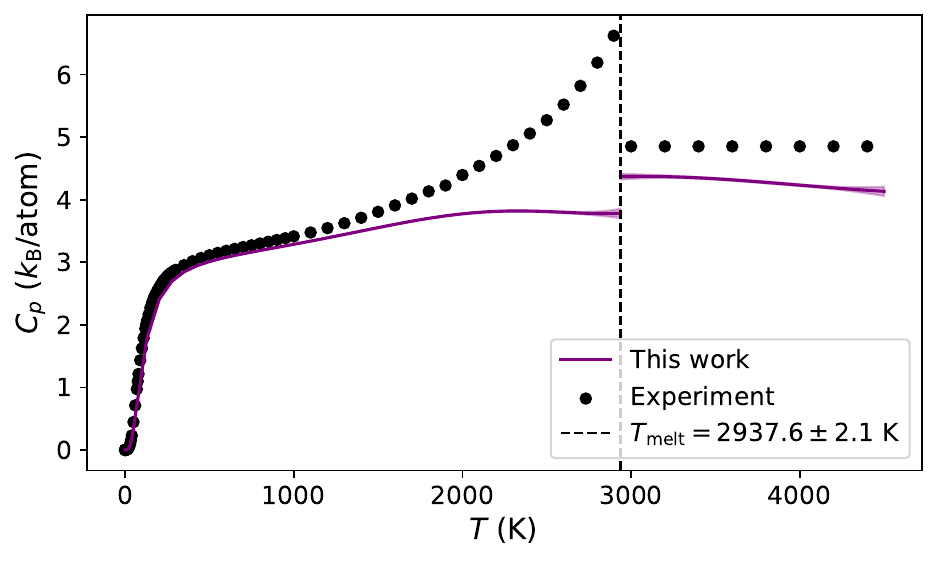}
        \caption{Mo (bcc), MEAM \cite{KIM2017131} vs exp. from \cite{desai1987molybdenum}}
        \label{fig:cp-mo}
    \end{subfigure}

    \vspace{0.1cm}
    
    \begin{subfigure}[t]{0.32\linewidth}
        \includegraphics[width=\linewidth]{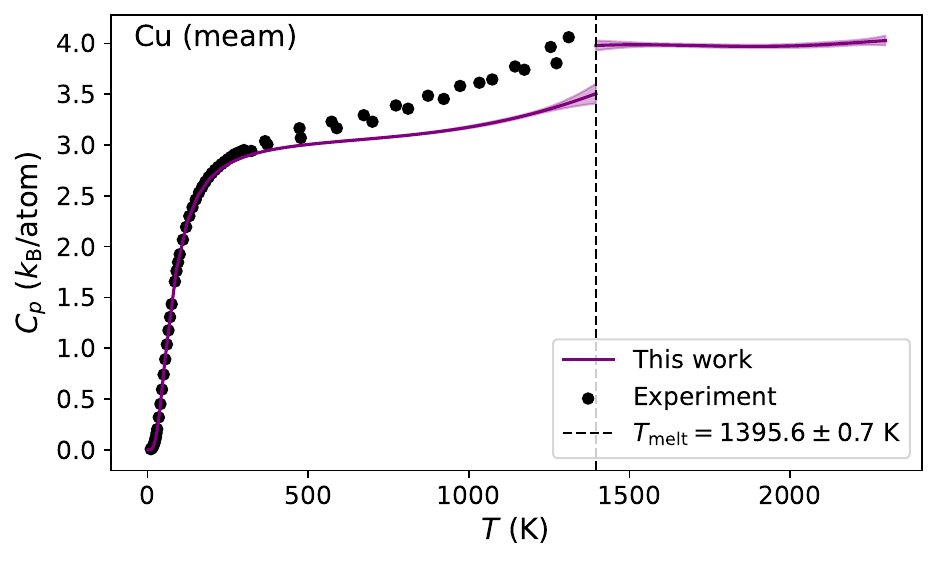}
        \caption{Cu (fcc), MEAM \cite{Asadi_meam} vs exp. from \cite{cu_marun1960_cu, cu_howse1961_cp, cu_lyustemik1959_cp, cu_franck1961_cp} as compiled in  \cite{touloukian1970_cp}}
        \label{fig:cp-cu-etesami}
    \end{subfigure}
    \hspace{0.001\linewidth}
    \begin{subfigure}[t]{0.32\linewidth}
        \includegraphics[width=\linewidth]{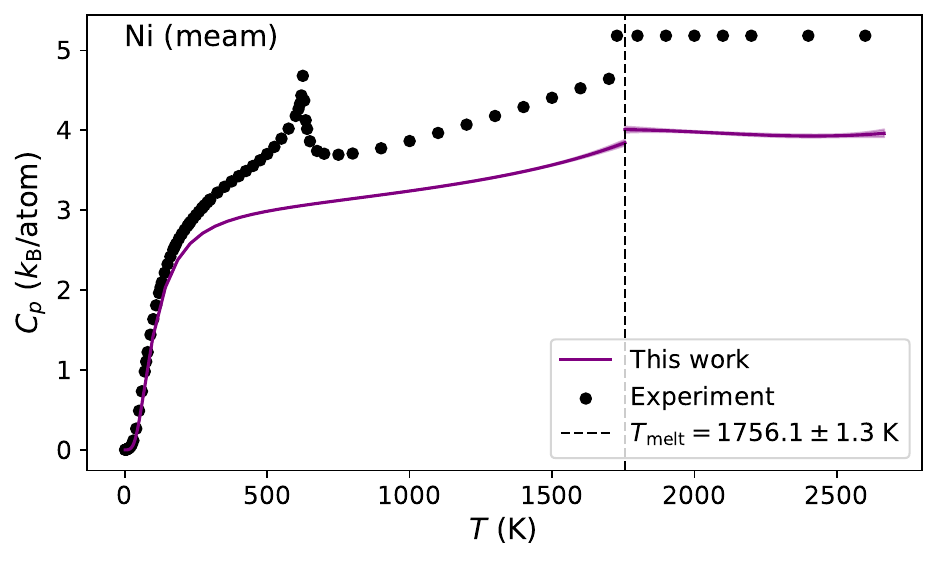}
        \caption{Ni (fcc), MEAM \cite{Asadi_meam} vs exp. from \cite{desai1987nickel}}
        \label{fig:cp-ni-etesami}
    \end{subfigure}
    \hspace{0.001\linewidth}
    \begin{subfigure}[t]{0.32\linewidth}
        \includegraphics[width=\linewidth]{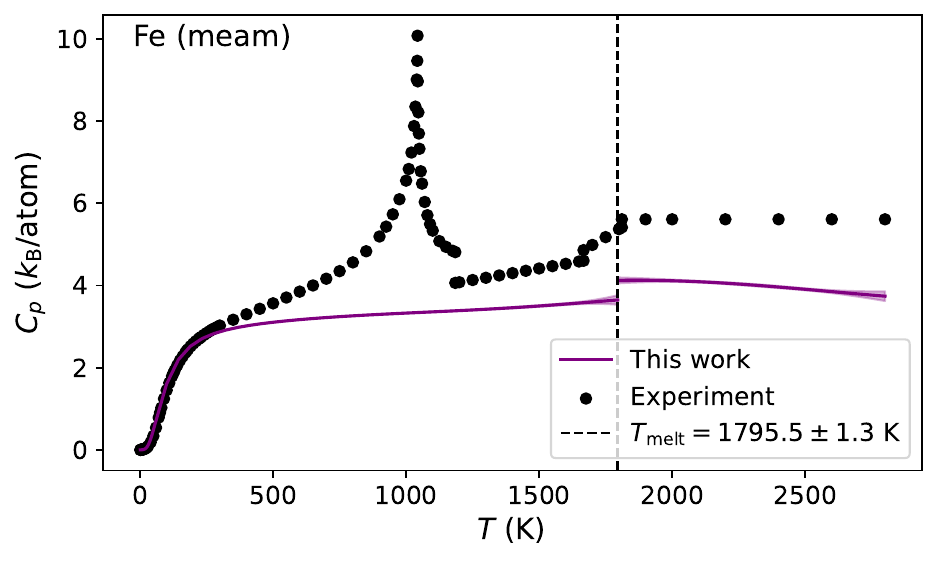}
        \caption{Fe (bcc), MEAM \cite{Asadi_meam} vs exp. from \cite{desai1986iron}}
        \label{fig:cp-fe-etesami}
    \end{subfigure}

    \vspace{0.1cm}

    \begin{subfigure}[t]{0.32\linewidth}
        \includegraphics[width=\linewidth]{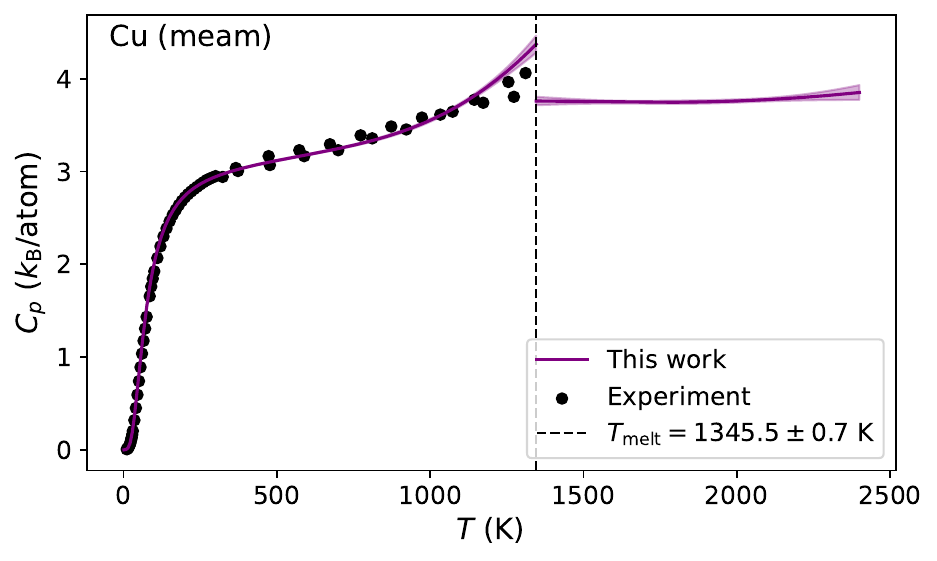}
        \caption{Cu (fcc), MEAM \cite{ETESAMI201861} vs exp. from \cite{cu_marun1960_cu, cu_howse1961_cp, cu_lyustemik1959_cp, cu_franck1961_cp} as compiled in  \cite{touloukian1970_cp}}
        \label{fig:cp-cu-asadi}
    \end{subfigure}
    \hspace{0.001\linewidth}
    \begin{subfigure}[t]{0.32\linewidth}
        \includegraphics[width=\linewidth]{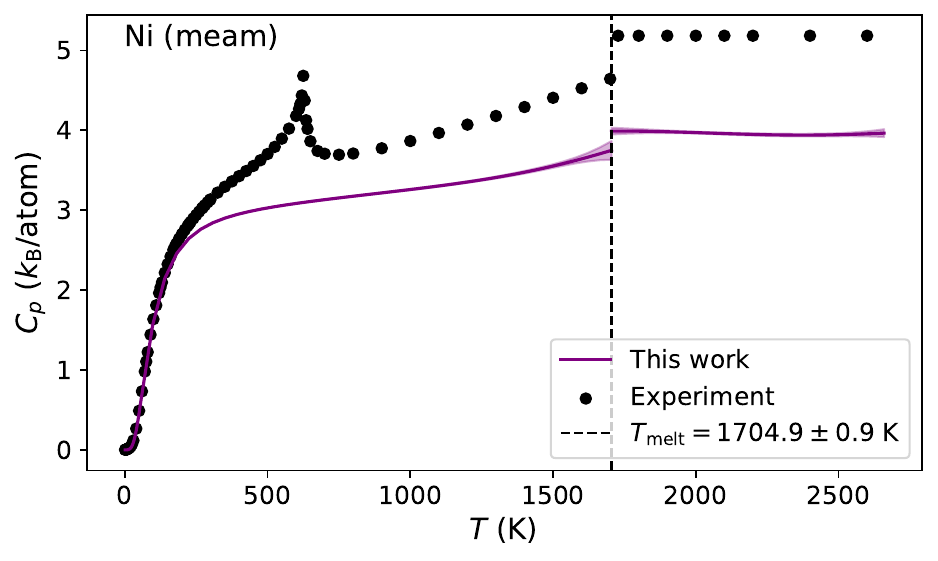}
        \caption{Ni (fcc), MEAM \cite{ETESAMI201861} vs exp. from \cite{desai1987nickel}}
        \label{fig:cp-ni-asadi}
    \end{subfigure}
    \hspace{0.001\linewidth}
    \begin{subfigure}[t]{0.32\linewidth}
        \includegraphics[width=\linewidth]{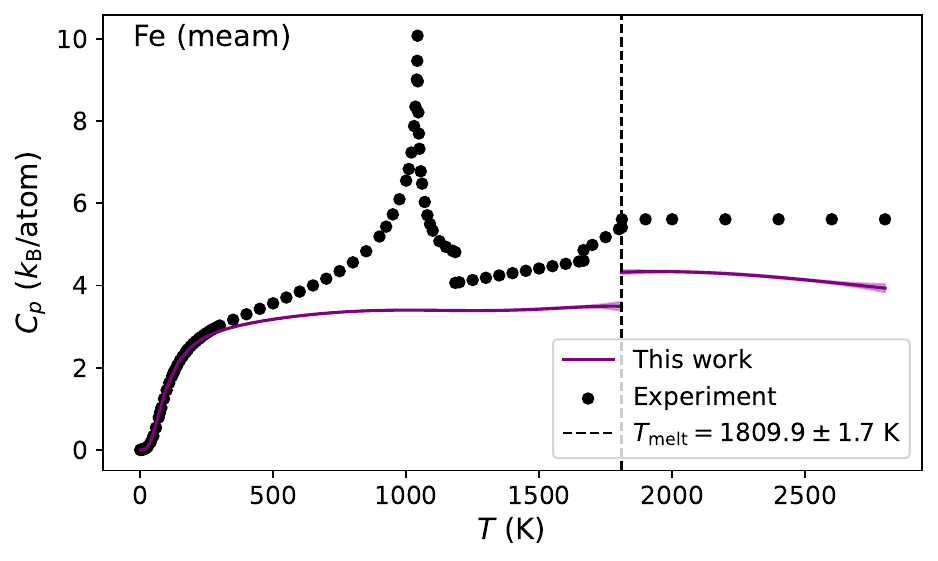}
        \caption{Fe (bcc), MEAM \cite{ETESAMI201861} vs exp. from \cite{desai1986iron}}
        \label{fig:cp-fe-asadi}
    \end{subfigure}

    \vspace{0.1cm}
    \begin{subfigure}[t]{0.32\linewidth}
        \includegraphics[width=\linewidth]{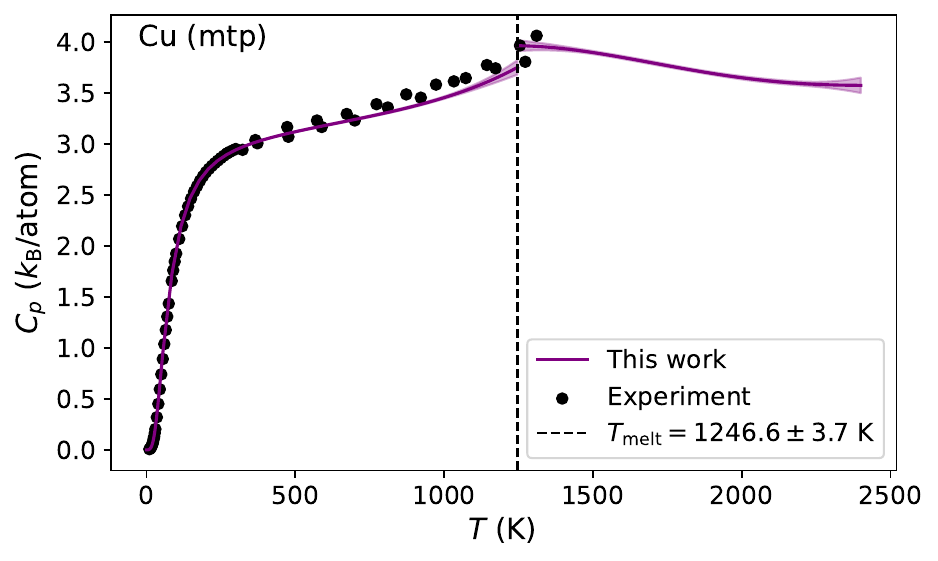}
        \caption{Cu (fcc), MTP vs exp. from \cite{cu_marun1960_cu, cu_howse1961_cp, cu_lyustemik1959_cp, cu_franck1961_cp} as compiled in  \cite{touloukian1970_cp}}
        \label{fig:cp-cu-mtp}
    \end{subfigure}   
    \hspace{0.001\linewidth}
    \begin{subfigure}[t]{0.32\linewidth}
        \includegraphics[width=\linewidth]{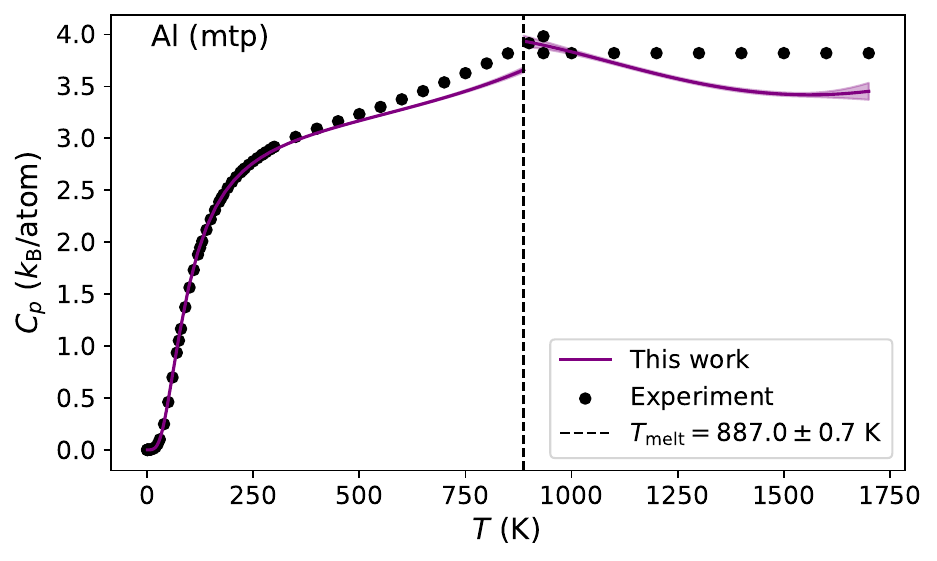}
        \caption{Al (fcc), MTP vs exp. from \cite{desai1987aluminium}}
        \label{fig:cp-al-mtp}
    \end{subfigure}
    \hspace{0.001\linewidth}
    \begin{subfigure}[t]{0.32\linewidth}
        \includegraphics[width=\linewidth]{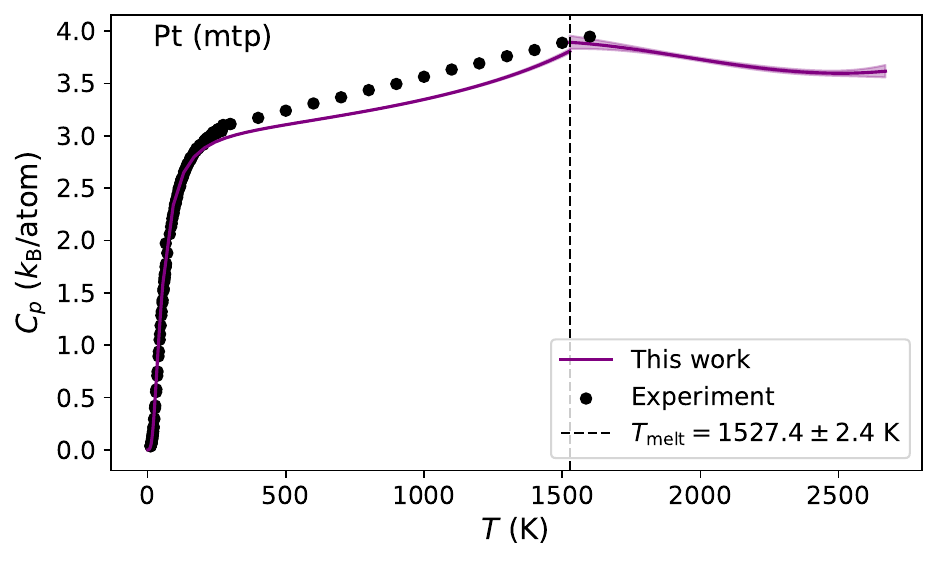}
        \caption{Pt (fcc), MTP vs exp. from \cite{clusius1957_pt, kendall1962_pt} as compiled in \cite{touloukian1970_cp}}
        \label{fig:cp-pt-mtp}
    \end{subfigure}
    
    \caption{Temperature dependence of the constant-pressure heat capacity $C_p(T)$ for FCC and BCC metals calculated using EAM, MEAM, and MTP interatomic potentials. Experimental data are shown for comparison where available.}
    \label{fig:cp-all}
\end{figure}

The reconstructed constant-pressure heat capacity $C_p(T)$, obtained from NVT simulations, is shown in Figure~\ref{fig:cp-all}.
The inclusion of the ZPE correction ensures the correct low-temperature limit, allowing the results to reproduce the expected quantum behavior. Although none of the considered models were explicitly fitted to heat-capacity data, most of the potentials capture the temperature dependence of $C_p(T)$ reasonably well. Unsurprisingly, large deviations occur for magnetic metals such as Fe and Ni, where classical molecular dynamics cannot account for magnetic entropy above the Curie temperature. Once again, the MTP potentials trained on the DFT data exhibit one of the most accurate results among all the models.
It is also worth noting that the MEAM potential for Mo~\cite{KIM2017131}, fitted only to 0~K properties, significantly underestimates $C_p(T)$ at high temperatures, consistent with its underestimated thermal expansion and melting enthalpy. In contrast, the EAM potentials of Zhou et al.~\cite{zhou2004misfit}, despite being parameterized only to 0~K structural and mechanical properties, show particularly good agreement with experiment for the heat capacity and maintain a qualitative agreement for other thermodynamic properties.

\subsection{Isothermal and Adiabatic Bulk Moduli}

\begin{figure}[H]
    \centering

    \begin{subfigure}[t]{0.32\linewidth}
        \includegraphics[width=\linewidth]{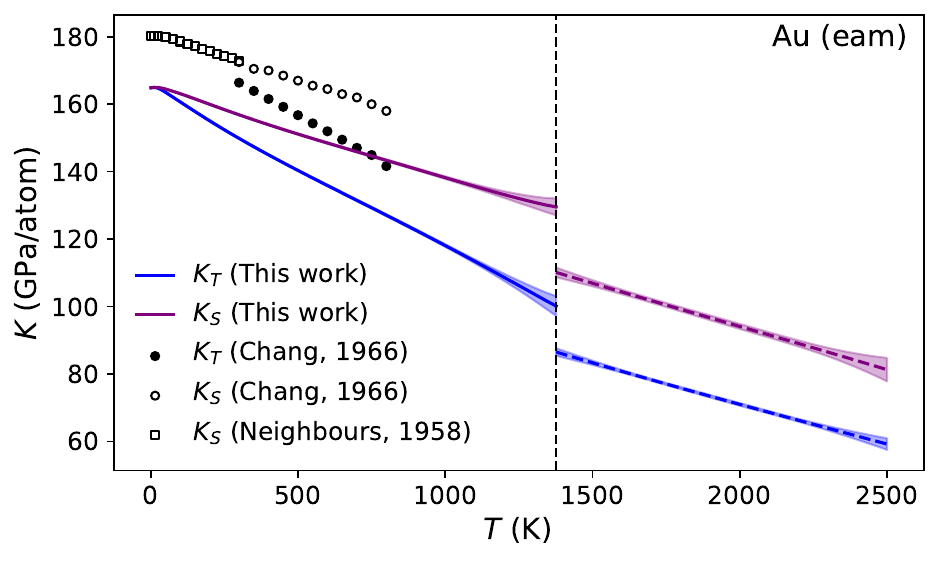}
        \caption{Au (fcc), EAM \cite{zhou2004misfit} vs exp. from \cite{chang1966_bulk_CuAgAu,neighbours_AgAu}}
        \label{fig:bulk-au-zhou}
    \end{subfigure}
    \hspace{0.001\linewidth}
    \begin{subfigure}[t]{0.32\linewidth}
        \includegraphics[width=\linewidth]{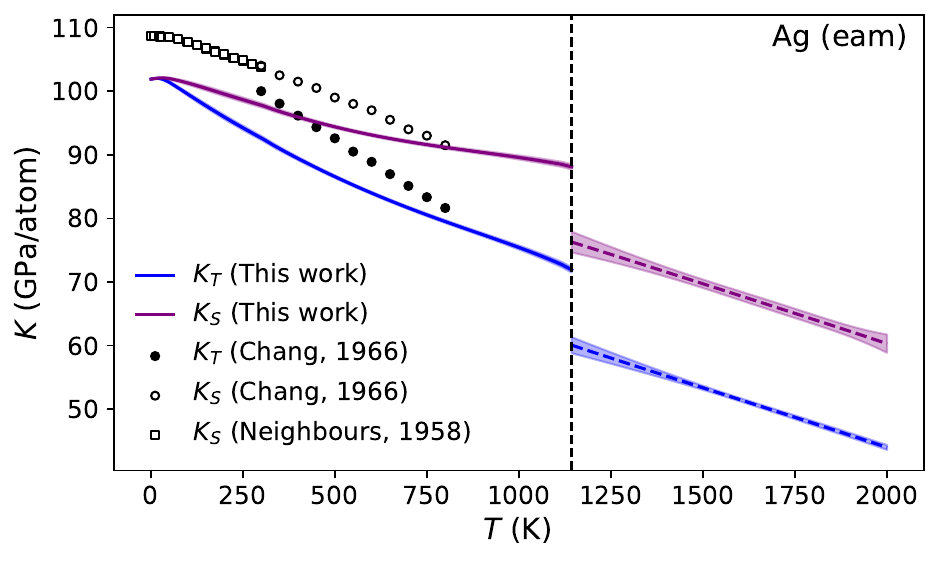}
        \caption{Ag (fcc), EAM \cite{zhou2004misfit} vs exp. from \cite{chang1966_bulk_CuAgAu,neighbours_AgAu}}
        \label{fig:bulk-ag-zhou}
    \end{subfigure}
    \hspace{0.001\linewidth}
    \begin{subfigure}[t]{0.32\linewidth}
        \includegraphics[width=\linewidth]{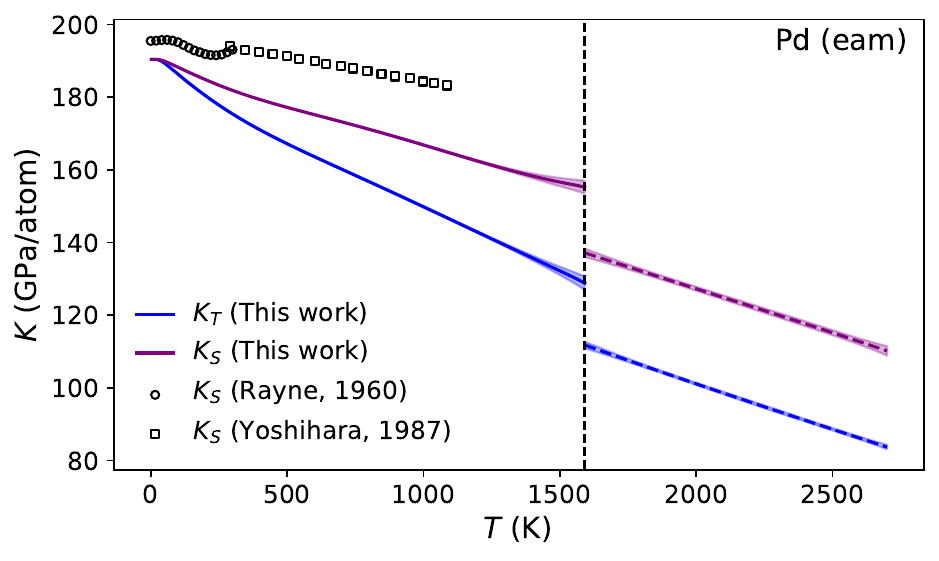}
        \caption{Pd (fcc), EAM \cite{zhou2004misfit} vs exp. from \cite{rayne_Pd,yoshihara_Pd}}
        \label{fig:bulk-pd-zhou}
    \end{subfigure}

    \vspace{0.1cm}

    \begin{subfigure}[t]{0.32\linewidth}
        \includegraphics[width=\linewidth]{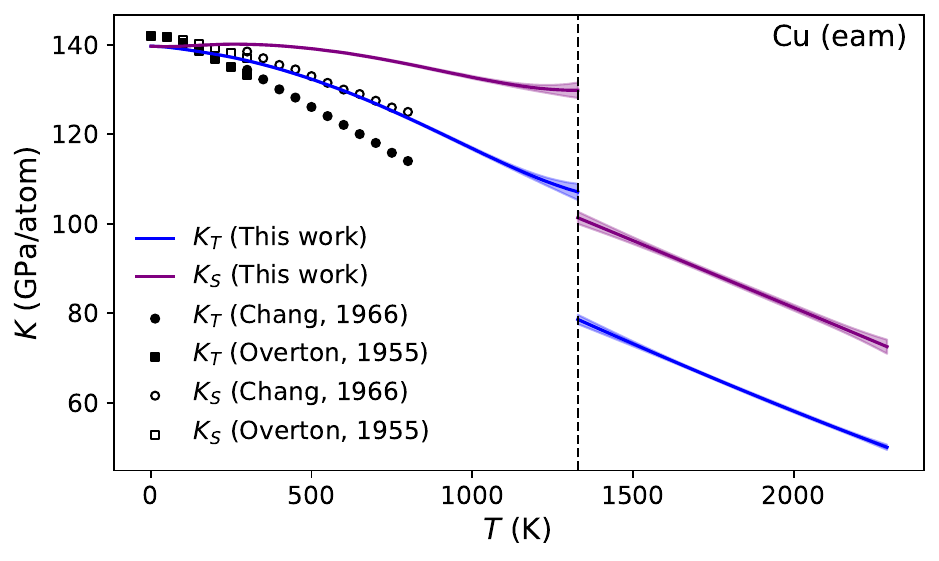}
        \caption{Cu (fcc), EAM \cite{Williams_2006} vs exp. from \cite{chang1966_bulk_CuAgAu, overton_Cu}}
        \label{fig:bulk-cu-williams}
    \end{subfigure}
    \hspace{0.001\linewidth}
    \begin{subfigure}[t]{0.32\linewidth}
        \includegraphics[width=\linewidth]{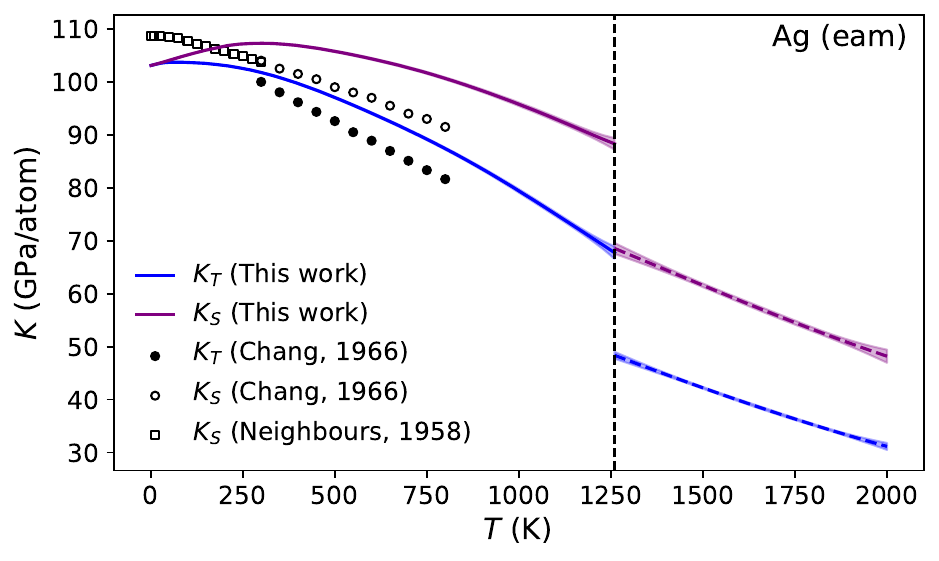}
        \caption{Ag (fcc), EAM \cite{Williams_2006} vs exp. from \cite{chang1966_bulk_CuAgAu,neighbours_AgAu}}
        \label{fig:bulk-ag-williams}
    \end{subfigure}
    \hspace{0.001\linewidth}
    \begin{subfigure}[t]{0.32\linewidth}
        \includegraphics[width=\linewidth]{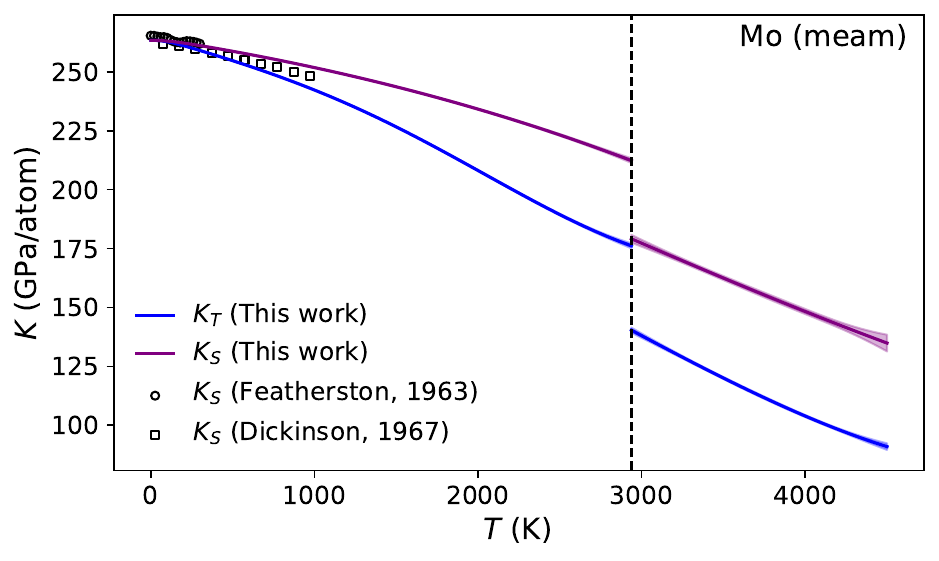}
        \caption{Mo (bcc), MEAM \cite{KIM2017131} vs exp. from \cite{featherston_Mo,dickinson_Mo}}
        \label{fig:bulk-mo}
    \end{subfigure}

    \vspace{0.1cm}

    \begin{subfigure}[t]{0.32\linewidth}
        \includegraphics[width=\linewidth]{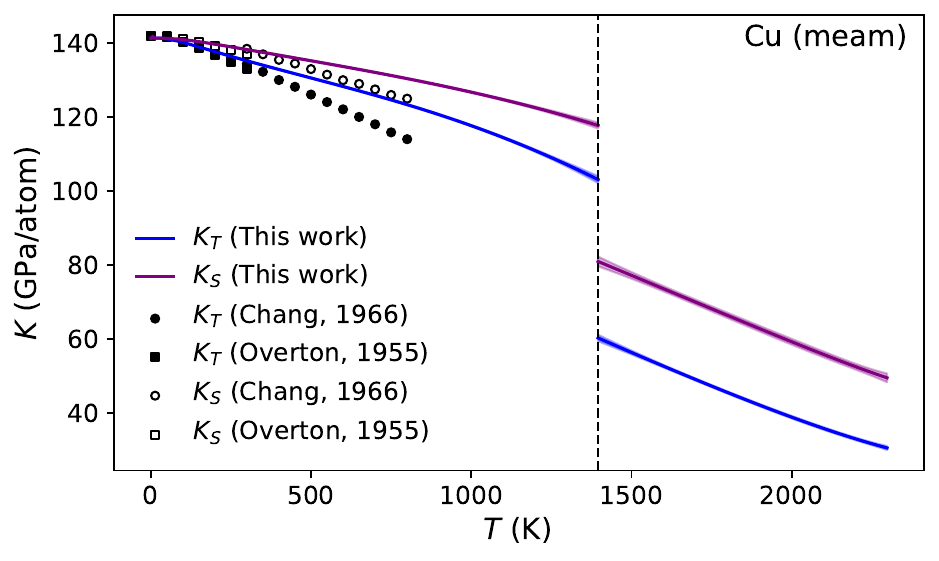}
        \caption{Cu (fcc), MEAM \cite{Asadi_meam} vs exp. from \cite{chang1966_bulk_CuAgAu, overton_Cu}}
        \label{fig:bulk-cu-asadi}
    \end{subfigure}
    \hspace{0.001\linewidth}
    \begin{subfigure}[t]{0.32\linewidth}
        \includegraphics[width=\linewidth]{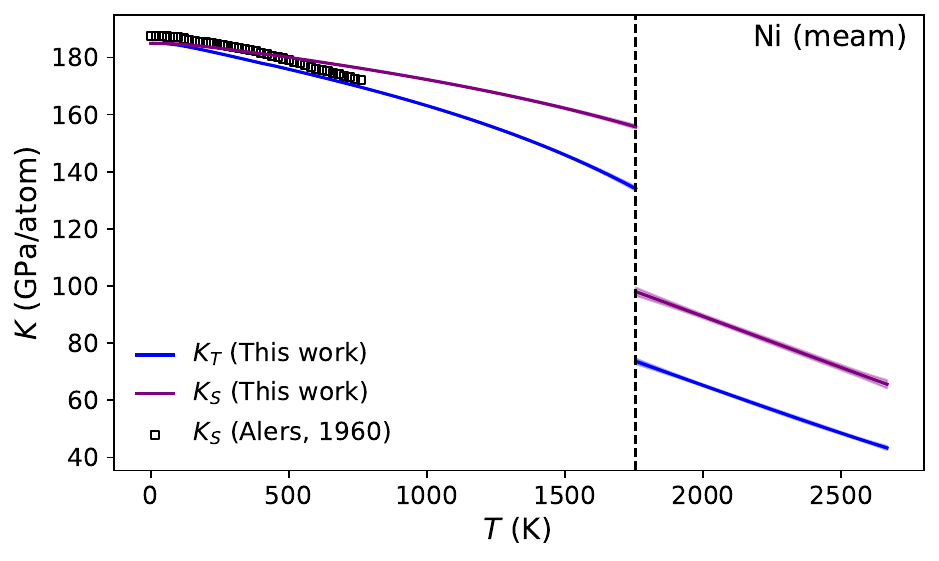}
        \caption{Ni (fcc), MEAM \cite{Asadi_meam} vs exp. from \cite{alers_Ni}}
        \label{fig:bulk-ni-asadi}
    \end{subfigure}
    \hspace{0.001\linewidth}
    \begin{subfigure}[t]{0.32\linewidth}
        \includegraphics[width=\linewidth]{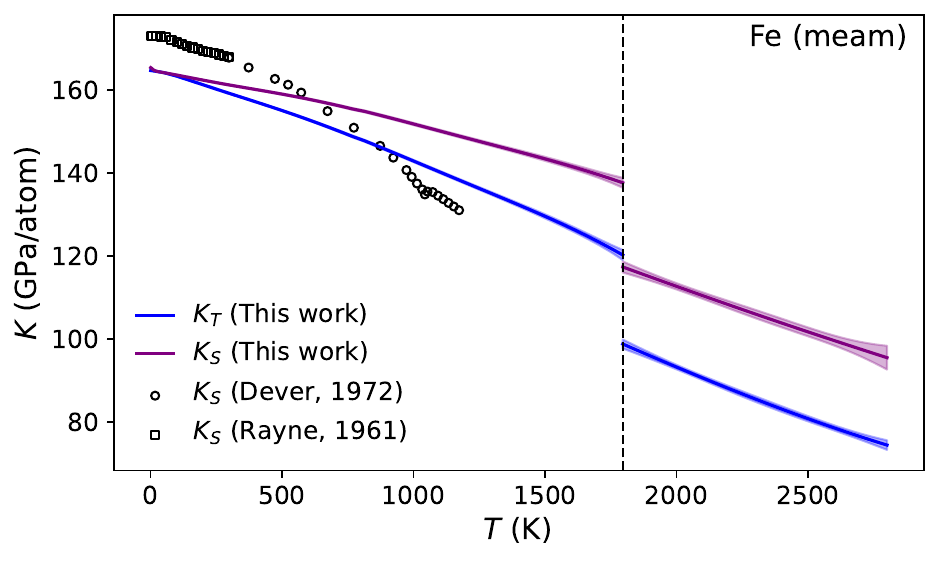}
        \caption{Fe (bcc), MEAM \cite{Asadi_meam} vs exp. from \cite{dever_Fe,rayne_Fe}}
        \label{fig:bulk-fe-asadi}
    \end{subfigure}    

    \vspace{0.1cm}

    \begin{subfigure}[t]{0.32\linewidth}
        \includegraphics[width=\linewidth]{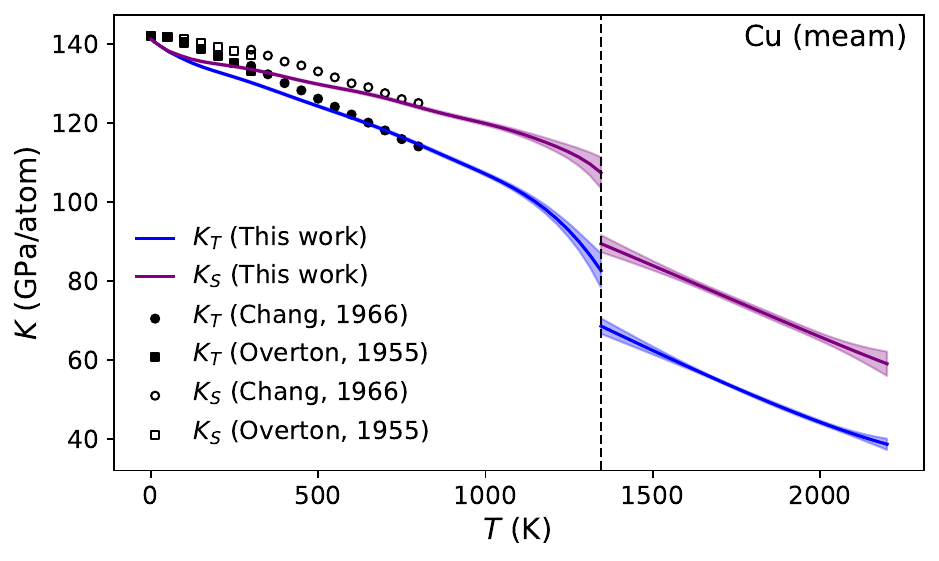}
        \caption{Cu (fcc), MEAM \cite{ETESAMI201861} vs exp. from \cite{chang1966_bulk_CuAgAu, overton_Cu}}
        \label{fig:bulk-cu-etesami}
    \end{subfigure}
    \hspace{0.001\linewidth}
    \begin{subfigure}[t]{0.32\linewidth}
        \includegraphics[width=\linewidth]{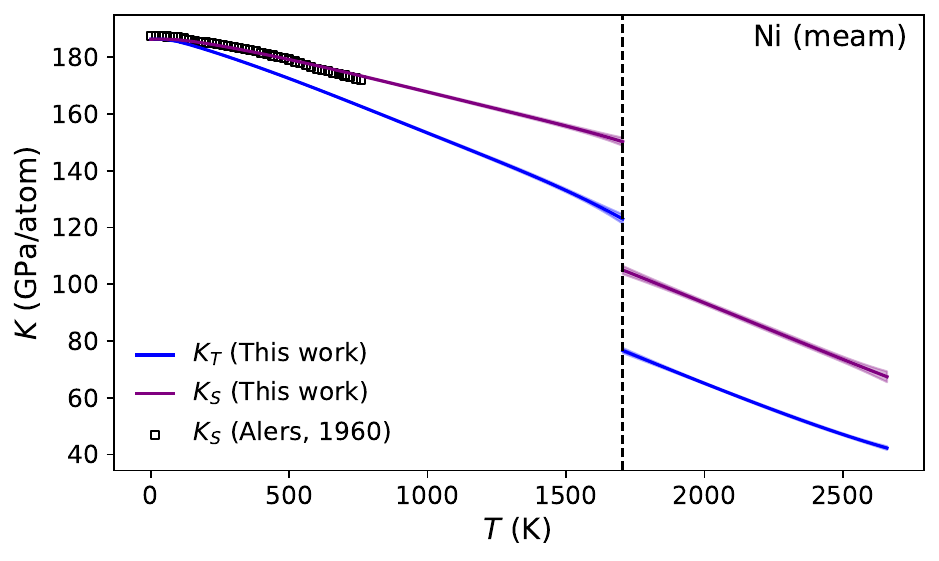}
        \caption{Ni (fcc), MEAM \cite{ETESAMI201861} vs exp. from \cite{alers_Ni}}
        \label{fig:bulk-ni-etesami}
    \end{subfigure}
    \hspace{0.001\linewidth}
    \begin{subfigure}[t]{0.32\linewidth}
        \includegraphics[width=\linewidth]{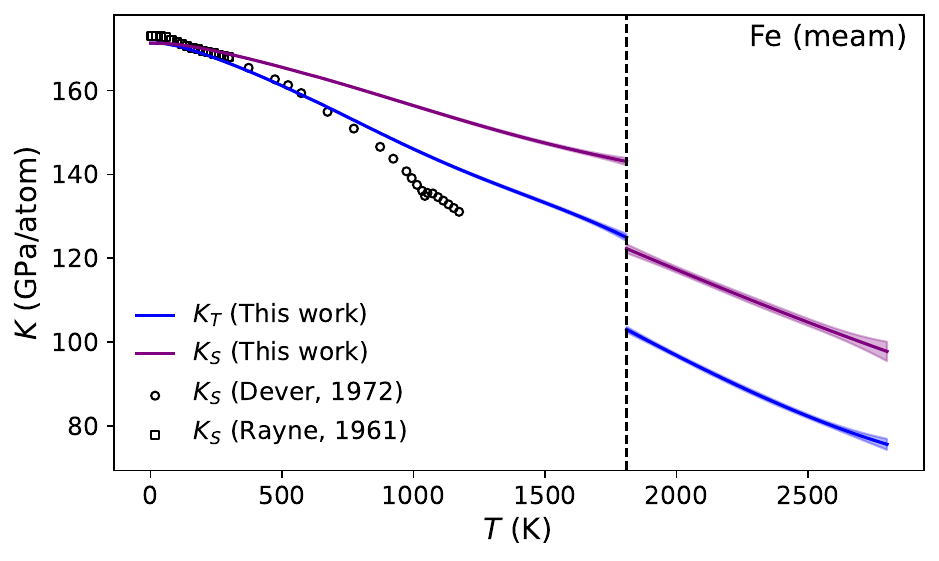}
        \caption{Fe (bcc), MEAM \cite{ETESAMI201861} vs exp. from \cite{dever_Fe,rayne_Fe}}
        \label{fig:bulk-fe-etesami}
    \end{subfigure}

    \vspace{0.1cm}
    \begin{subfigure}[t]{0.32\linewidth}
        \includegraphics[width=\linewidth]{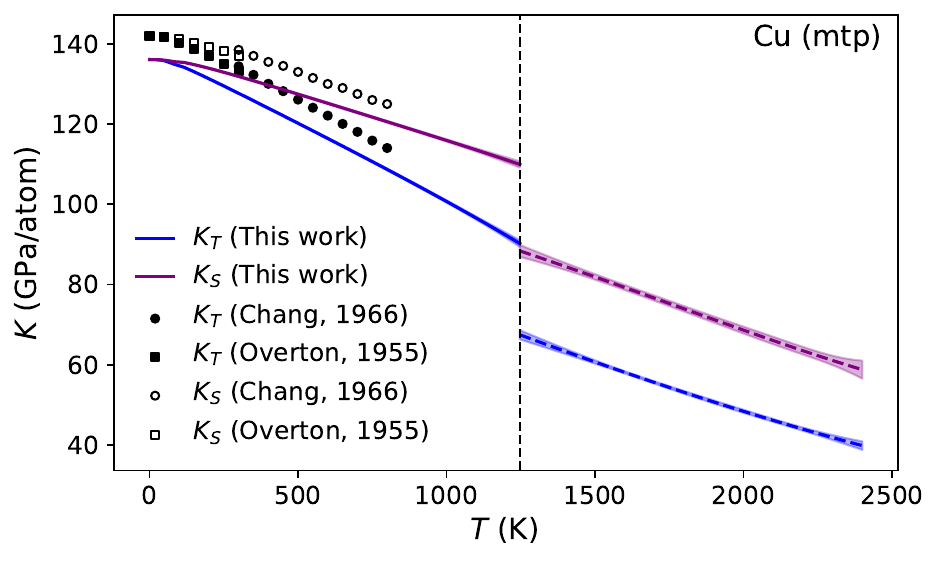}
        \caption{Cu (fcc), MTP vs exp. from \cite{chang1966_bulk_CuAgAu, overton_Cu}}
        \label{fig:bulk-cu-mtp}
    \end{subfigure} 
    \hspace{0.001\linewidth}
    \begin{subfigure}[t]{0.32\linewidth}
        \includegraphics[width=\linewidth]{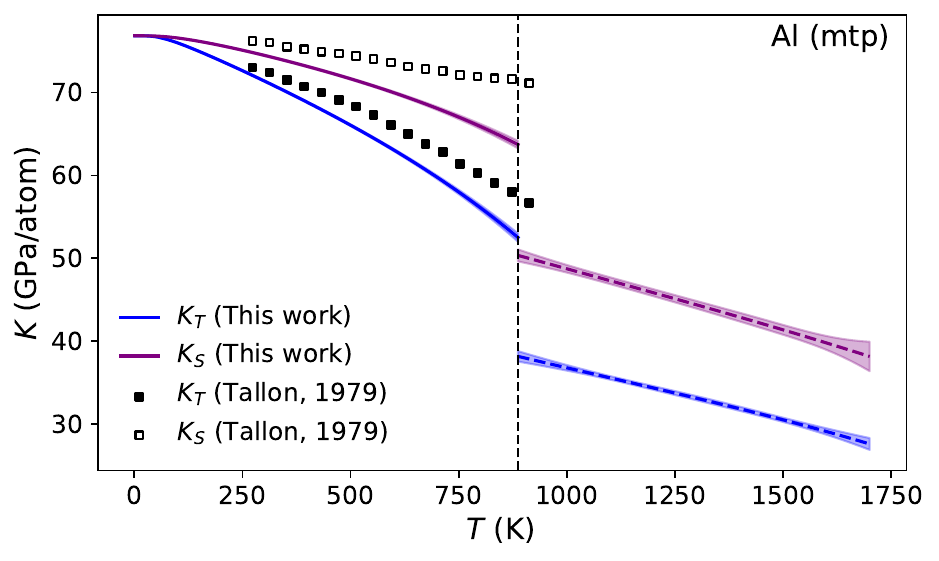}
        \caption{Al (fcc), MTP vs exp. from \cite{tallon1979temperature}}
        \label{fig:bulk-al-mtp}
    \end{subfigure}
    \hspace{0.001\linewidth}
    \begin{subfigure}[t]{0.32\linewidth}
        \includegraphics[width=\linewidth]{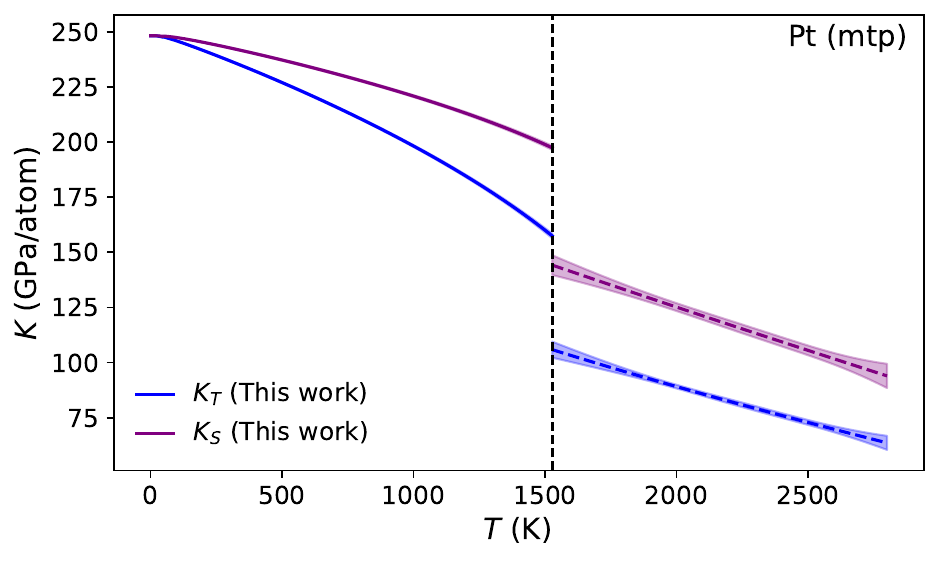}
        \caption{Pt (fcc), MTP}
        \label{fig:bulk-pt-mtp}
    \end{subfigure}
    
    \caption{Temperature dependence of the isothermal ($K_T$) and adiabatic ($K_S$) bulk moduli for FCC and BCC metals calculated using EAM, MEAM, and MTP interatomic potentials. Experimental data are shown for comparison where available.}
    \label{fig:bulk-modulus-other}
\end{figure}

Figure~\ref{fig:bulk-modulus-other} shows the temperature dependence of the isothermal ($K_T$) and adiabatic ($K_S$) bulk moduli for all the studied metals. Experimental data are plotted where available. Across all materials, $K_S$ is slightly higher than $K_T$, as expected from the $C_P/C_V$ ratio. The overall trends are reasonably reproduced by the simulations, although quantitative deviations depend on the interatomic potential. 
Notably, among the considered models, only the MEAM potential of Etesami~\cite{ETESAMI201861} included high-temperature elastic constants during fitting. Compared to the earlier MEAM of Asadi~\cite{Asadi_meam}, this addition improved the agreement with experiment, particularly for Fe and Ni.
The EAM potentials of Zhou~\cite{zhou2004misfit} and Williams~\cite{Williams_2006}, although fitted to reproduce both experimental and DFT elastic constants at 0~K, still show limited quantitative accuracy, even at low temperatures.
As already noted for other properties, the MTP potentials trained on DFT/PBE data reproduce the overall trends well, although in this case their results are less accurate than those of the empirical MEAM potentials~\cite{ETESAMI201861} that explicitly included elastic constants in their fitting procedure.

\subsection{Limitations of potentials in the Quasi-Harmonic Approximation: Demonstration for Platinum.}
\label{sec:limitations}

Some interatomic potentials, while performing well in direct molecular dynamics (MD) simulations, may yield unphysical results when used within the quasi-harmonic approximation (QHA). A notable example is provided by the embedded-atom method (EAM) potential for platinum developed by Zhou~\textit{et al.}~\cite{zhou2004misfit}, originally constructed as part of a generalized alloy EAM database. This potential was parameterized to reproduce static and defect-related properties --- such as the equation of state, elastic constants, cohesive energies, and vacancy formation energies --- and to model defect structures and interface phenomena in metallic multilayers. Importantly, no fitting to vibrational spectra or the volume dependence of phonon frequencies was performed, which can lead to inconsistencies when the potential is used in QHA.

\begin{figure}[H]
\centering
\begin{subfigure}[t]{0.48\textwidth}
    \includegraphics[width=\linewidth]{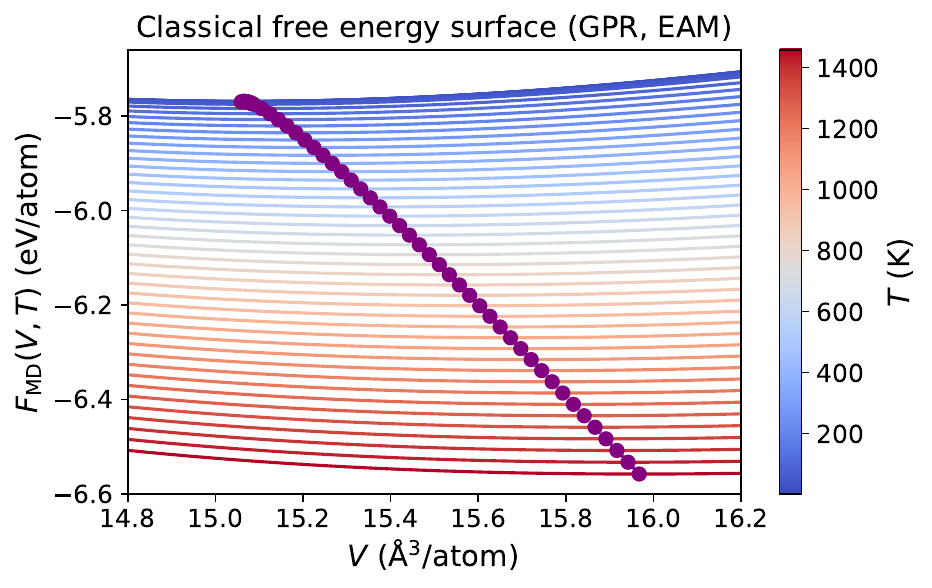}
    \caption{Classical free-energy surface $F_{\rm MD}(V,T)$ from MD with the EAM potential. Convex minima are present at all temperatures.}
\end{subfigure}
\hfill
\begin{subfigure}[t]{0.48\textwidth}
    \includegraphics[width=\linewidth]{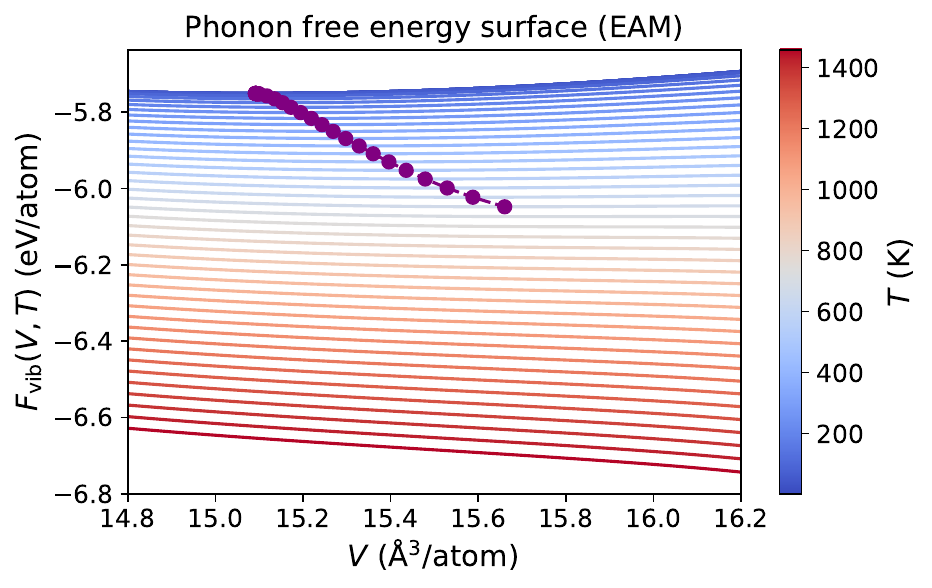}
    \caption{Phonon free-energy surface $F_{\rm vib}(V,T)$ from QHA with the same EAM potential. The equilibrium volume grows unphysically with $T$ and the minimum disappears at high temperatures.}
\end{subfigure}
\caption{Free-energy surfaces for platinum using the EAM potential. While the classical MD surface behaves physically, the QHA surface becomes non-convex at high $T$, preventing equilibrium volume determination.}
\label{fig:pt_fmd_fvib_eam}
\end{figure}

In classical MD, the EAM potential produces a smooth and convex free-energy surface $F_{\rm MD}(V,T)$ with a well-defined minimum at each temperature (Fig.~\ref{fig:pt_fmd_fvib_eam}a), allowing equilibrium volumes to be determined without difficulty. However, when the vibrational free energy $F_{\rm vib}(V,T)$ is computed within QHA using the same potential, the results become inconsistent: the predicted equilibrium volume increases much more rapidly with temperature than in the classical case, and at high $T$ the minimum disappears altogether, with $\partial F_{\rm vib}/\partial V$ never crossing zero (Fig.~\ref{fig:pt_fmd_fvib_eam}b). In this regime, the QHA correction cannot be applied.

\begin{figure}[H]
\centering
\begin{subfigure}[t]{0.48\textwidth}
    \includegraphics[width=\linewidth]{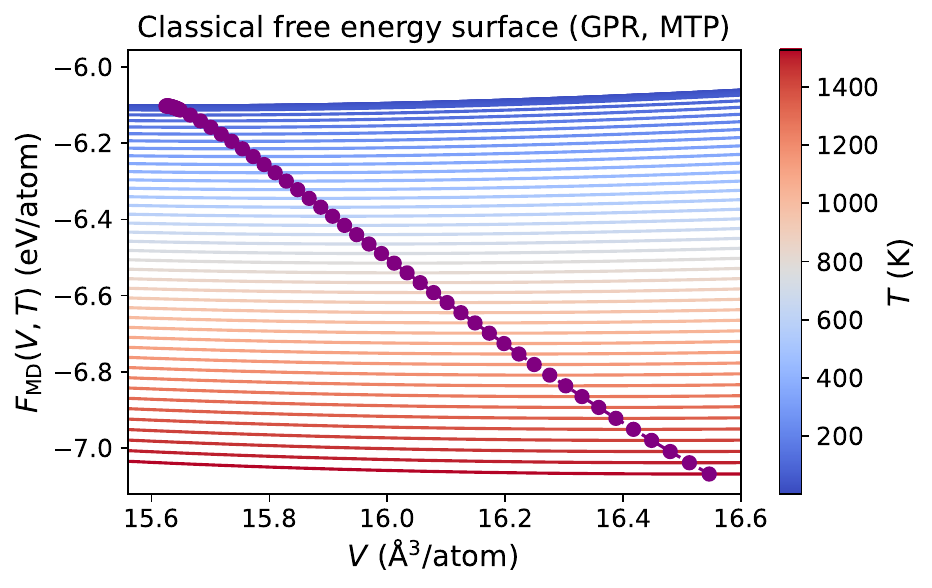}
    \caption{Classical free-energy surface $F_{\rm MD}(V,T)$ from MD with the MTP potential. Convex minima are present at all temperatures.}
\end{subfigure}
\hfill
\begin{subfigure}[t]{0.48\textwidth}
    \includegraphics[width=\linewidth]{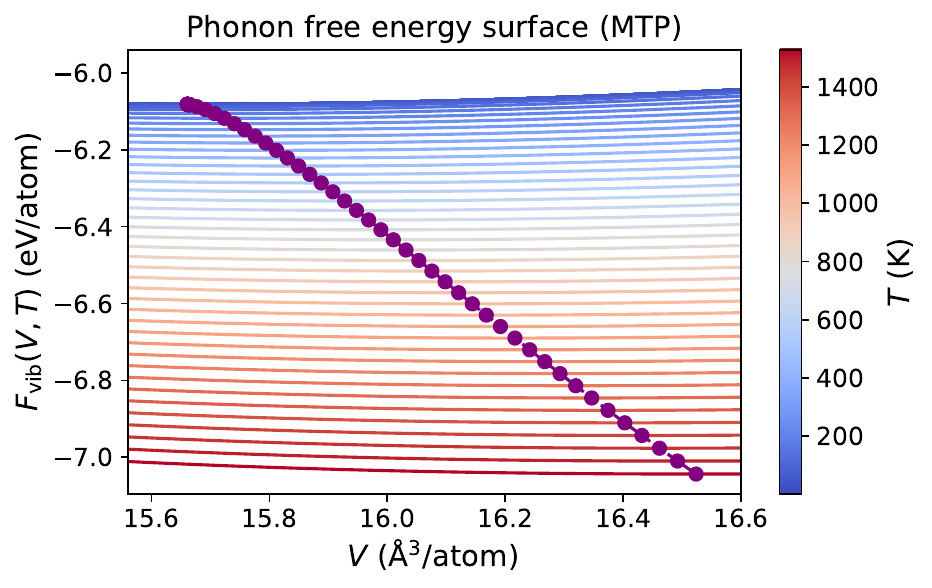}
    \caption{Phonon free-energy surface $F_{\rm vib}(V,T)$ from QHA with the MTP potential. The surface remains convex and physically consistent across all $T$.}
\end{subfigure}
\caption{Free-energy surfaces for platinum using the MTP potential. Both classical MD and QHA surfaces are convex and yield consistent equilibrium volumes.}
\label{fig:pt_fmd_fvib_mtp}
\end{figure}

To check whether this issue originates from the potential or from QHA itself, we repeated the analysis using a Moment Tensor Potential (MTP) trained on DFT data. In this case, both the classical $F_{\rm MD}(V,T)$ and the phonon $F_{\rm vib}(V,T)$ surfaces remain convex across the entire temperature range, with consistent trends in equilibrium volume (Fig.~\ref{fig:pt_fmd_fvib_mtp}).
All thermodynamic properties for platinum obtained with the MTP potential have been computed and presented earlier in the corresponding figures.

This confirms that the issue with QHA is not intrinsic to the method but arises due to limitations of the EAM potential. The MTP potential accurately captures the volume dependence of vibrational properties, validating the QHA approach in this case. This also demonstrates the ability of our framework to evaluate the internal consistency of thermodynamic models and detect when empirical potentials are not suitable for QHA-based corrections.

\section{Concluding Remarks}
\label{sec:conclusion}

We have proposed a unified computational methodology for reconstructing the Helmholtz free-energy surface $F(V,T)$ from molecular dynamics data and systematically calculating thermodynamic properties from its derivatives. By combining MD simulations with Gaussian Process Regression and incorporating a zero-point energy correction from the harmonic or quasi-harmonic approximation, the method captures both high-temperature anharmonic effects and low-temperature quantum behavior. The framework is general and applies to any interatomic potential, including empirical and machine-learned models. We demonstrated its capabilities by computing heat capacities, thermal expansion coefficients, isothermal and adiabatic bulk moduli, and melting properties for nine elemental FCC and BCC metals using 20 different potentials, with all results accompanied by quantified statistical uncertainties.

Apart from demonstrating the method's capabilities, our results reveal interesting trends in the performance of the interatomic potentials. The MTPs deliver the most balanced description across all the examined properties. Although trained exclusively on DFT data without any experimental input, they accurately reproduce the temperature dependence of thermodynamic quantities and capture the correct qualitative trends. Their remaining discrepancies with experiment are systematic and can be attributed to the well-known limitations of the PBE density functional employed in their training. In contrast, empirical potentials (EAM and MEAM) exhibit substantially larger deviations overall. When these potentials are parameterized using experimental data for specific properties, they reproduce those target properties with high fidelity; however, this improvement does not generally extend to other thermodynamic observables.

Beyond accuracy and efficiency, our methodology provides flexibility: it can treat crystalline and liquid phases, and it includes an NPT-based variant suitable for rapid screening when only melting-related properties and heat capacities are required. The Bayesian nature of the approach further enables active learning, allowing simulation points to be selected adaptively for maximal information gain. Importantly, because the workflow is fully automated and transferable across systems, it aligns with broader community efforts to standardize and accelerate materials simulations. In this role, it can also act as a validation tool, systematically benchmarking interatomic potentials against reference or experimental data.

Looking forward, the framework opens the door to automated exploration of phase diagrams, high-throughput screening of candidate materials, and integration with existing materials-design platforms. Its combination of generality, automation, and quantified uncertainty makes it a powerful step toward reliable, data-driven prediction of thermodynamic properties across the materials domain.

\section*{Acknowledgments}
This work was supported by the Russian Science Foundation (grant number 23-13-00332, 
\url{https://rscf.ru/project/23-13-00332/}).

\bibliographystyle{unsrt}
\bibliography{bibliography}

\end{document}


\begin{center}
{\LARGE \textbf{Supplementary Information}} \\[6pt]
for \\[6pt]
\textit{Automated Prediction of Thermodynamic Properties via Bayesian Free-Energy Reconstruction from Molecular Dynamics} \\[12pt]
Ekaterina Spirande, Timofei Miryashkin, Andrei Kolmakov, Alexander Shapeev
\end{center}

\section{Free Energy Asymptotics}
\label{ap:asympotics}

As discussed in the Main Text (Section 2.2), we represent the free energy in the form
\begin{equation*}
F(T, V, N) = F_{\text{ref}}(T, V, N) - T S(T, V, N),
\end{equation*}
where $F_{\text{ref}}$ is the reference free energy and $S$ we call the `entropy'. 

We reconstruct the free energy by fitting the entropy derivatives to the data obtained in molecular dynamics simulations. Using the expression for the reference energy, we may obtain the derivatives of the entropy. Following Ladygin~\textit{et al.}~\cite{Ladygin}, we specify $F_{\text{ref}}$ separately for the solid and liquid phases.

For the solid phase, the reference free energy:
\begin{equation*}
F_{\text{ref}}^{(\rm s)} (T, V, N) := E_0(V) + T \left( -\log(NV) + 1 - {\textstyle\frac{3}{2}} \log(2 \pi T) \right),
\end{equation*}
where $E_0(V)$ is the potential energy at zero temperature for the given volume. 
The corresponding entropy derivatives, obtained from MD averages, are
\begin{equation*}
\frac{\partial S^{\rm (s)}}{\partial T} = T^{-2} \langle E - E_0 \rangle - {\textstyle\frac{3}{2}} T^{-1}, 
\end{equation*}
\begin{equation*}
\frac{\partial S^{\rm (s)}}{\partial V} = T^{-1} \langle P - P_0 \rangle, 
\end{equation*}
where $P_0 := -\frac{\partial E_0}{\partial V}$ is the pressure at zero temperature. 

For the liquid phase we take
\begin{equation*}
F_{\text{ref}}^{(\mathrm l)} (T, V, N) = -T \log(NV),
\end{equation*}
which yields the following entropy derivatives:
\begin{equation*}
\frac{\partial S^{(\mathrm l)}}{\partial T} = \frac{\langle E \rangle}{T^{2}}, 
\qquad
\frac{\partial S^{(\mathrm l)}}{\partial V} = \frac{\langle P \rangle}{T}.
\end{equation*}

\section{Gaussian Process Regression}
\label{ap:gpr}

In the Main Text (Section 2.3) we introduced the use of Gaussian Process Regression (GPR) for reconstructing the entropy from molecular dynamics data.  
Here we provide the detailed formulation of the method: how the data are represented as linear functionals, how the joint Gaussian distribution is constructed, and how prediction and active learning are carried out.

In Gaussian Process Regression, the key idea is that any data point can be viewed as a linear functional $X$ acting on $S$, such as $ \<S | X\> = S(T, V) $ or $ \<S | X\> = \frac{\partial S}{\partial T}(T, V) $. Accordingly, we generalize the kernel definition to arbitrary functionals $X_1$ and $X_2$:
\[
k(X_1, X_2) := {\rm Cov}[\< S| X_1 \>, \< S| X_2 \>].
\]

We now discuss the Gaussian Process Regression algorithm in which we use the kernels defined above.
We perform calculations in the data point (functional) $X_i$ for which we obtain the target value $Y_i$ with the Gaussian noise $\Delta Y_i$, and hence require that the input point has the Gaussian distribution:
\[
\< S | X_i\> \sim \mathcal{N} \big( Y_i, (\Delta Y_i)^2 \big).    
\]
Thus, the input data to the regression algorithm has the form of tuples \((X_i, Y_i, \Delta Y_i)\), and we denote the collection of all such points as \(\mathcal{D} = \{(X_i, Y_i, \Delta Y_i)\}_{i=1}^n\).
To predict the value $Y_* = \<S | X_*\>$ at a new point $X_*$, we form a joint Gaussian distribution over $\{Y_i\}$ and $Y_*$, with zero mean and covariance:
\[
	{\rm cov}
	\begin{pmatrix}
		\boldsymbol{Y} \\
		Y_*
	\end{pmatrix}
=
	\begin{pmatrix}
		K(\boldsymbol{X}, \boldsymbol{X})+\operatorname{diag}(\boldsymbol{\Delta} \boldsymbol{Y}^2) & K\left(\boldsymbol{X}, X_*\right) \\
		K\left(X_*, \boldsymbol{X}\right) & K\left(X_*, X_*\right)
	\end{pmatrix},
\]
where ${\bm X}$ and ${\bm Y}$ are the vectors composed of $X_i$ and $Y_i$, $\operatorname{diag}(\boldsymbol{\Delta Y}^2)$ is the matrix with $(\Delta Y_i)^2$ on the diagonal, and  $K(\bm{X}, \bm{X})$ is the matrix composed of $k(X_i, X_j)$. 

Following \cite{rasmussen2006-book-gaussian}, the posterior distribution of $Y_*$ is Gaussian with:
\begin{align} \notag
\mathbb{E}[Y_*] &= K(X_*, \bm{X}) K_y^{-1} \bm{Y}, \qquad \text{and}
\\ \label{eq:var_gp}
\mathbb{V}[Y_*] &= K(X_*, X_*) - K(X_*, \bm{X}) K_y^{-1} K(\bm{X}, X_*),
\end{align}
where $K_y = K(\bm{X}, \bm{X}) + \operatorname{diag}(\bm{\Delta Y}^2)$.

The predicted confidence intervals of thermodynamic properties can be minimized in an optimal manner using an active learning approach. We define the information function for a candidate point $X_*$:
\begin{equation}
\mathcal{H}(X_*) = \sum_i \mathcal{H}(\mathcal{Q}_i \mid X_*) = -\sum_i \log \frac{ \mathbb{V}[\mathcal{Q}_i \mid \mathcal{D} \cup X_*] }{ \mathbb{V}[\mathcal{Q}_i \mid  \mathcal{D}]},
\end{equation}
where \(\mathcal{Q}_i\) are functionals of the entropy whose uncertainties we aim to minimize. In our study, these functionals represent thermodynamic properties. For example, to reduce the uncertainty of the heat-capacity curve, we discretize it at temperatures \(\{T_i\}\), so that \(\mathcal{Q}_i = C_P(T_i)\). The variances \(\mathbb{V}[\mathcal{Q}_i \mid \mathcal{D} \cup X_*]\) and \(\mathbb{V}[\mathcal{Q}_i \mid \mathcal{D}]\) are computed from the Gaussian process posterior using \eqref{eq:var_gp}. The next simulation point is chosen as 
\begin{align*}
    X_{\text{next}} = \argmax_{X_*} \mathcal{H}(X_*),
\end{align*}
and the greedy selection is repeated until the desired convergence is achieved.

The covariance of a nonlinear functional $\mathcal{Q}[S]$ of a Gaussian process $S$ can be estimated by first linearizing $\mathcal{Q}$ around the process mean $\overline{S}=\mathbb{E}[S]$:
\begin{equation}
\mathcal{Q}[S] \approx \mathcal{Q}[\overline{S}] 
+ \langle S-\overline{S}, J \rangle,
\label{eq:linear_of_nonlinear_functional}
\end{equation}
where
\[
  J \;=\; \left.\frac{\delta \mathcal{Q}}{\delta S}\right|_{S=\overline{S}}
\]
is the functional derivative of $\mathcal{Q}$ evaluated at $\overline{S}$.
After accounting for the divergent terms in the reference free energy, the entropy $S$ is a smooth function of its arguments. With sufficient data, the posterior fluctuations $S - \overline{S}$ are small, so the first-order approximation is well justified.
Under this first-order approximation, $\mathcal{Q}[S]$ is
linear in $S$, and its covariance follows directly from the linear-functional expressions given above.

Gaussian Process Regression is a non-parametric machine learning method that is defined by its kernel and a set of hyperparameters, such as ${\bm \theta} = (\theta_0, \theta_f, \theta_T, \theta_V, \theta_N)$ in (3) in the Main Text. These hyperparameters are learned by maximizing the marginal likelihood $p({\bm Y} \mid {\bm X}, {\bm \theta})$, which represents the probability of observing the targets ${\bm Y}$ given the inputs ${\bm X}$ and hyperparameters ${\bm \theta}$. The log marginal likelihood is given by
\begin{align*}
    \log p({\bm Y} \mid {\bm X}, {\bm \theta}) 
    &= -\tfrac{1}{2} {\bm Y}^{\mathsf{T}} K_y^{-1} {\bm Y}
    - \tfrac{1}{2} \log |K_y|
    - \tfrac{n}{2} \log(2\pi),
\end{align*}
where $n$ is the number of input points.

\section{Thermodynamic Properties}
\label{appendix:td_properties}

In the Main Text (Section 2.5), we summarized the thermodynamic quantities that can be obtained from the free-energy surface $F(V,T)$. All these quantities are derived from the GPR model of $F(V,T)$ together with their associated uncertainties. In this appendix, we focus on deriving the thermodynamic relations, while in the following appendix we describe in detail how the GPR uncertainties are propagated to each property. 

All properties are evaluated at the equilibrium volume $V_{\mathrm{eq}}(T)$, defined by minimizing the free energy with respect to $V$:
\begin{equation}
\left.\frac{\partial F}{\partial V}\right|_{V = V_{\text{eq}}(T)} = 0.
\label{eq_ap:equilibrium_condition}
\end{equation}

\subsection{Thermodynamic Properties at Melting Point}
\label{appendix:melting_properties}

At the melting temperature, we compute the thermodynamic differences in volume and enthalpy between the solid and liquid phases. The volume change upon melting is
\begin{equation}
\Delta V = V_{\rm liq}(T_{\rm melt}) - V_{\rm sol}(T_{\rm melt}),
\end{equation}
where each volume is obtained by minimizing the corresponding free-energy surface at $T_{\rm melt}$.

The enthalpy $H$ at temperature $T$ and zero pressure is evaluated from the free energy as
\begin{equation*}
    H(T)= F(T, V_{\rm eq}) - T \left. \frac{\partial F}{\partial T} \right|_{V = V_{\rm eq}(T)} ,
\end{equation*}
so that the enthalpy of fusion is
\begin{equation}
\label{eq_ap:delta_h}
\Delta H = H_{\rm liq}(T_{\rm melt}) - H_{\rm sol}(T_{\rm melt}).
\end{equation}

\subsection{Thermal Expansion}
\label{appendix:thermal_expansion}

The linear thermal expansion coefficient $\alpha(T)$ is defined as
\begin{equation}
\alpha(T) = \frac{1}{3V_\mathrm{eq}(T)} \left( \frac{\partial V_{\mathrm{eq}}(T)}{\partial T} \right),
\end{equation}
for an isotropic material. To evaluate $\partial V_{\mathrm{eq}}/\partial T$, we differentiate the equilibrium condition \eqref{eq_ap:equilibrium_condition} for $\beta F$ with respect to $\beta=1/T$:
\begin{equation*}
\frac{\partial}{\partial\beta}
\left[
\frac{\partial(\beta F)}{\partial V}
\right]
=
\frac{\partial^{2}(\beta F)}{\partial V^{2}}\,
\frac{\partial V}{\partial\beta}
+
\frac{\partial^{2}(\beta F)}{\partial\beta\,\partial V}.
\end{equation*}
Solving for $\partial V_\mathrm{eq}/\partial \beta$ gives
\[
\frac{\partial V_\mathrm{eq}}{\partial \beta}
=
-
\frac{
\displaystyle
\frac{\partial^{2}(\beta F)}{\partial \beta\,\partial V}\Big|_{V = V_\mathrm{eq}}
}{
\displaystyle
\frac{\partial^{2}(\beta F)}{\partial V^{2}}\Big|_{V = V_\mathrm{eq}}
}.
\]
Since $\beta = 1/T$, we have $d\beta/dT = -1/T^2$, and therefore
$\partial V_{\mathrm{eq}}/\partial T $ gives
\begin{equation}
\frac{\partial V_{\mathrm{eq}}}{\partial T}
=
-
\frac{
\displaystyle
\left.
\dfrac{\partial^{2}(\beta F)}{\partial T\,\partial V}
\right|_{V = V_{\mathrm{eq}}}
}{
\displaystyle
\left.
\dfrac{\partial^{2}(\beta F)}{\partial V^{2}}
\right|_{V = V_{\mathrm{eq}}}
}.
\label{eq_ap:dv_eq_dt}
\end{equation}

\subsection{Heat Capacity}
\label{appendix:heat_capcity}
The constant-pressure heat capacity at zero pressure is defined as:
\begin{equation}
\label{eq_ap:cp_def}
C_P (T) = -T \left( \frac{\partial^2 G}{\partial T^2} \right)_{P=0}.
\end{equation}

To evaluate $C_P(T)$, we first note that in the thermodynamic limit at $P=0$ the Gibbs free energy  equals the Helmholtz free energy at the equilibrium volume:
\begin{equation}
\label{eq_ap:gibbs_helmholtz}
G(T, P=0) = F (T, V_{\text{eq}}(T)).
\end{equation}

Taking a total derivative of \eqref{eq_ap:gibbs_helmholtz} with respect to $T$ twice gives
\begin{equation}
\label{eq_ap:cp}
    C_{p}(T) = C_V(T) - T \frac{\partial^{2} F}{\partial T\,\partial V} (T, V_{\text{eq}}(T)) \,
    \frac{\partial V_{{\rm eq}}}{\partial T} (T),
\end{equation}
where the constant-volume heat capacity is defined as
\begin{equation}
\label{eq_ap:cv}
C_V(T) = -T \left( \frac{\partial^2 F}{\partial T^2} \right)_{V=V_{\mathrm eq}}.
\end{equation}

\subsection{Bulk Modulus}
\label{appendix:bulk}

The isothermal bulk modulus at zero pressure is obtained from the second derivative of the Helmholtz free energy with respect to volume, evaluated at the equilibrium volume $V_{{\rm eq}}(T)$:
\begin{equation}
\label{eq_ap:kt}
K_T(T) = V_{{\rm eq}}(T) \left. \frac{\partial^2 F}{\partial V^2} \right|_{V = V_{{\rm eq}}(T)}.
\end{equation}

This expression follows directly from the thermodynamic relation $K_T = -V(\partial P/\partial V)_T$ together with the identity $P = -(\partial F/\partial V)_T$. We evaluate $K_T(T)$ by taking the second derivative of our fitted free-energy surface at the equilibrium volume. 

The adiabatic bulk modulus $K_S(T)$ is obtained from the standard relation
\begin{equation}
\label{eq_ap:ks}
K_S(T) = \left. K_T(T) \, \frac{C_P(T)}{C_V(T)} \right|_{V = V_{{\rm eq}}(T)},
\end{equation}
where the heat capacities $C_P$ and $C_V$ are evaluated from the free-energy derivatives according to \eqref{eq_ap:cp} and \eqref{eq_ap:cv}.

\section{Variance of Thermodynamic Properties}
\label{appendix:thermodynamic_variance}

In this appendix, we present the calculation of uncertainties for thermodynamic properties obtained from the GPR model. The reconstructed free-energy surface carries statistical uncertainty, and therefore all thermodynamic quantities derived from its derivatives also inherit uncertainty. To quantify this, we linearize each functional with respect to the free energy.

In practice, the GP is trained on the entropy surface $S(T,V)$ rather than directly on the Helmholtz free energy $F(T,V)$. Since
\[
S(T, V) = -\beta F(T, V) + Q(T, V),
\]
where $Q(T,V)$ is a chosen reference contribution with $\mathbb{V}[Q(T,V)] = 0$, any linear functional $A[\,]$ satisfies
\[
\mathbb{V}[A[S]] = \mathbb{V}[A[-\beta F + Q]] 
= \mathbb{V}[-A[\beta F] + A[Q]] 
= \mathbb{V}[A[\beta F]],
\]
by the linearity of $A$ and the vanishing variance of $Q$.  
Therefore, for the purposes of variance analysis, all expressions can be equivalently written in terms of $\beta F$, even though the GP fit is performed on $S$.

\subsection{Variance of the equilibrium volume}
\label{appendix:var_of_V_eq}

Here we show how to analytically track the variance of the equilibrium volume $V_{\mathrm eq}$. 
The equilibrium volume is defined implicitly by  
\begin{equation}
\Phi  \bigl[ V_{\mathrm{eq}}, \beta F \bigr]
:=
\left.\frac{\partial (\beta F)}{\partial V}\right|_{V=V_{\mathrm{eq}}}=0,
\label{eq:eq-cond}
\end{equation}
where $\beta F(V)$ is the scaled free energy given by the Gaussian process; the explicit temperature dependence is omitted to lighten notation.  
Because $\beta F$ is random, $V_{\mathrm{eq}}$ inherits uncertainty.  
To quantify it, expand $V_{\mathrm{eq}}$ around the mean scaled free energy $\beta\overline{F}$:
\begin{equation}
V_{\mathrm{eq}}[\beta F]
=
V_{\mathrm{eq}}[\beta\overline{F}]
+
\langle J,\,\beta F-\beta\overline{F}\rangle
+
\mathcal{O}\!\bigl(\|\beta F-\beta\overline{F}\|^{2}\bigr),
\label{eq:Veq-lin}
\end{equation}
with  
\[
\langle J,\,\beta F-\beta\overline{F}\rangle=\int dV\,J(V)\,[\beta F(V)-\beta\overline{F}(V)]
\] 
and 
\[
J(V)= \frac{\delta V_{\mathrm{eq}}}{\delta (\beta F)(V)}\bigg|_{\beta F = \beta\overline{F}}.
\]

The $J$ is obtained by differentiating $\Phi$ with respect to $\beta F$:
\begin{equation}
\frac{\delta\Phi}{\delta V_{\mathrm{eq}}}\,
\frac{\delta V_{\mathrm{eq}}}{\delta (\beta F)(V)}
+
\frac{\delta\Phi}{\delta (\beta F)(V)} = 0.
\label{eq:dPhi}
\end{equation}
The first functional derivative evaluates to
\begin{equation}
\frac{\delta\Phi}{\delta V_{\mathrm{eq}}}
=
\left.
\frac{\partial^{2} (\beta F) }{\partial V^{2}}
\right|_{V=V_{\mathrm{eq}}},
\label{eq:dPhi_dVeq}
\end{equation}
while the second is obtained by rewriting $\Phi$ in integral form,
\[
\Phi(V_{\mathrm{eq}},\beta F)=\int d\tilde V\,
\delta\!\bigl(V_{\mathrm{eq}}-\tilde V\bigr)\,
\partial_{\tilde V}(\beta F)(\tilde V),
\]
and then performing integration by parts,
\begin{align}
\frac{\delta\Phi}{\delta (\beta F)(V)}
&=
\int d\tilde V\,
\delta\!\bigl(V_{\mathrm{eq}}-\tilde V\bigr)\,
\partial_{\tilde V}\delta\!\bigl(V-\tilde V\bigr)\nonumber\\
&=
\underbrace{\Bigl[\delta(V-\tilde V)\,\delta(V_{\mathrm{eq}}-\tilde V)\Bigr]_{-\infty}^{+\infty}}_{0}
-
\int d\tilde V\,
\delta\!\bigl(V-\tilde V\bigr)\,
\partial_{\tilde V}\delta\!\bigl(V_{\mathrm{eq}}-\tilde V\bigr)\nonumber\\
&=
-\partial_{V}\delta\!\bigl(V_{\mathrm{eq}}-V\bigr).
\label{eq:dPhi_dF}
\end{align}

Substituting \eqref{eq:dPhi_dVeq}-\eqref{eq:dPhi_dF} into \eqref{eq:dPhi} and solving for $J$ gives
\begin{equation}
J(V)=\frac{\delta V_{\mathrm{eq}}}{\delta (\beta F)(V)} \bigg|_{\beta F = \beta\overline{F}}
=
\left(
\left.
\frac{\partial^{2}(\beta\overline{F})}{\partial V^{2}} 
\right|_{V_{\mathrm{eq}}[\beta\overline{F}]}
\right)^{-1}
\partial_{V}\delta\!\bigl(V_{\mathrm{eq}}[\beta\overline{F}]-V\bigr).
\label{eq:J-final}
\end{equation}

Inserting \eqref{eq:J-final} into \eqref{eq:Veq-lin},
\begin{align}
\langle J, \beta F-\beta\overline{F}\rangle 
&=
\left(\frac{\partial^2 (\beta\overline{F})}{\partial V^2}\Big|_{V_{\mathrm eq}[\beta\overline{F}]}\right)^{-1}
\,
\left[ \int d\tilde{V} (\beta F(\tilde{V}) - \beta\overline{F}(\tilde{V})) 
\frac{\partial}{\partial \tilde{V}} \delta(V_{\mathrm eq}[\beta\overline{F}] - \tilde{V}) \right] \nonumber \\
&=
\left(\frac{\partial^2 (\beta\overline{F})}{\partial V^2}\Big|_{V_{\mathrm eq}[\beta\overline{F}]}\right)^{-1}
\,
\bigg[\big[ (\beta F(\tilde{V}) - \beta\overline{F}(\tilde{V})) \delta(V_{\mathrm eq}[\beta\overline{F}] - \tilde{V})\big]_{-\infty}^{+\infty} \nonumber\\
&\phantom{= \left(\frac{\partial^2 (\beta\overline{F})}{\partial V^2}\Big|_{V_{\mathrm eq}[\beta\overline{F}]}\right)^{-1}} -\int d\tilde{V} \delta(V_{\mathrm eq}[\beta\overline{F}] - \tilde{V})
\frac{\partial}{\partial \tilde{V}} 
 (\beta F(\tilde{V}) - \beta\overline{F}(\tilde{V})) \bigg] \nonumber\\
&=
- \left(\frac{\partial^2 (\beta\overline{F})}{\partial V^2}\Big|_{V_{\mathrm eq}[\beta\overline{F}]}\right)^{-1}
\,
\frac{\partial(\beta F(V) - \beta\overline{F}(V))}{\partial V}\Big|_{V=V_{\mathrm eq}[\beta\overline{F}]},
\label{eq:inner-product}
\end{align}
where another integration by parts transfers the derivative from the delta-function onto $\beta F-\beta\overline{F}$, and the boundary term again vanishes.

Combining \eqref{eq:Veq-lin} and \eqref{eq:inner-product} yields the linear response of the equilibrium volume to fluctuations in the scaled free energy,
\begin{align*}
\label{eq:V_eq_derived}
V_{{\rm eq}}[\beta F] = V_{{\rm eq}}[\,\beta\overline{F}\,] 
&- 
\left(\frac{\partial^2 (\beta\overline{F})}{\partial V^2}\bigg|_{V_{\mathrm eq}[\beta\overline{F}]}\right)^{-1}
\,
\frac{\partial(\beta F(V) - \beta\overline{F}(V))}{\partial V}\bigg|_{V=V_{\mathrm eq}[\beta\overline{F}]} \\
&+ O((\beta F-\beta\overline{F})^2).
\end{align*}

Because $\beta\overline{F}$ is deterministic, only $\beta F$ contributes to the variance.  
Retaining only the linear term in the above equation---exactly as in the general linearization used to estimate the variance of a nonlinear functional \eqref{eq:linear_of_nonlinear_functional}---and taking the variance therefore gives
\begin{equation}
\mathbb{V}[V_{\mathrm eq}] \approx  \left(\frac{\partial^2 (\beta\overline{F})}{\partial V^2}\bigg|_{V_{\mathrm eq}[\beta\overline{F}]}\right)^{-2} 
\,
\mathbb{V}\left[\frac{\partial (\beta F)}{\partial V}\bigg|_{V_{\mathrm eq}[\beta\overline{F}]}\right].
\label{eq:V_eq_approx}
\end{equation}

\subsection{Variance of melting volume difference}
\label{appendix:vol_melt_variance}

We want to estimate the variance of the functional
\[
Q_1(V_{\mathrm eq}[\beta F], T_{\mathrm m}) := V_{\mathrm eq}[\beta F](T_{\mathrm m}),
\]
where $T_{\mathrm m}$ is a normal random variable with mean $\overline{T}_{\mathrm m} = \mathbb{E}[T_{\mathrm m}]$ and $\beta F$ is a Gaussian process with mean $\beta\overline{F} = \mathbb{E}[\beta F]$.  
Linearizing around the mean $\overline{T}_{\mathrm m}$ gives:
\[
Q_1 = V_{\mathrm eq}[\beta F](\overline{T}_{\mathrm m}) 
+ 
\left.
\frac{\partial V_{\mathrm eq}[\beta F]}{\partial T}\right|_{\overline{T}_{\mathrm m}} (T_{\mathrm m} - \overline{T}_{\mathrm m}) 
+ 
\mathcal{O}\bigl(|T_{\mathrm m} - \overline{T}_{\mathrm m}|^{2}\bigr).
\]
Linearizing around the mean $(\overline{T}_{\mathrm m}, \beta\overline{F})$ gives:
\begin{align*} 
Q_1 &= V_{\mathrm eq}[\beta\overline{F}](\overline{T}_{\mathrm m}) 
+ 
\left.
\langle J_1, \beta F-\beta\overline{F} \rangle \right|_{\overline{T}_{\mathrm m}}
+ 
\left.
\frac{\partial V_{\mathrm eq}[\beta\overline{F}]}{\partial T}\right|_{\overline{T}_{\mathrm m}} (T_{\mathrm m} - \overline{T}_{\mathrm m}) \\
&\quad + 
\underbrace{\left.\frac{\partial \langle J_1, \beta F-\beta\overline{F} \rangle}{\partial T} \right|_{\overline{T}_{\mathrm m}} (T_{\mathrm m} - \overline{T}_{\mathrm m})}_{\mathcal{O}\bigl(\|\beta F-\beta\overline{F}\|\,|T_{\mathrm m} - \overline{T}_{\mathrm m}|\bigr)}
+ 
\mathcal{O}\bigl(\|\beta F-\beta\overline{F}\|^{2} + |T_{\mathrm m} - \overline{T}_{\mathrm m}|^{2}\bigr) \\ 
&= 
V_{\mathrm eq}[\beta\overline{F}](\overline{T}_{\mathrm m}) 
+ 
\left. \langle J_1, \beta F-\beta\overline{F} \rangle \right|_{\overline{T}_{\mathrm m}} 
+ 
\left. \frac{\partial V_{\mathrm eq}[\beta\overline{F}]}{\partial T} \right|_{\overline{T}_{\mathrm m}} (T_{\mathrm m} - \overline{T}_{\mathrm m}) \\
&\quad + 
\mathcal{O}\bigl(\|\beta F-\beta\overline{F}\|^{2} + |T_{\mathrm m} - \overline{T}_{\mathrm m}|^{2} + \|\beta F-\beta\overline{F}\|\,|T_{\mathrm m} - \overline{T}_{\mathrm m}|\bigr).
\end{align*}
Retaining only the linear terms in the above equation and noting that $\mathrm{Cov}[T_{\mathrm m}, \beta F] = 0$, we have therefore
\[
\mathbb{V}[Q_1] \approx \mathbb{V}[{V_{\mathrm eq}[\beta\overline{F}]}(T_\mathrm{m})] 
+
\left( \left.\frac{\partial V_{\mathrm eq}[\beta\overline{F}]}{\partial T} \right|_{\overline{T}_{\mathrm m}}\right)^2 \mathbb{V}[T_{\mathrm m}], 
\]
where we used \eqref{eq:V_eq_approx} for the variance of the equilibrium volume.

\subsection{Variance of the melting enthalpy difference}
\label{appendix:enthalpy_melt_variance}

The enthalpy $H$ at temperature $T$ and zero pressure is evaluated from the free energy via
\begin{equation*}
    H(T)= F(T, V_{\rm eq}) - T \left. \frac{\partial F}{\partial T} \right|_{V = V_{\rm eq}(T)},
\end{equation*}
where $\partial F/ {\partial T}$, using $\beta F = F/T$, can be written as
\begin{equation*}
\frac{\partial F}{\partial T}
= \frac{\partial (T \, \beta F)}{\partial T}
= T \frac{\partial (\beta F)}{\partial T} + \beta F.
\end{equation*}

Therefore the enthalpy equation can be simplified to
\begin{equation*}
H(T) = -T^2 \left. \frac{\partial (\beta F)}{\partial T} \right|_{V = V_{\rm eq}(T)}.
\end{equation*}

We consider the auxiliary functional
\begin{equation*}
Q_2[\beta F, V_{\mathrm{eq}}[\beta F, T_{\mathrm{m}}], T_{\mathrm{m}}]
:= -\,T_{\mathrm{m}}^2 \, \frac{\partial (\beta F)}{\partial T} \bigg|_{V_{\mathrm{eq}}[\beta F](T_{\mathrm{m}})},
\end{equation*}

To simplify the notations, we denote $V_{\star} = V_{\mathrm eq}[\beta \overline{F}](\overline{T}_{\mathrm m})$, 
$\kappa = \frac{\partial^2 \beta \overline{F}}{\partial V^2}\big|_{V_{\mathrm eq}[\beta \overline{F}]}$, 
$\zeta = \frac{\partial V_{\mathrm eq}[\beta \overline{F}]}{\partial T} \big|_{\overline{T}_{\mathrm m}}$, 
$\Delta \beta F = \beta F - \beta \overline{F}$, $\Delta T_{\mathrm m} = T_{\mathrm m} - \overline{T}_{\mathrm m}$.

We first expand around the equilibrium volume $V_\star$:
\begin{align*}
Q_2
&= - T_{\mathrm{m}}^2
\Bigg[
\frac{\partial (\beta F)}{\partial T} \Big|_{V_\star}
+ \frac{\partial^{2} (\beta F)}{\partial V\,\partial T} \Big|_{V_\star}
\bigg( -\kappa^{-1} \frac{\partial}{\partial V} \Delta(\beta F) \Big|_{V_\star} + \zeta \,\Delta T_{\mathrm{m}} \bigg) \Bigg] \\
&\quad + \mathcal{O}\!\big(\|\Delta \beta F\|^2 + \Delta T^2 + \|\Delta \beta F\|\,|\Delta T|\big).
\end{align*}

Next, we linearize around $\beta \overline{F}$:
\begin{align*}
Q_2
&= - T_{\mathrm{m}}^2
\Bigg[
\frac{\partial (\beta \overline{F})}{\partial T} \Big|_{V_\star}
+ \frac{\partial}{\partial T} \Delta(\beta F) \Big|_{V_\star}
- \kappa^{-1} \frac{\partial^{2} (\beta \overline{F})}{\partial V\,\partial T} \Big|_{V_\star}
\frac{\partial}{\partial V} \Delta(\beta F) \Big|_{V_\star} \nonumber \\
&\quad
+ \zeta \frac{\partial^{2} (\beta \overline{F})}{\partial V\,\partial T} \Big|_{V_\star} \,\Delta T_{\mathrm{m}} 
\Bigg]
+ \mathcal{O}\!\big(\|\Delta \beta F\|^2 + \Delta T^2 + \|\Delta \beta F\|\,|\Delta T|\big)
.
\end{align*}

And, finally, linearizing around $\overline{T}_{\mathrm m}$ gives:
\begin{align*}
Q_2
&=
-\overline{T}_{\mathrm m}^2
\Bigg[
      \frac{\partial (\beta \overline{F})}{\partial T}(V_{\star}, \overline{T}_{\mathrm m})
      + \left( \partial_T 
      - \kappa^{-1} \frac{\partial^{2} (\beta \overline{F})}{\partial V\,\partial T}(V_{\star}, \overline{T}_{\mathrm m}) \partial_V \right) 
        \Delta (\beta F)\Big|_{(V_{\star}, \overline{T}_{\mathrm m})}
\Bigg] \\
&\quad
- \overline{T}_{\mathrm m}^2
\left(
      \frac{\partial^2 (\beta \overline{F})}{\partial T^2}(V_{\star}, \overline{T}_{\mathrm m})
      + \zeta \frac{\partial^{2} (\beta \overline{F})}{\partial V\,\partial T}(V_{\star}, \overline{T}_{\mathrm m})
\right) \Delta T_{\mathrm m} \\
&\quad
- 2 \overline{T}_{\mathrm m} \left( \frac{\partial (\beta \overline{F})}{\partial T}(V_{\star}, \overline{T}_{\mathrm m}) \right) \, \Delta T_{\mathrm m}
+ \mathcal{O}\!\Bigl(
        \|\Delta (\beta F)\|^{2}
        + |\Delta T_{\mathrm m}|^{2}
        + \|\Delta (\beta F)\|\,|\Delta T_{\mathrm m}|
    \Bigr).
\end{align*}

Retaining only the linear terms in the above equation and noting that $\text{Cov}[T_{\mathrm m}, F] = 0$, we have therefore
\begin{align*}
\mathbb{V}[Q_2]
&\approx 
\overline{T}_{\mathrm{m}}^4
\mathbb{V}\left[\left( \partial_T 
- \kappa^{-1} \frac{\partial^{2}(\beta \overline{F})}{\partial V\,\partial T}(V_{\star}) \partial_V \right) (\beta F(V)) \big|_{(V_{\star}, \overline{T}_{\mathrm m})} \right]\\
&+
\left(\overline{T}^2_{\mathrm m} \frac{\partial^2 (\beta \overline{F})}{\partial T^2}(V_{\star}, \overline{T}_{\mathrm m}) 
+ \overline{T}^2_{\mathrm m} \zeta \frac{\partial^{2}(\beta \overline{F})}{\partial V\,\partial T}(V_{\star}, \overline{T}_{\mathrm m}) 
+ 2 \overline{T}_{\mathrm m} \frac{\partial (\beta \overline{F})}{\partial T}(V_{\star}, \overline{T}_{\mathrm m})
\right)^2 \mathbb{V} [T_{\mathrm{m}}].
\end{align*}

\subsection{Variance of Thermal Expansion}
\label{appendix:alpha_variance}

We want to estimate the variance of the auxiliary functional
\[
Q_\alpha[\beta F, V_{\mathrm eq}[\beta F]] := 
-\frac{1}{3}
\underbrace{(V_\mathrm{eq}[\beta F])^{-1}}_{=A_1}
\underbrace{\frac{\partial^{2}(\beta F)}{\partial T\,\partial V}(V_{\mathrm{eq}}[\beta F])}_{=A_2}
\underbrace{\left(\frac{\partial^{2}(\beta F)}{\partial V^{2}}(V_{\mathrm{eq}}[\beta F])\right)^{-1}}_{=A_3}.
\]

From previous computations, we have:
\[
\langle J,\, \beta F - \beta \overline{F} \rangle 
= - \kappa^{-1} \frac{\partial}{\partial V} (\beta F - \beta \overline{F}) \Big|_{V_{\star}}.
\]

For the equilibrium volume:
\begin{align*}
V_{{\rm eq}}[\beta F] 
&= V_\star
- \left( \frac{\partial^2 (\beta\overline{F})}{\partial V^2} \Big|_{V_\star} \right)^{-1}
  \frac{\partial(\beta F - \beta \overline{F})}{\partial V} \Big|_{V_\star} + O\big((\beta F - \beta \overline{F})^2\big).
\end{align*}

Therefore term $A_1$:
\begin{align*}
A_1 
&= (V_\mathrm{eq}[\beta F])^{-1} 
= (V_{\star} + \delta V_\mathrm{eq}[\beta F])^{-1} \\
&= \frac{1}{V_{\star}}\left(1 - \frac{\delta V_\mathrm{eq}[\beta F]}{V_{\star}} \right)
+ \mathcal{O}\!\big(|\delta V_\mathrm{eq}[\beta F]|^2\big) \\ 
&= \frac{1}{V_{\star}}
+ \frac{\kappa^{-1}}{V_{\star}^2}
  \frac{\partial}{\partial V} (\beta F - \beta \overline{F}) \Big|_{V_{\star}}
+ \mathcal{O}\!\big(\|\beta F-\beta \overline{F}\|^{2}\big).
\end{align*}

Term $A_2$:
\begin{align*}
A_2 
&= \frac{\partial^{2}(\beta F)}{\partial T\,\partial V}(V_{\mathrm{eq}}[\beta F])
= \frac{\partial^{2}(\beta F)}{\partial T\,\partial V}(V_{\star} + \delta V_\mathrm{eq}[\beta F]) \\
&= \frac{\partial^{2}(\beta F)}{\partial T\,\partial V}(V_{\star})
+ \frac{\partial^{3}(\beta F)}{\partial T\,\partial^2 V}(V_{\star})\delta V_\mathrm{eq}[\beta F] 
+ \mathcal{O}\!\Bigl(|\delta V_\mathrm{eq}[\beta F]|^2\Bigr) \\
&= \underbrace{\frac{\partial^{2}(\beta F)}{\partial T\,\partial V}(V_{\star})}_{=\tilde{A}_2}
+ \frac{\partial^{3}(\beta \overline{F})}{\partial T\,\partial^2 V}(V_{\star})
\langle J,\,\beta F-\beta \overline{F}\rangle 
+ \mathcal{O}\!\bigl(\|\beta F-\beta \overline{F}\|^{2}\bigr)
\end{align*}
\begin{align*}
\tilde{A}_2 
&= \frac{\partial^{2}(\beta \overline{F})}{\partial T\,\partial V}(V_{\star})
+ \left\langle 
\frac{\delta }{\delta \beta F(\tilde{V}, \tilde{T})} 
\frac{\partial^{2}(\beta  F)}{\partial T\,\partial V}(V_{\star}, T_{\star}), 
\beta F - \beta \overline{F}
\right\rangle \\
&= \frac{\partial^{2}(\beta \overline{F})}{\partial T\,\partial V}(V_{\star})
+ \frac{\partial^2}{\partial V \partial T} 
  \big(\beta F(T, V) - \beta \overline{F}(T, V)\big) \Big|_{V_\star, T_\star}  
\end{align*}

Therefore,
\begin{align*}
A_2 
&= \frac{\partial^{2}(\beta \overline{F})}{\partial T\,\partial V}\Big|_{V_{\star}} \\
&\quad + \left[ \frac{\partial^2}{\partial V \partial T} 
- \kappa^{-1} \frac{\partial^{3}(\beta \overline{F})}{\partial T\,\partial^2 V} \Big|_{V_{\star}} 
  \frac{\partial}{\partial V} \right]
  \big(\beta F - \beta \overline{F}\big) \Big|_{V_\star, T_\star} \\
&\quad + \mathcal{O}\!\big(\|\beta F-\beta \overline{F}\|^{2}\big).
\end{align*}

Term $A_3$:
\begin{align*}
A_3 
&= \left(\frac{\partial^{2} \beta F}{\partial V^{2}}(V_{\mathrm{eq}}[\beta F])\right)^{-1} \\
&= \kappa^{-1}
\left[ 1 - \kappa^{-1} \delta\left(\frac{\partial^{2} \beta F}{\partial V^{2}}(V_{\mathrm{eq}}[\beta F])\right) +
\mathcal{O}\!\bigl(|\delta(\ldots)|^{2}\bigr) \right] \\
&= \kappa^{-1} 
- \kappa^{-2}\delta\bigg( \underbrace{\frac{\partial^{2} \beta F}{\partial V^{2}}(V_{\mathrm{eq}}[\beta F])}_{=\tilde{A_3}}\bigg) 
+ \mathcal{O}\!\bigl(|\delta(\ldots)|^{2}\bigr) 
\end{align*}

\begin{align*}
\tilde{A}_3 
&= \frac{\partial^{2} \beta F}{\partial V^{2}}(V_{\mathrm{eq}}[\beta F]) 
= \frac{\partial^{2} \beta F}{\partial V^{2}}(V_{\star} + \delta V_\mathrm{eq}[\beta F]) \\
&= \frac{\partial^{2} \beta F}{\partial V^{2}}(V_{\star}) 
+ \frac{\partial^{3} \beta F}{\partial V^{3}}(V_{\star})\delta V_\mathrm{eq}[\beta F] 
+ \mathcal{O}\!\Bigl(|\delta V_\mathrm{eq}[\beta F]|^2\Bigr) \\
&= \frac{\partial^{2} \beta \overline{F}}{\partial V^{2}}\Big|_{V_{\star}} 
+ \frac{\partial^{2}}{\partial V^{2}} (\beta F - \beta \overline{F}) \Big|_{V_{\star}} \\
&\phantom{=} \quad 
+ \frac{\partial^{3} \beta \overline{F}}{\partial V^{3}}\Big|_{V_{\star}}
\langle J,\,\beta F-\beta \overline{F}\rangle 
+ \mathcal{O}\!\bigl(\|\beta F-\beta \overline{F}\|^{2}\bigr) \\
&= \frac{\partial^{2} \beta \overline{F}}{\partial V^{2}}\Big|_{V_{\star}} 
+ \left[ 
\frac{\partial^{2}}{\partial V^{2}}
- \kappa^{-1} \frac{\partial^{3} \beta \overline{F}}{\partial V^{3}}\Big|_{V_{\star}}
\frac{\partial}{\partial V}
\right]
\big(\beta F - \beta \overline{F}\big) \Big|_{V_\star, T_\star} \\
&\phantom{=} \quad + \mathcal{O}\!\bigl(\|\beta F-\beta \overline{F}\|^{2}\bigr)
\end{align*}

Therefore,
\begin{align*}
A_3 
&= \kappa^{-1} 
+ \kappa^{-2} \left[ -\frac{\partial^{2}}{\partial V^{2}}
+ \kappa^{-1} \frac{\partial^{3} (\beta \overline{F})}{\partial V^{3}}\Big|_{V_{\star}}
  \frac{\partial }{\partial V} \right]
  \big(\beta F - \beta \overline{F}\big) \Big|_{V_\star, T_\star} \\
&\phantom{=} \quad + \mathcal{O}\!\big(\|\beta F-\beta \overline{F}\|^{2}\big).
\end{align*}

We therefore have:
\begin{align*}
Q_\alpha[\beta F, V_{\mathrm eq}[\beta F]] &= 
-\frac13 
\left(
\frac{1}{V_{\star}}
+ \frac{\kappa^{-1}}{V_{\star}^2}
  \frac{\partial}{\partial V} \Delta (\beta F) 
\right) \, \\
&\phantom{=\frac13}
\,
\left(
\frac{\partial^{2}(\beta \overline{F})}{\partial T\,\partial V}\Big|_{V_{\star}}
+ \left[ \frac{\partial^2}{\partial V \partial T} 
- \kappa^{-1} \frac{\partial^{3}(\beta \overline{F})}{\partial T\,\partial^2 V} \Big|_{V_{\star}} 
  \frac{\partial}{\partial V} \right] \Delta (\beta F) 
\right) \, \\
&\phantom{=\frac13}
\,
\left(
\kappa^{-1} 
+ \kappa^{-2} \left[ -\frac{\partial^{2}}{\partial V^{2}}
+ \kappa^{-1} \frac{\partial^{3} (\beta \overline{F})}{\partial V^{3}}\Big|_{V_{\star}}
  \frac{\partial }{\partial V} \right] \Delta (\beta F) 
\right) \\
&+ \mathcal{O}\!\bigl(\|\beta F-\beta\overline{F}\|^{2}\bigr)
\end{align*}

Finally for variance of thermal expansion,
\begin{align*}
\mathbb{V}[Q_\alpha] 
&\approx \frac19 \mathbb{V}
\Bigg[
\frac{\kappa^{-1}}{V_\star} \left( \frac{\partial^2}{\partial V \partial T} 
- \kappa^{-1} \frac{\partial^{3}(\beta \overline{F})}{\partial T \, \partial^{2} V} \Big|_{V_{\star}} \frac{\partial}{\partial V} \right) (\beta F) \Big|_{V_\star, T_\star} \\
&\quad + \frac{\kappa^{-2}}{V_\star} \frac{\partial^{2}(\beta \overline{F})}{\partial T \, \partial V} \Big|_{V_{\star}}
 \left( -\frac{\partial^{2}}{\partial V^{2}}
+ \kappa^{-1} \frac{\partial^{3} (\beta \overline{F})}{\partial V^{3}}\Big|_{V_{\star}}
\frac{\partial }{\partial V} \right) (\beta F) \Big|_{V_\star, T_\star} \\
&\quad + \frac{\kappa^{-2}}{V_\star^2} \frac{\partial^{2}(\beta \overline{F})}{\partial T \, \partial V} \Big|_{V_{\star}} \frac{\partial }{\partial V} (\beta F) \Big|_{V_\star, T_\star}
\Bigg].
\end{align*}

\subsection{Variance of Heat Capacity}
\label{appendix:cp_variance}

We define the auxiliary functional
\[
Q_C[\beta F, V_{\mathrm{eq}}[\beta F]] := 
- T \left( \frac{\partial^2 F}{\partial T^2} \right)_{V = V_{\mathrm{eq}}[\beta F]}
- T \left( \frac{\partial^2 F}{\partial T \, \partial V} \right)_{V = V_{\mathrm{eq}}[\beta F]} \, \frac{dV_{\mathrm{eq}}}{dT}.
\]

From previous computations, we have
\[
\frac{dV_{\mathrm{eq}}}{dT} = - A_2 \, A_3,
\]
with
\begin{align*}
A_2 
&= \frac{\partial^{2}(\beta \overline{F})}{\partial T\,\partial V}\bigg|_{V_{\star}} 
+ \left[ \frac{\partial^2}{\partial V \partial T} 
- \kappa^{-1} \frac{\partial^{3}(\beta \overline{F})}{\partial T\,\partial^2 V} \bigg|_{V_{\star}} 
  \frac{\partial}{\partial V} \right]
  \Delta (\beta F) + \mathcal{O}\!\big(\|\Delta (\beta F)\|^{2}\big); \\
A_3 
&= \kappa^{-1} 
+ \kappa^{-2} \left[ -\frac{\partial^{2}}{\partial V^{2}}
+ \kappa^{-1} \frac{\partial^{3} (\beta \overline{F})}{\partial V^{3}}\bigg|_{V_{\star}}
  \frac{\partial }{\partial V} \right]
  \Delta (\beta F) + \mathcal{O}\!\big(\|\Delta (\beta F)\|^{2}\big).
\end{align*}

Their product becomes:
\begin{align*}
\frac{dV_{\mathrm{eq}}}{dT}  
&= - \kappa^{-1} \frac{\partial^{2}(\beta \overline{F})}{\partial T\,\partial V}\bigg|_{V_{\star}} \\
&\quad - \Bigg[\kappa^{-1} \left( \frac{\partial^2}{\partial V \partial T} 
- \kappa^{-1} \frac{\partial^{3}(\beta \overline{F})}{\partial T\,\partial^2 V} \bigg|_{V_{\star}} 
  \frac{\partial}{\partial V} \right) \\
&\quad + \kappa^{-2} \frac{\partial^{2}(\beta \overline{F})}{\partial T\,\partial V}\bigg|_{V_{\star}} 
\left( -\frac{\partial^{2}}{\partial V^{2}}
+ \kappa^{-1} \frac{\partial^{3} (\beta \overline{F})}{\partial V^{3}}\bigg|_{V_{\star}}
  \frac{\partial }{\partial V} \right) \Bigg] \Delta (\beta F) \\
&\quad + \mathcal{O}(\|\Delta F\|^2).
\end{align*}

We now expand the derivative terms
\[
\frac{\partial^2 F}{\partial T^2}, 
\quad
\frac{\partial^2 F}{\partial T \, \partial V}
\]
appearing in $Q_C[\beta F]$ around the equilibrium volume 
$V = V_\star + \delta V_{\mathrm{eq}}[\beta F]$, 
retaining only the terms linear in $\delta V_{\mathrm{eq}}$.
Since $F = T \, \beta F$, these derivatives can be expressed as
\[
\frac{\partial^{2}F}{\partial T^{2}} = \frac{\partial^{2}(T \, \beta F)}{\partial T^{2}} = T \frac{\partial^{2}(\beta F)}{\partial T^{2}} + 2\frac{\partial(\beta F)}{\partial T}
\]
\[
\frac{\partial^{2}F}{\partial T\, \partial V} = \frac{\partial^{2}(T \, \beta F)}{\partial T \, \partial V} = T \underbrace{\frac{\partial^{2}(\beta F)}{\partial T \, \partial V}}_{=A_2} + \frac{\partial(\beta F)}{\partial V}
\]

Proceeding in the same way as for $A_2$ and $A_3$ in the previous section,  
the derivatives becomes
\begin{align*}
\frac{\partial^{2}(\beta F)}{\partial T^{2}}(V_{\mathrm{eq}}[\beta F])
&= \frac{\partial^{2}(\beta F)}{\partial T^{2}}(V_{\star} + \delta V_{\mathrm{eq}}[\beta F]) \\
&= \frac{\partial^{2}(\beta \overline{F})}{\partial T^2}\bigg|_{V_{\star}} 
+ \left[ \frac{\partial^2}{\partial T^2} 
- \kappa^{-1} \frac{\partial^{3}(\beta \overline{F})}{\partial T^2 \,\partial V} \bigg|_{V_{\star}} 
  \frac{\partial}{\partial V} \right]
  \Delta (\beta F) \\
&+ \mathcal{O}\!\big(\|\Delta (\beta F)\|^{2}\big);
\end{align*}

\begin{align*}
\frac{\partial(\beta F)}{\partial T}(V_{\mathrm{eq}}[\beta F])
&= \frac{\partial(\beta F)}{\partial T}(V_{\star} + \delta V_{\mathrm{eq}}[\beta F]) \\
&= \frac{\partial(\beta \overline{F})}{\partial T}\bigg|_{V_{\star}} 
+ \left[ \frac{\partial}{\partial T} 
- \kappa^{-1} \frac{\partial^{2}(\beta \overline{F})}{\partial T \,\partial V} \bigg|_{V_{\star}} 
  \frac{\partial}{\partial V} \right]
  \Delta (\beta F) \\
&+ \mathcal{O}\!\big(\|\Delta (\beta F)\|^{2}\big);
\end{align*}

\begin{align*}
\frac{\partial(\beta F)}{\partial V}(V_{\mathrm{eq}}[\beta F])
&= \frac{\partial(\beta F)}{\partial V}(V_{\star} + \delta V_{\mathrm{eq}}[\beta F]) 
= \frac{\partial(\beta \overline{F})}{\partial V}\bigg|_{V_{\star}} \\
&+ \Bigg[ \frac{\partial}{\partial V} 
- \kappa^{-1} \underbrace{\frac{\partial^{2}(\beta \overline{F})}{\partial V^2} \bigg|_{V_{\star}} }_{=\kappa}
  \frac{\partial}{\partial V} \Bigg]
  \Delta (\beta F) + \mathcal{O}\!\big(\|\Delta (\beta F)\|^{2}\big) \\
&= \frac{\partial(\beta \overline{F})}{\partial V}\bigg|_{V_{\star}} 
+ \mathcal{O}\!\big(\|\Delta (\beta F)\|^{2}\big).
\end{align*}

Combining these results, the full expansion of $\partial^2 F / \partial T^2$ at $V_{\mathrm{eq}}$ is
\begin{align*}
\frac{\partial^{2}F}{\partial T^{2}}(V_{\mathrm{eq}}[\beta F])
&= \frac{\partial^2\overline{F}}{\partial T^2}\bigg|_{V_{\star}} 
+ T \left[ \frac{\partial^2}{\partial T^2} 
- \kappa^{-1} \frac{\partial^{3}(\beta \overline{F})}{\partial T^2 \,\partial V} \bigg|_{V_{\star}} 
  \frac{\partial}{\partial V} \right]
  \Delta (\beta F) \\
&\phantom{= \frac{\partial^2\overline{F}}{\partial T^2}\bigg|_{V_{\star}}} \,
+ 2 \left[ \frac{\partial}{\partial T} 
- \kappa^{-1} \frac{\partial^{2}(\beta \overline{F})}{\partial T \,\partial V} \bigg|_{V_{\star}} 
  \frac{\partial}{\partial V} \right]
  \Delta (\beta F) \\
&\phantom{= \frac{\partial^2\overline{F}}{\partial T^2}\bigg|_{V_{\star}}} \,
+ \mathcal{O}\!\big(\|\Delta (\beta F)\|^{2}\big),
\end{align*}
while the mixed derivative expansion reads
\begin{align*}
\frac{\partial^{2}F}{\partial T \, \partial V}(V_{\mathrm{eq}}[\beta F])
&= \frac{\partial^2\overline{F}}{\partial T \, \partial V}\bigg|_{V_{\star}} 
+ T \left[ \frac{\partial^2}{\partial V \partial T} 
- \kappa^{-1} \frac{\partial^{3}(\beta \overline{F})}{\partial T\,\partial^2 V} \bigg|_{V_{\star}} 
  \frac{\partial}{\partial V} \right]
  \Delta (\beta F) \\
&\phantom{=\frac{\partial^2\overline{F}}{\partial T \, \partial V}\bigg|_{V_{\star}}} \,
+ \mathcal{O}\!\big(\|\Delta (\beta F)\|^{2}\big).
\end{align*}

Substituting everything into the definition of $Q_C$, we obtain:
\begin{align*}
Q_C[\beta F, V_{\mathrm{eq}}[\beta F]] =\; 
& - T_\star \Bigg\{ \frac{\partial^2\overline{F}}{\partial T^2}\bigg|_{V_{\star}} 
- \kappa^{-1} \frac{\partial^{2}(\beta \overline{F})}{\partial T\,\partial V}\bigg|_{V_{\star}} 
\frac{\partial^2\overline{F}}{\partial T \, \partial V}\bigg|_{V_{\star}} \\
& + T_\star \left[ \frac{\partial^2}{\partial T^2} 
- \kappa^{-1} \frac{\partial^{3}(\beta \overline{F})}{\partial T^2 \,\partial V} \bigg|_{V_{\star}} 
  \frac{\partial}{\partial V} \right]
  \Delta (\beta F) \\
& + 2 \left[ \frac{\partial}{\partial T} 
- \kappa^{-1} \frac{\partial^{2}(\beta \overline{F})}{\partial T \,\partial V} \bigg|_{V_{\star}} 
  \frac{\partial}{\partial V} \right]
  \Delta (\beta F) \\
& - \frac{\partial^2\overline{F}}{\partial T \, \partial V}\bigg|_{V_{\star}}  
\Bigg[
\kappa^{-1} \left( \frac{\partial^2}{\partial V \partial T} 
- \kappa^{-1} \frac{\partial^{3}(\beta \overline{F})}{\partial T\,\partial^2 V} \bigg|_{V_{\star}} 
  \frac{\partial}{\partial V} \right) \\
&\quad + \kappa^{-2} \frac{\partial^{2}(\beta \overline{F})}{\partial T\,\partial V}\bigg|_{V_{\star}} 
\left( -\frac{\partial^{2}}{\partial V^{2}}
+ \kappa^{-1} \frac{\partial^{3} (\beta \overline{F})}{\partial V^{3}}\bigg|_{V_{\star}}
  \frac{\partial }{\partial V} \right) 
\Bigg] \Delta (\beta F) \\
&- T \kappa^{-1} \frac{\partial^{2}(\beta \overline{F})}{\partial T\,\partial V}\bigg|_{V_{\star}} 
\left( \frac{\partial^2}{\partial V \partial T} 
- \kappa^{-1} \frac{\partial^{3}(\beta \overline{F})}{\partial T\,\partial^2 V} \bigg|_{V_{\star}} 
  \frac{\partial}{\partial V} \right) \Delta (\beta F) \Bigg\} \\
& + \mathcal{O}\!\big(\|\Delta (\beta F)\|^{2}\big).
\end{align*}

\subsection{Variance of Isothermal Bulk Modulus}
\label{appendix:kt_variance}

We define the auxiliary functional
\[
Q_K[\beta F, V_{\mathrm eq}[\beta F]] := 
V_{\mathrm eq}[\beta F] \, \frac{\partial^2 F}{\partial V^2}\big(V_{\mathrm eq}[\beta F]\big).
\]

We now expand both factors in $Q_K[F]$ about $V_\star$ to linear order in $\Delta F$. First, the volume factor is:
\begin{align*}
V_{{\rm eq}}[\beta F] 
&= V_\star - \kappa^{-1}
\frac{\partial}{\partial V} \Delta (\beta F)  \Big|_{V_\star} 
+ \mathcal{O}\!\big(\|\Delta (\beta F)\|^{2}\big).
\end{align*}

The second factor is the second derivative of $F$:
\begin{align*}
\frac{\partial^{2}F}{\partial V^{2}}(V_{\mathrm{eq}}[\beta F])
&= T \frac{\partial^{2}(\beta F)}{\partial V^{2}}(V_{\star} + \delta V_{\mathrm{eq}}[\beta F]) \\
&= \frac{\partial^{2}\overline{F}}{\partial V^2}\bigg|_{V_{\star}} 
+ T_\star \left[ \frac{\partial^2}{\partial V^2} 
- \kappa^{-1} \frac{\partial^{3}(\beta \overline{F})}{\partial V^3} \bigg|_{V_{\star}} 
  \frac{\partial}{\partial V} \right]
  \Delta (\beta F) \\
&+ \mathcal{O}\!\big(\|\Delta (\beta F)\|^{2}\big);
\end{align*}

Combining both, the total linearized expression for $Q_K$ becomes:
\begin{align*}
Q_K[\beta F, V_{\mathrm eq}[\beta F]]
&= V_\star \frac{\partial^{2}\overline{F}}{\partial V^2}\bigg|_{V_{\star}} - 
\kappa^{-1} \frac{\partial^{2}\overline{F}}{\partial V^2}\bigg|_{V_{\star}} \frac{\partial}{\partial V} \Delta (\beta F)  \Big|_{V_\star} \\
&\, + T_\star V_\star \left[ \frac{\partial^2}{\partial V^2} 
- \kappa^{-1} \frac{\partial^{3}(\beta \overline{F})}{\partial V^3} \bigg|_{V_{\star}} 
  \frac{\partial}{\partial V} \right]
  \Delta (\beta F) \\
&+ \mathcal{O}\!\big(\|\Delta (\beta F)\|^{2}\big);
\end{align*}

Therefore, the variance of the bulk modulus is given by
\begin{align*}
\mathbb{V}[K_T] \approx 
\mathbb{V} \Bigg[ 
\Bigg( T_\star V_\star \left( \frac{\partial^2}{\partial V^2}
- \kappa^{-1} \frac{\partial^{3}(\beta \overline{F})}{\partial V^3} \bigg|_{V_{\star}} 
\frac{\partial}{\partial V} \right)
- \kappa^{-1} \frac{\partial^{2}\overline{F}}{\partial V^2}\bigg|_{V_{\star}} \frac{\partial}{\partial V}
\Bigg) \Delta (\beta F)  \Big|_{V_\star} 
\Bigg].
\end{align*}

\subsection{Variance of Adiabatic Bulk Modulus}
\label{appendix:ks_variance}

We define the adiabatic bulk modulus as
\[
Q_S[\beta F, V_{\mathrm eq}[\beta F]]:= K_S(T) = \frac{K_T[F, V_\mathrm{eq}[F]] \, C_P[F, V_\mathrm{eq}[F]]}{C_V[F, V_\mathrm{eq}[F]]},
\]
where all functionals are evaluated at the temperature-dependent equilibrium volume $V_\mathrm{eq}[F]$. 

From previous derivations, the linearized expressions for each quantity read:
\begin{align*}
K_T[\beta F] &= K_T^\star + \delta K_T[\beta F], \\
C_V[\beta F] &= C_V^\star + \delta C_V[\beta F], \\
C_P[\beta F] &= C_P^\star + \delta C_P[\beta F], \\
\end{align*}
where
\begin{align*}
K_T^\star &= T_\star V_\star \frac{\partial^2}{\partial V^2}, 
\quad C_V^\star = - T_\star \frac{\partial^2\overline{F}}{\partial T^2}\bigg|_{V_{\star}} , \\ 
C_P^\star &= - T_\star \Bigg( \frac{\partial^2\overline{F}}{\partial T^2}\bigg|_{V_{\star}} 
- \kappa^{-1} \frac{\partial^{2}(\beta \overline{F})}{\partial T\,\partial V}\bigg|_{V_{\star}} 
\frac{\partial^2\overline{F}}{\partial T \, \partial V}\bigg|_{V_{\star}} \Bigg)
\end{align*}
We compute the linearized variation of the product:
\begin{align*}
K_S[\beta F] &= \frac{K_T[\beta F] \, C_P[\beta F]}{C_V[\beta F]} 
= \frac{K_T^\star C_P^\star}{C_V^\star} + \delta K_S[\beta F] + \mathcal{O}(\|\Delta (\beta F)\|^2),
\end{align*}
with
\begin{align*}
\delta K_S[\beta F] &= \frac{C_P^\star}{C_V^\star} \, \delta K_T[\beta F]
+ \frac{K_T^\star}{C_V^\star} \, \delta C_P[\beta F] 
- \frac{K_T^\star C_P^\star}{(C_V^\star)^2} \, \delta C_V[\beta F].
\end{align*}

\bibliographystyle{unsrt}
\bibliography{bibliography}